\documentclass[apjl,ip,twocolumn]{aastex61}
\usepackage{comment}
\usepackage{ifthen}
\usepackage[switch]{lineno}
\modulolinenumbers[5]

\newcommand{\forloop}[5][1]%
{%
\setcounter{#2}{#3}%
\ifthenelse{#4}%
	{%
	#5%
	\addtocounter{#2}{#1}%
	\forloop[#1]{#2}{\value{#2}}{#4}{#5}%
	}%
	{%
	}%
}%

\newcommand{\ctbd}[1]{}

\newcommand{\Lc}{Light curve}

\newcommand{\masy}{\ensuremath{\rm mas\,yr^{-1}}}
\newcommand{\kms}{\ensuremath{\rm km\,s^{-1}}}
\newcommand{\ms}{\ensuremath{\rm m\,s^{-1}}}

\newcommand{\gcmc}{\ensuremath{\rm g\,cm^{-3}}}

\newcommand{\teff}{\ensuremath{T_{\rm eff}}}
\newcommand{\logg}{\ensuremath{\log{g}}}
\newcommand{\vsini}{\ensuremath{v \sin{i}}}
\newcommand{\feh}{\ensuremath{\rm [Fe/H]}}

\newcommand{\vmac}{\ensuremath{v_{\rm mac}}}
\newcommand{\vmic}{\ensuremath{v_{\rm mic}}}

\newcommand{\rsun}{\ensuremath{R_\sun}}
\newcommand{\msun}{\ensuremath{M_\sun}}
\newcommand{\lsun}{\ensuremath{L_\sun}}

\newcommand{\rstar}{\ensuremath{R_\star}}
\newcommand{\mstar}{\ensuremath{M_\star}}
\newcommand{\lstar}{\ensuremath{L_\star}}

\newcommand{\teffstar}{\ensuremath{T_{\rm eff\star}}}

\newcommand{\loggstar}{\ensuremath{\log{g_{\star}}}}

\newcommand{\rpl}{\ensuremath{R_{p}}}
\newcommand{\mpl}{\ensuremath{M_{p}}}

\newcommand{\rhopl}{\ensuremath{\rho_{p}}}

\newcommand{\arstar}{\ensuremath{a/\rstar}}
\newcommand{\zrstar}{\ensuremath{\zeta/\rstar}}

\newcommand{\rjup}{\ensuremath{R_{\rm J}}}
\newcommand{\mjup}{\ensuremath{M_{\rm J}}}

\newcommand{\refsecl}[1]{\mbox{Section \ref{sec:#1}}}

\newcommand{\reftabl}[1]{Table~\ref{tab:#1}}

\newcommand{\loopand}{\ifnum\value{planetcounter}=2 and \else\fi}
\newcommand{\loopcomma}{\ifnum\value{planetcounter}<2 ,\else. \fi}
\newcommand{\loopcommanoperiod}{\ifnum\value{planetcounter}<2 ,\else \space\fi}
\newcommand{\loopcommanospace}{\ifnum\value{planetcounter}<2 ,\else \fi}

\newcommand{\hatcurhtrxxxxxA}{HATS567-001}                      
\newcommand{\hatcurfieldxxxxxA}{\ensuremath{string}}            
\newcommand{\hatcurCCraxxxxxA}{\ensuremath{13^{\mathrm h}19^{\mathrm m}12.4637{\mathrm s}}}                   
\newcommand{\hatcurCCdecxxxxxA}{\ensuremath{-22{\arcdeg}59{\arcmin}12.7306{\arcsec}}}                 
\newcommand{\hatcurCCmagxxxxxA}{12.266}                         
\newcommand{\hatcurCCtwomassxxxxxA}{2MASS~13191246-2259127}     
\newcommand{\hatcurCCgscxxxxxA}{GSC~6700-00149}                 
\newcommand{\hatcurCCgaiaxxxxxA}{GAIA~6194574671813047424}      
\newcommand{\hatcurCCgaiadrtwoxxxxxA}{GAIA~DR2~6194574671813047424} 
\newcommand{\hatcurCCtassmvxxxxxA}{\ensuremath{12.266\pm0.030}} 
\newcommand{\hatcurCCtassmvshortxxxxxA}{\ensuremath{12.3}}      
\newcommand{\hatcurCCtassmBxxxxxA}{\ensuremath{13.222\pm0.060}} 
\newcommand{\hatcurCCtassmBshortxxxxxA}{\ensuremath{13.2}}      
\newcommand{\hatcurCCtassmIxxxxxA}{\ensuremath{nff\pmnff}}      
\newcommand{\hatcurCCtassmIshortxxxxxA}{\ensuremath{0.0}}       
\newcommand{\hatcurCCtassmgxxxxxA}{\ensuremath{12.733\pm0.060}} 
\newcommand{\hatcurCCtassmgshortxxxxxA}{\ensuremath{12.7}}      
\newcommand{\hatcurCCtassmrxxxxxA}{\ensuremath{11.906\pm0.030}} 
\newcommand{\hatcurCCtassmrshortxxxxxA}{\ensuremath{11.9}}      
\newcommand{\hatcurCCtassmixxxxxA}{\ensuremath{11.616\pm0.030}} 
\newcommand{\hatcurCCtassmishortxxxxxA}{\ensuremath{11.6}}      
\newcommand{\hatcurCCparallaxxxxxxA}{\ensuremath{4.692\pm0.061}} 
\newcommand{\hatcurCCgaiamGxxxxxA}{\ensuremath{11.99780\pm0.00020}} 
\newcommand{\hatcurCCgaiamBPxxxxxA}{\ensuremath{12.5309\pm0.0023}} 
\newcommand{\hatcurCCgaiamRPxxxxxA}{\ensuremath{11.3387\pm0.0017}} 
\newcommand{\hatcurCCtwomassJmagxxxxxA}{\ensuremath{10.528\pm0.024}} 
\newcommand{\hatcurCCtwomassHmagxxxxxA}{\ensuremath{10.038\pm0.022}} 
\newcommand{\hatcurCCtwomassKmagxxxxxA}{\ensuremath{9.947\pm0.021}} 
\newcommand{\hatcurCCcitJmagxxxxxA}{\ensuremath{10.534\pm0.025}} 
\newcommand{\hatcurCCcitHmagxxxxxA}{\ensuremath{10.032\pm0.023}} 
\newcommand{\hatcurCCcitKmagxxxxxA}{\ensuremath{9.971\pm0.021}} 
\newcommand{\hatcurCCbbJmagxxxxxA}{\ensuremath{10.600\pm0.027}} 
\newcommand{\hatcurCCbbHmagxxxxxA}{\ensuremath{10.054\pm0.024}} 
\newcommand{\hatcurCCbbKmagxxxxxA}{\ensuremath{9.991\pm0.022}}  
\newcommand{\hatcurCCesoJmagxxxxxA}{\ensuremath{10.605\pm0.029}} 
\newcommand{\hatcurCCesoHmagxxxxxA}{\ensuremath{10.049\pm0.027}} 
\newcommand{\hatcurCCesoKmagxxxxxA}{\ensuremath{9.989\pm0.022}} 
\newcommand{\hatcurCCesoJHmagxxxxxA}{\ensuremath{0.555\pm0.037}} 
\newcommand{\hatcurCCesoJKmagxxxxxA}{\ensuremath{0.617\pm0.036}} 
\newcommand{\hatcurCCesoHKmagxxxxxA}{\ensuremath{0.061\pm0.035}} 
\newcommand{\hatcurCCWonemagxxxxxA}{\ensuremath{9.866\pm0.022}} 
\newcommand{\hatcurCCWtwomagxxxxxA}{\ensuremath{9.942\pm0.021}} 
\newcommand{\hatcurCCWthreemagxxxxxA}{\ensuremath{9.896\pm0.047}} 
\newcommand{\hatcurLCdipxxxxxA}{\ensuremath{4.6}}               
\newcommand{\hatcurLCrprstarxxxxxA}{\ensuremath{0.0668\pm0.0015}} 
\newcommand{\hatcurLCbsqxxxxxA}{\ensuremath{0.161_{-0.026}^{+0.027}}} 
\newcommand{\hatcurLCimpxxxxxA}{\ensuremath{0.401_{-0.034}^{+0.032}}} 
\newcommand{\hatcurLCzetaxxxxxA}{\ensuremath{17.65\pm0.25}}     
\newcommand{\hatcurLCdurxxxxxA}{\ensuremath{0.1223\pm0.0014}}   
\newcommand{\hatcurLCdurshortxxxxxA}{\ensuremath{0.1223}}       
\newcommand{\hatcurLCdurhrxxxxxA}{\ensuremath{2.936\pm0.034}}   
\newcommand{\hatcurLCdurhrshortxxxxxA}{\ensuremath{2.936}}      
\newcommand{\hatcurLCqxxxxxA}{\ensuremath{0.02820\pm0.00033}}   
\newcommand{\hatcurLCqshortxxxxxA}{\ensuremath{0.028}}          
\newcommand{\hatcurLCingdurxxxxxA}{\ensuremath{0.00903\pm0.00031}} 
\newcommand{\hatcurLCPxxxxxA}{\ensuremath{4.3315379\pm0.0000039}} 
\newcommand{\hatcurLCPprecxxxxxA}{\ensuremath{4.3315379}}       
\newcommand{\hatcurLCPshortxxxxxA}{\ensuremath{4.3315}}         
\newcommand{\hatcurLCTxxxxxA}{\ensuremath{2458002.46968\pm0.00052}} 
\newcommand{\hatcurLCTAxxxxxA}{\ensuremath{2455646.1132\pm0.0021}} 
\newcommand{\hatcurLCTBxxxxxA}{\ensuremath{2458214.71503\pm0.00059}} 
\newcommand{\hatcurLChatnetmxxxxxA}{\ensuremath{12.038930\pm0.000056}} 
\newcommand{\hatcurLCiblendxxxxxA}{\ensuremath{0.844\pm0.082}}  
\newcommand{\hatcurLCrhoxxxxxA}{\ensuremath{1.394\pm0.029}}     
\newcommand{\hatcurSMEiteffxxxxxA}{\ensuremath{5247\pm50}}      
\newcommand{\hatcurSMEizfehxxxxxA}{\ensuremath{0.040\pm0.030}}  
\newcommand{\hatcurSMEizfehshortxxxxxA}{\ensuremath{0.04}}      
\newcommand{\hatcurSMEiloggxxxxxA}{\ensuremath{4.70\pm0.10}}    
\newcommand{\hatcurSMEivsinxxxxxA}{\ensuremath{3.98\pm0.30}}    
\newcommand{\hatcurSMEivmacxxxxxA}{\ensuremath{nff\pmnff}}      
\newcommand{\hatcurSMEivmicxxxxxA}{\ensuremath{nff\pmnff}}      
\newcommand{\hatcurSMEiiteffxxxxxA}{\ensuremath{5188\pm40}}     
\newcommand{\hatcurSMEiizfehxxxxxA}{\ensuremath{0.040\pm0.030}} 
\newcommand{\hatcurSMEiizfehshortxxxxxA}{\ensuremath{0.04}}     
\newcommand{\hatcurSMEiiloggxxxxxA}{\ensuremath{4.371\pm0.018}} 
\newcommand{\hatcurSMEiivsinxxxxxA}{\ensuremath{3.38\pm0.30}}   
\newcommand{\hatcurLBizxxxxxA}{\ensuremath{0.3754}}             
\newcommand{\hatcurLBiizxxxxxA}{\ensuremath{0.1769}}            
\newcommand{\hatcurLBiixxxxxA}{\ensuremath{0.4491}}             
\newcommand{\hatcurLBiiixxxxxA}{\ensuremath{0.1683}}            
\newcommand{\hatcurLBiIxxxxxA}{\ensuremath{0.4245}}             
\newcommand{\hatcurLBiiIxxxxxA}{\ensuremath{0.1712}}            
\newcommand{\hatcurLBigxxxxxA}{\ensuremath{0.7785}}             
\newcommand{\hatcurLBiigxxxxxA}{\ensuremath{0.0224}}            
\newcommand{\hatcurLBirxxxxxA}{\ensuremath{0.5594}}             
\newcommand{\hatcurLBiirxxxxxA}{\ensuremath{0.1497}}            
\newcommand{\hatcurLBiRxxxxxA}{\ensuremath{0.5328}}             
\newcommand{\hatcurLBiiRxxxxxA}{\ensuremath{0.1491}}            
\newcommand{\hatcurLBikepxxxxxA}{\ensuremath{0.5332}}           
\newcommand{\hatcurLBiikepxxxxxA}{\ensuremath{0.1620}}          
\newcommand{\hatcurISOmxxxxxA}{\ensuremath{0.8810\pm0.0043}}    
\newcommand{\hatcurISOmshortxxxxxA}{\ensuremath{0.88}}          
\newcommand{\hatcurISOmlongxxxxxA}{\ensuremath{0.8810\pm0.0043}} 
\newcommand{\hatcurISOrxxxxxA}{\ensuremath{0.9621\pm0.0081}}    
\newcommand{\hatcurISOrshortxxxxxA}{\ensuremath{0.96}}          
\newcommand{\hatcurISOrlongxxxxxA}{\ensuremath{0.9621\pm0.0081}} 
\newcommand{\hatcurISOrhoxxxxxA}{\ensuremath{1.394\pm0.029}}    
\newcommand{\hatcurISOrholongxxxxxA}{\ensuremath{1.394\pm0.029}} 
\newcommand{\hatcurISOloggxxxxxA}{\ensuremath{4.4165\pm0.0055}} 
\newcommand{\hatcurISOlumxxxxxA}{\ensuremath{0.667\pm0.013}}    
\newcommand{\hatcurISOlumshortxxxxxA}{\ensuremath{0.67}}        
\newcommand{\hatcurISOteffxxxxxA}{\ensuremath{5328.2\pm9.4}}    
\newcommand{\hatcurISOzfehxxxxxA}{\ensuremath{0.208\pm0.018}}   
\newcommand{\hatcurISOagexxxxxA}{\ensuremath{12.452_{-0.169}^{+0.084}}} 
\newcommand{\hatcurISOspecxxxxxA}{K}                            
\newcommand{\hatcurRVKxxxxxA}{\ensuremath{11.6\pm4.0}}          
\newcommand{\hatcurRVrkxxxxxA}{\ensuremath{0\pm0}}              
\newcommand{\hatcurRVrhxxxxxA}{\ensuremath{0\pm0}}              
\newcommand{\hatcurRVkxxxxxA}{\ensuremath{0\pm0}}               
\newcommand{\hatcurRVhxxxxxA}{\ensuremath{0\pm0}}               
\newcommand{\hatcurRVtronexxxxxA}{\ensuremath{0.4539\pm0.0015}} 
\newcommand{\hatcurRVtrtwoxxxxxA}{\ensuremath{0\pm0}}           
\newcommand{\hatcurRVgammaAxxxxxA}{\ensuremath{6417\pm0}}       
\newcommand{\hatcurRVjitterAxxxxxA}{\ensuremath{1\pm39}}        
\newcommand{\hatcurRVjittertwosiglimAxxxxxA}{\ensuremath{<72.8}} 
\newcommand{\hatcurRVfitrmsAxxxxxA}{\ensuremath{0.0}}           
\newcommand{\hatcurRVgammaBxxxxxA}{\ensuremath{-893\pm0}}       
\newcommand{\hatcurRVjitterBxxxxxA}{\ensuremath{8.0\pm3.0}}     
\newcommand{\hatcurRVjittertwosiglimBxxxxxA}{\ensuremath{<14.0}} 
\newcommand{\hatcurRVfitrmsBxxxxxA}{\ensuremath{0.0}}           
\newcommand{\hatcurRVeccenxxxxxA}{\ensuremath{0\pm0}}           
\newcommand{\hatcurRVeccentwosiglimxxxxxA}{\ensuremath{<0.000}} 
\newcommand{\hatcurRVomegaxxxxxA}{\ensuremath{0\pm0}}           
\newcommand{\hatcurPPixxxxxA}{\ensuremath{87.94\pm0.18}}        
\newcommand{\hatcurPPgxxxxxA}{\ensuremath{5.4\pm1.9}}           
\newcommand{\hatcurPPloggxxxxxA}{\ensuremath{2.74\pm0.17}}      
\newcommand{\hatcurPParxxxxxA}{\ensuremath{11.142\pm0.078}}     
\newcommand{\hatcurPParelxxxxxA}{\ensuremath{0.049854\pm0.000080}} 
\newcommand{\hatcurPPrhoxxxxxA}{\ensuremath{0.44\pm0.15}}       
\newcommand{\hatcurPPmxxxxxA}{\ensuremath{0.086\pm0.030}}       
\newcommand{\hatcurPPmshortxxxxxA}{\ensuremath{0.09}}           
\newcommand{\hatcurPPmlongxxxxxA}{\ensuremath{0.086\pm0.030}}   
\newcommand{\hatcurPPmexxxxxA}{\ensuremath{27.3\pm9.4}}         
\newcommand{\hatcurPPmeshortxxxxxA}{\ensuremath{27.3}}          
\newcommand{\hatcurPPmelongxxxxxA}{\ensuremath{27.3\pm9.4}}     
\newcommand{\hatcurPPrxxxxxA}{\ensuremath{0.626\pm0.015}}       
\newcommand{\hatcurPPrshortxxxxxA}{\ensuremath{0.63}}           
\newcommand{\hatcurPPrlongxxxxxA}{\ensuremath{0.626\pm0.015}}   
\newcommand{\hatcurPPrexxxxxA}{\ensuremath{7.01\pm0.17}}        
\newcommand{\hatcurPPreshortxxxxxA}{\ensuremath{7.0}}           
\newcommand{\hatcurPPrelongxxxxxA}{\ensuremath{7.01\pm0.17}}    
\newcommand{\hatcurPPmrcorrxxxxxA}{\ensuremath{0.03}}           
\newcommand{\hatcurPPteffxxxxxA}{\ensuremath{1128.9\pm5.0}}     
\newcommand{\hatcurPPthetaxxxxxA}{\ensuremath{0.0156\pm0.0054}} 
\newcommand{\hatcurPPfluxperixxxxxA}{\ensuremath{3.656\pm0.065}} 
\newcommand{\hatcurPPfluxperidimxxxxxA}{\ensuremath{8}}         
\newcommand{\hatcurPPfluxapxxxxxA}{\ensuremath{3.656\pm0.065}}  
\newcommand{\hatcurPPfluxapdimxxxxxA}{\ensuremath{8}}           
\newcommand{\hatcurPPfluxavgxxxxxA}{\ensuremath{3.656\pm0.065}} 
\newcommand{\hatcurPPfluxavgdimxxxxxA}{\ensuremath{8}}          
\newcommand{\hatcurPPfluxavglogxxxxxA}{\ensuremath{8.5630\pm0.0077}} 
\newcommand{\hatcurXsecphasexxxxxA}{\ensuremath{0\pm0}}         
\newcommand{\hatcurXsecondaryxxxxxA}{\ensuremath{2458004.63545\pm0.00052}} 
\newcommand{\hatcurXsecdurxxxxxA}{\ensuremath{0.1223\pm0.0014}} 
\newcommand{\hatcurXsecingdurxxxxxA}{\ensuremath{0.00903\pm0.00031}} 
\newcommand{\hatcurPPphiconjxxxxxA}{\ensuremath{0\pm0}}         
\newcommand{\hatcurPPperixxxxxA}{\ensuremath{2458001.38679\pm0.00052}} 
\newcommand{\hatcurPPaequivxxxxxA}{\ensuremath{0.06100\pm0.00054}} 
\newcommand{\hatcurPPtcircxxxxxA}{\ensuremath{1440\pm530}}      
\newcommand{\hatcurPPtinfallxxxxxA}{\ensuremath{72000_{-18000}^{+31000}}} 
\newcommand{\hatcurXdistxxxxxA}{\ensuremath{200.4\pm2.0}}       
\newcommand{\hatcurXAvxxxxxA}{\ensuremath{0.4470_{-0.0040}^{+0.0020}}} 
\newcommand{\hatcurXdistredxxxxxA}{\ensuremath{200.4\pm2.0}}    
\newcommand{\hatcurXEBVxxxxxA}{\ensuremath{0.1440\pm0.0012}}    
\newcommand{\hatcurCCpmraxxxxxA}{\ensuremath{-21.78\pm0.11}}    
\newcommand{\hatcurCCpmdecxxxxxA}{\ensuremath{6.15\pm0.11}}     
\newcommand{\hatcurCCpmxxxxxA}{\ensuremath{22.64\pm0.15}}       

\newcommand{\hatcurhtrxxxxxB}{HATS561-006}                      
\newcommand{\hatcurfieldxxxxxB}{\ensuremath{string}}            
\newcommand{\hatcurCCraxxxxxB}{\ensuremath{10^{\mathrm h}17^{\mathrm m}05.0796{\mathrm s}}}                   
\newcommand{\hatcurCCdecxxxxxB}{\ensuremath{-25{\arcdeg}16{\arcmin}34.5568{\arcsec}}}                 
\newcommand{\hatcurCCmagxxxxxB}{12.411}                         
\newcommand{\hatcurCCtwomassxxxxxB}{2MASS~10170509-2516345}     
\newcommand{\hatcurCCgscxxxxxB}{GSC~6622-00794}                 
\newcommand{\hatcurCCgaiaxxxxxB}{GAIA~5472386847387498496}      
\newcommand{\hatcurCCgaiadrtwoxxxxxB}{GAIA~DR2~5472386851683941376} 
\newcommand{\hatcurCCtassmvxxxxxB}{\ensuremath{12.411\pm0.030}} 
\newcommand{\hatcurCCtassmvshortxxxxxB}{\ensuremath{12.4}}      
\newcommand{\hatcurCCtassmBxxxxxB}{\ensuremath{13.22\pm0.11}}   
\newcommand{\hatcurCCtassmBshortxxxxxB}{\ensuremath{13.2}}      
\newcommand{\hatcurCCtassmIxxxxxB}{\ensuremath{nff\pmnff}}      
\newcommand{\hatcurCCtassmIshortxxxxxB}{\ensuremath{0.0}}       
\newcommand{\hatcurCCtassmgxxxxxB}{\ensuremath{12.780\pm0.037}} 
\newcommand{\hatcurCCtassmgshortxxxxxB}{\ensuremath{12.8}}      
\newcommand{\hatcurCCtassmrxxxxxB}{\ensuremath{12.220\pm0.057}} 
\newcommand{\hatcurCCtassmrshortxxxxxB}{\ensuremath{12.2}}      
\newcommand{\hatcurCCtassmixxxxxB}{\ensuremath{12.26\pm0.19}}   
\newcommand{\hatcurCCtassmishortxxxxxB}{\ensuremath{12.3}}      
\newcommand{\hatcurCCparallaxxxxxxB}{\ensuremath{2.883\pm0.043}} 
\newcommand{\hatcurCCgaiamGxxxxxB}{\ensuremath{12.27810\pm0.00020}} 
\newcommand{\hatcurCCgaiamBPxxxxxB}{\ensuremath{12.6494\pm0.0012}} 
\newcommand{\hatcurCCgaiamRPxxxxxB}{\ensuremath{11.76070\pm0.00060}} 
\newcommand{\hatcurCCtwomassJmagxxxxxB}{\ensuremath{11.184\pm0.026}} 
\newcommand{\hatcurCCtwomassHmagxxxxxB}{\ensuremath{10.850\pm0.024}} 
\newcommand{\hatcurCCtwomassKmagxxxxxB}{\ensuremath{10.768\pm0.024}} 
\newcommand{\hatcurCCcitJmagxxxxxB}{\ensuremath{11.198\pm0.026}} 
\newcommand{\hatcurCCcitHmagxxxxxB}{\ensuremath{10.845\pm0.025}} 
\newcommand{\hatcurCCcitKmagxxxxxB}{\ensuremath{10.792\pm0.024}} 
\newcommand{\hatcurCCbbJmagxxxxxB}{\ensuremath{11.252\pm0.028}} 
\newcommand{\hatcurCCbbHmagxxxxxB}{\ensuremath{10.866\pm0.025}} 
\newcommand{\hatcurCCbbKmagxxxxxB}{\ensuremath{10.812\pm0.024}} 
\newcommand{\hatcurCCesoJmagxxxxxB}{\ensuremath{11.254\pm0.029}} 
\newcommand{\hatcurCCesoHmagxxxxxB}{\ensuremath{10.862\pm0.027}} 
\newcommand{\hatcurCCesoKmagxxxxxB}{\ensuremath{10.811\pm0.025}} 
\newcommand{\hatcurCCesoJHmagxxxxxB}{\ensuremath{0.392\pm0.038}} 
\newcommand{\hatcurCCesoJKmagxxxxxB}{\ensuremath{0.443\pm0.038}} 
\newcommand{\hatcurCCesoHKmagxxxxxB}{\ensuremath{0.050\pm0.010}} 
\newcommand{\hatcurCCWonemagxxxxxB}{\ensuremath{10.714\pm0.023}} 
\newcommand{\hatcurCCWtwomagxxxxxB}{\ensuremath{10.783\pm0.022}} 
\newcommand{\hatcurCCWthreemagxxxxxB}{\ensuremath{10.736\pm0.091}} 
\newcommand{\hatcurCCWfourmagxxxxxB}{\ensuremath{0\pm0}}        
\newcommand{\hatcurLCdipxxxxxB}{\ensuremath{4.1}}               
\newcommand{\hatcurLCrprstarxxxxxB}{\ensuremath{0.0570\pm0.0012}} 
\newcommand{\hatcurLCbsqxxxxxB}{\ensuremath{0.227_{-0.027}^{+0.027}}} 
\newcommand{\hatcurLCimpxxxxxB}{\ensuremath{0.476_{-0.030}^{+0.027}}} 
\newcommand{\hatcurLCzetaxxxxxB}{\ensuremath{16.02\pm0.25}}     
\newcommand{\hatcurLCdurxxxxxB}{\ensuremath{0.1340\pm0.0019}}   
\newcommand{\hatcurLCdurshortxxxxxB}{\ensuremath{0.1340}}       
\newcommand{\hatcurLCdurhrxxxxxB}{\ensuremath{3.215\pm0.046}}   
\newcommand{\hatcurLCdurhrshortxxxxxB}{\ensuremath{3.215}}      
\newcommand{\hatcurLCqxxxxxB}{\ensuremath{0.03060\pm0.00044}}   
\newcommand{\hatcurLCqshortxxxxxB}{\ensuremath{0.031}}          
\newcommand{\hatcurLCingdurxxxxxB}{\ensuremath{0.00924\pm0.00035}} 
\newcommand{\hatcurLCPxxxxxB}{\ensuremath{4.375021\pm0.000010}} 
\newcommand{\hatcurLCPprecxxxxxB}{\ensuremath{4.3750209}}       
\newcommand{\hatcurLCPshortxxxxxB}{\ensuremath{4.3750}}         
\newcommand{\hatcurLCTxxxxxB}{\ensuremath{2457725.16042\pm0.00072}} 
\newcommand{\hatcurLCTAxxxxxB}{\ensuremath{2457012.0321\pm0.0019}} 
\newcommand{\hatcurLCTBxxxxxB}{\ensuremath{2457847.66104\pm0.00077}} 
\newcommand{\hatcurLChatnetmAxxxxxB}{\ensuremath{12.354740\pm0.000038}} 
\newcommand{\hatcurLCiblendAxxxxxB}{\ensuremath{0.964\pm0.036}} 
\newcommand{\hatcurLChatnetmBxxxxxB}{\ensuremath{12.35470\pm0.00016}} 
\newcommand{\hatcurLCiblendBxxxxxB}{\ensuremath{0.90\pm0.10}}   
\newcommand{\hatcurLCrhoxxxxxB}{\ensuremath{0.933\pm0.039}}     
\newcommand{\hatcurSMEiteffxxxxxB}{\ensuremath{5740\pm50}}      
\newcommand{\hatcurSMEizfehxxxxxB}{\ensuremath{0.060\pm0.026}}  
\newcommand{\hatcurSMEizfehshortxxxxxB}{\ensuremath{0.06}}      
\newcommand{\hatcurSMEiloggxxxxxB}{\ensuremath{4.550\pm0.095}}  
\newcommand{\hatcurSMEivsinxxxxxB}{\ensuremath{3.10\pm0.27}}    
\newcommand{\hatcurSMEivmacxxxxxB}{\ensuremath{3.934\pm0.076}}  
\newcommand{\hatcurSMEivmicxxxxxB}{\ensuremath{1.059\pm0.028}}  
\newcommand{\hatcurSMEiiteffxxxxxB}{\ensuremath{5696\pm50}}     
\newcommand{\hatcurSMEiizfehxxxxxB}{\ensuremath{0.02\pm0.32}}   
\newcommand{\hatcurSMEiizfehshortxxxxxB}{\ensuremath{0.02}}     
\newcommand{\hatcurSMEiiloggxxxxxB}{\ensuremath{4.296\pm0.013}} 
\newcommand{\hatcurSMEiivsinxxxxxB}{\ensuremath{3.19\pm0.17}}   
\newcommand{\hatcurSMEiivmacxxxxxB}{\ensuremath{3.76\pm0.10}}   
\newcommand{\hatcurSMEiivmicxxxxxB}{\ensuremath{0.999\pm0.035}} 
\newcommand{\hatcurLBiBxxxxxB}{\ensuremath{0.7272}}             
\newcommand{\hatcurLBiiBxxxxxB}{\ensuremath{0.1002}}            
\newcommand{\hatcurLBiVxxxxxB}{\ensuremath{0.5631}}             
\newcommand{\hatcurLBiiVxxxxxB}{\ensuremath{0.1695}}            
\newcommand{\hatcurLBiRxxxxxB}{\ensuremath{0.4660}}             
\newcommand{\hatcurLBiiRxxxxxB}{\ensuremath{0.1829}}            
\newcommand{\hatcurLBiIxxxxxB}{\ensuremath{0.3693}}             
\newcommand{\hatcurLBiiIxxxxxB}{\ensuremath{0.1915}}            
\newcommand{\hatcurLBiuxxxxxB}{\ensuremath{0.8957}}             
\newcommand{\hatcurLBiiuxxxxxB}{\ensuremath{-0.0545}}           
\newcommand{\hatcurLBigxxxxxB}{\ensuremath{0.6706}}             
\newcommand{\hatcurLBiigxxxxxB}{\ensuremath{0.1178}}            
\newcommand{\hatcurLBirxxxxxB}{\ensuremath{0.23\pm0.13}}        
\newcommand{\hatcurLBiirxxxxxB}{\ensuremath{0.35\pm0.16}}       
\newcommand{\hatcurLBiixxxxxB}{\ensuremath{0.37_{-0.14}^{+0.11}}} 
\newcommand{\hatcurLBiiixxxxxB}{\ensuremath{0.34\pm0.15}}       
\newcommand{\hatcurLBizxxxxxB}{\ensuremath{0.3244}}             
\newcommand{\hatcurLBiizxxxxxB}{\ensuremath{0.1946}}            
\newcommand{\hatcurLBiJxxxxxB}{\ensuremath{0.2130}}             
\newcommand{\hatcurLBiiJxxxxxB}{\ensuremath{0.2265}}            
\newcommand{\hatcurLBiHxxxxxB}{\ensuremath{0.1233}}             
\newcommand{\hatcurLBiiHxxxxxB}{\ensuremath{0.2628}}            
\newcommand{\hatcurLBiKxxxxxB}{\ensuremath{0.1173}}             
\newcommand{\hatcurLBiiKxxxxxB}{\ensuremath{0.2079}}            
\newcommand{\hatcurLBiTxxxxxB}{\ensuremath{0.35\pm0.15}}        
\newcommand{\hatcurLBiiTxxxxxB}{\ensuremath{0.31\pm0.15}}       
\newcommand{\hatcurLBikepxxxxxB}{\ensuremath{0.4878}}           
\newcommand{\hatcurLBiikepxxxxxB}{\ensuremath{0.1842}}          
\newcommand{\hatcurLBiCxxxxxB}{\ensuremath{0.4784}}             
\newcommand{\hatcurLBiiCxxxxxB}{\ensuremath{0.1823}}            
\newcommand{\hatcurLBiMxxxxxB}{\ensuremath{0.5727}}             
\newcommand{\hatcurLBiiMxxxxxB}{\ensuremath{0.1573}}            
\newcommand{\hatcurISOmxxxxxB}{\ensuremath{0.890_{-0.012}^{+0.016}}} 
\newcommand{\hatcurISOmshortxxxxxB}{\ensuremath{0.89}}          
\newcommand{\hatcurISOmlongxxxxxB}{\ensuremath{0.890_{-0.012}^{+0.016}}} 
\newcommand{\hatcurISOrxxxxxB}{\ensuremath{1.105\pm0.016}}      
\newcommand{\hatcurISOrshortxxxxxB}{\ensuremath{1.10}}          
\newcommand{\hatcurISOrlongxxxxxB}{\ensuremath{1.105\pm0.016}}  
\newcommand{\hatcurISOrhoxxxxxB}{\ensuremath{0.933\pm0.039}}    
\newcommand{\hatcurISOrholongxxxxxB}{\ensuremath{0.933\pm0.039}} 
\newcommand{\hatcurISOloggxxxxxB}{\ensuremath{4.301\pm0.013}}   
\newcommand{\hatcurISOlumxxxxxB}{\ensuremath{1.179\pm0.037}}    
\newcommand{\hatcurISOlumshortxxxxxB}{\ensuremath{1.18}}        
\newcommand{\hatcurISOteffxxxxxB}{\ensuremath{5732\pm25}}       
\newcommand{\hatcurISOzfehxxxxxB}{\ensuremath{-0.102\pm0.043}}  
\newcommand{\hatcurISOagexxxxxB}{\ensuremath{11.89\pm0.60}}     
\newcommand{\hatcurISOspecxxxxxB}{G}                            
\newcommand{\hatcurRVKxxxxxB}{\ensuremath{9.9\pm1.5}}           
\newcommand{\hatcurRVrkxxxxxB}{\ensuremath{0\pm0}}              
\newcommand{\hatcurRVrhxxxxxB}{\ensuremath{0\pm0}}              
\newcommand{\hatcurRVkxxxxxB}{\ensuremath{0\pm0}}               
\newcommand{\hatcurRVhxxxxxB}{\ensuremath{0\pm0}}               
\newcommand{\hatcurRVtronexxxxxB}{\ensuremath{0\pm0}}           
\newcommand{\hatcurRVtrtwoxxxxxB}{\ensuremath{0\pm0}}           
\newcommand{\hatcurRVgammaAxxxxxB}{\ensuremath{4130.6\pm4.9}}   
\newcommand{\hatcurRVjitterAxxxxxB}{\ensuremath{15.3\pm5.3}}    
\newcommand{\hatcurRVjittertwosiglimAxxxxxB}{\ensuremath{<25.6}} 
\newcommand{\hatcurRVfitrmsAxxxxxB}{\ensuremath{0.0}}           
\newcommand{\hatcurRVgammaBxxxxxB}{\ensuremath{4144.0\pm1.5}}   
\newcommand{\hatcurRVjitterBxxxxxB}{\ensuremath{0.06\pm0.93}}   
\newcommand{\hatcurRVjittertwosiglimBxxxxxB}{\ensuremath{<2.4}} 
\newcommand{\hatcurRVfitrmsBxxxxxB}{\ensuremath{0.0}}           
\newcommand{\hatcurRVgammaCxxxxxB}{\ensuremath{-4.0\pm1.6}}     
\newcommand{\hatcurRVjitterCxxxxxB}{\ensuremath{0.8\pm2.0}}     
\newcommand{\hatcurRVjittertwosiglimCxxxxxB}{\ensuremath{<5.4}} 
\newcommand{\hatcurRVfitrmsCxxxxxB}{\ensuremath{0.0}}           
\newcommand{\hatcurRVeccenxxxxxB}{\ensuremath{0\pm0}}           
\newcommand{\hatcurRVeccentwosiglimxxxxxB}{\ensuremath{<0.000}} 
\newcommand{\hatcurRVomegaxxxxxB}{\ensuremath{0\pm0}}           
\newcommand{\hatcurPPixxxxxB}{\ensuremath{87.21\pm0.18}}        
\newcommand{\hatcurPPgxxxxxB}{\ensuremath{4.91\pm0.81}}         
\newcommand{\hatcurPPloggxxxxxB}{\ensuremath{2.691\pm0.075}}    
\newcommand{\hatcurPParxxxxxB}{\ensuremath{9.81\pm0.14}}        
\newcommand{\hatcurPParelxxxxxB}{\ensuremath{0.05036_{-0.00023}^{+0.00030}}} 
\newcommand{\hatcurPPrhoxxxxxB}{\ensuremath{0.403\pm0.071}}     
\newcommand{\hatcurPPmxxxxxB}{\ensuremath{0.074\pm0.011}}       
\newcommand{\hatcurPPmshortxxxxxB}{\ensuremath{0.07}}           
\newcommand{\hatcurPPmlongxxxxxB}{\ensuremath{0.074\pm0.011}}   
\newcommand{\hatcurPPmexxxxxB}{\ensuremath{23.7\pm3.7}}         
\newcommand{\hatcurPPmeshortxxxxxB}{\ensuremath{23.7}}          
\newcommand{\hatcurPPmelongxxxxxB}{\ensuremath{23.7\pm3.7}}     
\newcommand{\hatcurPPrxxxxxB}{\ensuremath{0.614\pm0.017}}       
\newcommand{\hatcurPPrshortxxxxxB}{\ensuremath{0.61}}           
\newcommand{\hatcurPPrlongxxxxxB}{\ensuremath{0.614\pm0.017}}   
\newcommand{\hatcurPPrexxxxxB}{\ensuremath{6.88\pm0.19}}        
\newcommand{\hatcurPPreshortxxxxxB}{\ensuremath{6.9}}           
\newcommand{\hatcurPPrelongxxxxxB}{\ensuremath{6.88\pm0.19}}    
\newcommand{\hatcurPPmrcorrxxxxxB}{\ensuremath{-0.06}}          
\newcommand{\hatcurPPteffxxxxxB}{\ensuremath{1294\pm10}}        
\newcommand{\hatcurPPthetaxxxxxB}{\ensuremath{0.0136\pm0.0022}} 
\newcommand{\hatcurPPfluxperixxxxxB}{\ensuremath{6.33\pm0.20}}  
\newcommand{\hatcurPPfluxperidimxxxxxB}{\ensuremath{8}}         
\newcommand{\hatcurPPfluxapxxxxxB}{\ensuremath{6.33\pm0.20}}    
\newcommand{\hatcurPPfluxapdimxxxxxB}{\ensuremath{8}}           
\newcommand{\hatcurPPfluxavgxxxxxB}{\ensuremath{6.33\pm0.20}}   
\newcommand{\hatcurPPfluxavgdimxxxxxB}{\ensuremath{8}}          
\newcommand{\hatcurPPfluxavglogxxxxxB}{\ensuremath{8.801\pm0.014}} 
\newcommand{\hatcurXsecphasexxxxxB}{\ensuremath{0\pm0}}         
\newcommand{\hatcurXsecondaryxxxxxB}{\ensuremath{2457727.34793\pm0.00072}} 
\newcommand{\hatcurXsecdurxxxxxB}{\ensuremath{0.1340\pm0.0019}} 
\newcommand{\hatcurXsecingdurxxxxxB}{\ensuremath{0.00924\pm0.00035}} 
\newcommand{\hatcurPPphiconjxxxxxB}{\ensuremath{0\pm0}}         
\newcommand{\hatcurPPperixxxxxB}{\ensuremath{2457724.06667\pm0.00072}} 
\newcommand{\hatcurPPaequivxxxxxB}{\ensuremath{0.04640\pm0.00075}} 
\newcommand{\hatcurPPtcircxxxxxB}{\ensuremath{1450\pm310}}      
\newcommand{\hatcurPPtinfallxxxxxB}{\ensuremath{45300_{-6700}^{+8800}}} 
\newcommand{\hatcurXdistxxxxxB}{\ensuremath{347.7\pm5.1}}       
\newcommand{\hatcurXAvxxxxxB}{\ensuremath{0.122\pm0.024}}       
\newcommand{\hatcurXdistredxxxxxB}{\ensuremath{347.7\pm5.1}}    
\newcommand{\hatcurXEBVxxxxxB}{\ensuremath{0.0390\pm0.0076}}    
\newcommand{\hatcurCCpmraxxxxxB}{\ensuremath{-21.752\pm0.066}}  
\newcommand{\hatcurCCpmdecxxxxxB}{\ensuremath{-7.540\pm0.070}}  
\newcommand{\hatcurCCpmxxxxxB}{\ensuremath{23.022\pm0.096}}     

\newcommand{\hatcurCCbbHmag}[1]{\ifnum#1=37 %
\hatcurCCbbHmagxxxxxA
\else
\ifnum#1=38 %
\hatcurCCbbHmagxxxxxB
\else
??????\fi
\fi
}
\newcommand{\hatcurCCbbJmag}[1]{\ifnum#1=37 %
\hatcurCCbbJmagxxxxxA
\else
\ifnum#1=38 %
\hatcurCCbbJmagxxxxxB
\else
??????\fi
\fi
}
\newcommand{\hatcurCCbbKmag}[1]{\ifnum#1=37 %
\hatcurCCbbKmagxxxxxA
\else
\ifnum#1=38 %
\hatcurCCbbKmagxxxxxB
\else
??????\fi
\fi
}
\newcommand{\hatcurCCcitHmag}[1]{\ifnum#1=37 %
\hatcurCCcitHmagxxxxxA
\else
\ifnum#1=38 %
\hatcurCCcitHmagxxxxxB
\else
??????\fi
\fi
}
\newcommand{\hatcurCCcitJmag}[1]{\ifnum#1=37 %
\hatcurCCcitJmagxxxxxA
\else
\ifnum#1=38 %
\hatcurCCcitJmagxxxxxB
\else
??????\fi
\fi
}
\newcommand{\hatcurCCcitKmag}[1]{\ifnum#1=37 %
\hatcurCCcitKmagxxxxxA
\else
\ifnum#1=38 %
\hatcurCCcitKmagxxxxxB
\else
??????\fi
\fi
}
\newcommand{\hatcurCCdec}[1]{\ifnum#1=37 %
\hatcurCCdecxxxxxA
\else
\ifnum#1=38 %
\hatcurCCdecxxxxxB
\else
??????\fi
\fi
}
\newcommand{\hatcurCCesoHKmag}[1]{\ifnum#1=37 %
\hatcurCCesoHKmagxxxxxA
\else
\ifnum#1=38 %
\hatcurCCesoHKmagxxxxxB
\else
??????\fi
\fi
}
\newcommand{\hatcurCCesoHmag}[1]{\ifnum#1=37 %
\hatcurCCesoHmagxxxxxA
\else
\ifnum#1=38 %
\hatcurCCesoHmagxxxxxB
\else
??????\fi
\fi
}
\newcommand{\hatcurCCesoJHmag}[1]{\ifnum#1=37 %
\hatcurCCesoJHmagxxxxxA
\else
\ifnum#1=38 %
\hatcurCCesoJHmagxxxxxB
\else
??????\fi
\fi
}
\newcommand{\hatcurCCesoJKmag}[1]{\ifnum#1=37 %
\hatcurCCesoJKmagxxxxxA
\else
\ifnum#1=38 %
\hatcurCCesoJKmagxxxxxB
\else
??????\fi
\fi
}
\newcommand{\hatcurCCesoJmag}[1]{\ifnum#1=37 %
\hatcurCCesoJmagxxxxxA
\else
\ifnum#1=38 %
\hatcurCCesoJmagxxxxxB
\else
??????\fi
\fi
}
\newcommand{\hatcurCCesoKmag}[1]{\ifnum#1=37 %
\hatcurCCesoKmagxxxxxA
\else
\ifnum#1=38 %
\hatcurCCesoKmagxxxxxB
\else
??????\fi
\fi
}
\newcommand{\hatcurCCgaia}[1]{\ifnum#1=37 %
\hatcurCCgaiaxxxxxA
\else
\ifnum#1=38 %
\hatcurCCgaiaxxxxxB
\else
??????\fi
\fi
}
\newcommand{\hatcurCCgaiadrtwo}[1]{\ifnum#1=37 %
\hatcurCCgaiadrtwoxxxxxA
\else
\ifnum#1=38 %
\hatcurCCgaiadrtwoxxxxxB
\else
??????\fi
\fi
}
\newcommand{\hatcurCCgaiamBP}[1]{\ifnum#1=37 %
\hatcurCCgaiamBPxxxxxA
\else
\ifnum#1=38 %
\hatcurCCgaiamBPxxxxxB
\else
??????\fi
\fi
}
\newcommand{\hatcurCCgaiamG}[1]{\ifnum#1=37 %
\hatcurCCgaiamGxxxxxA
\else
\ifnum#1=38 %
\hatcurCCgaiamGxxxxxB
\else
??????\fi
\fi
}
\newcommand{\hatcurCCgaiamRP}[1]{\ifnum#1=37 %
\hatcurCCgaiamRPxxxxxA
\else
\ifnum#1=38 %
\hatcurCCgaiamRPxxxxxB
\else
??????\fi
\fi
}
\newcommand{\hatcurCCgsc}[1]{\ifnum#1=37 %
\hatcurCCgscxxxxxA
\else
\ifnum#1=38 %
\hatcurCCgscxxxxxB
\else
??????\fi
\fi
}
\newcommand{\hatcurCCmag}[1]{\ifnum#1=37 %
\hatcurCCmagxxxxxA
\else
\ifnum#1=38 %
\hatcurCCmagxxxxxB
\else
??????\fi
\fi
}
\newcommand{\hatcurCCparallax}[1]{\ifnum#1=37 %
\hatcurCCparallaxxxxxxA
\else
\ifnum#1=38 %
\hatcurCCparallaxxxxxxB
\else
??????\fi
\fi
}
\newcommand{\hatcurCCpm}[1]{\ifnum#1=37 %
\hatcurCCpmxxxxxA
\else
\ifnum#1=38 %
\hatcurCCpmxxxxxB
\else
??????\fi
\fi
}
\newcommand{\hatcurCCpmdec}[1]{\ifnum#1=37 %
\hatcurCCpmdecxxxxxA
\else
\ifnum#1=38 %
\hatcurCCpmdecxxxxxB
\else
??????\fi
\fi
}
\newcommand{\hatcurCCpmra}[1]{\ifnum#1=37 %
\hatcurCCpmraxxxxxA
\else
\ifnum#1=38 %
\hatcurCCpmraxxxxxB
\else
??????\fi
\fi
}
\newcommand{\hatcurCCra}[1]{\ifnum#1=37 %
\hatcurCCraxxxxxA
\else
\ifnum#1=38 %
\hatcurCCraxxxxxB
\else
??????\fi
\fi
}
\newcommand{\hatcurCCtassmB}[1]{\ifnum#1=37 %
\hatcurCCtassmBxxxxxA
\else
\ifnum#1=38 %
\hatcurCCtassmBxxxxxB
\else
??????\fi
\fi
}
\newcommand{\hatcurCCtassmBshort}[1]{\ifnum#1=37 %
\hatcurCCtassmBshortxxxxxA
\else
\ifnum#1=38 %
\hatcurCCtassmBshortxxxxxB
\else
??????\fi
\fi
}
\newcommand{\hatcurCCtassmg}[1]{\ifnum#1=37 %
\hatcurCCtassmgxxxxxA
\else
\ifnum#1=38 %
\hatcurCCtassmgxxxxxB
\else
??????\fi
\fi
}
\newcommand{\hatcurCCtassmgshort}[1]{\ifnum#1=37 %
\hatcurCCtassmgshortxxxxxA
\else
\ifnum#1=38 %
\hatcurCCtassmgshortxxxxxB
\else
??????\fi
\fi
}
\newcommand{\hatcurCCtassmi}[1]{\ifnum#1=37 %
\hatcurCCtassmixxxxxA
\else
\ifnum#1=38 %
\hatcurCCtassmixxxxxB
\else
??????\fi
\fi
}
\newcommand{\hatcurCCtassmI}[1]{\ifnum#1=37 %
\hatcurCCtassmIxxxxxA
\else
\ifnum#1=38 %
\hatcurCCtassmIxxxxxB
\else
??????\fi
\fi
}
\newcommand{\hatcurCCtassmishort}[1]{\ifnum#1=37 %
\hatcurCCtassmishortxxxxxA
\else
\ifnum#1=38 %
\hatcurCCtassmishortxxxxxB
\else
??????\fi
\fi
}
\newcommand{\hatcurCCtassmIshort}[1]{\ifnum#1=37 %
\hatcurCCtassmIshortxxxxxA
\else
\ifnum#1=38 %
\hatcurCCtassmIshortxxxxxB
\else
??????\fi
\fi
}
\newcommand{\hatcurCCtassmr}[1]{\ifnum#1=37 %
\hatcurCCtassmrxxxxxA
\else
\ifnum#1=38 %
\hatcurCCtassmrxxxxxB
\else
??????\fi
\fi
}
\newcommand{\hatcurCCtassmrshort}[1]{\ifnum#1=37 %
\hatcurCCtassmrshortxxxxxA
\else
\ifnum#1=38 %
\hatcurCCtassmrshortxxxxxB
\else
??????\fi
\fi
}
\newcommand{\hatcurCCtassmv}[1]{\ifnum#1=37 %
\hatcurCCtassmvxxxxxA
\else
\ifnum#1=38 %
\hatcurCCtassmvxxxxxB
\else
??????\fi
\fi
}
\newcommand{\hatcurCCtassmvshort}[1]{\ifnum#1=37 %
\hatcurCCtassmvshortxxxxxA
\else
\ifnum#1=38 %
\hatcurCCtassmvshortxxxxxB
\else
??????\fi
\fi
}
\newcommand{\hatcurCCtwomass}[1]{\ifnum#1=37 %
\hatcurCCtwomassxxxxxA
\else
\ifnum#1=38 %
\hatcurCCtwomassxxxxxB
\else
??????\fi
\fi
}
\newcommand{\hatcurCCtwomassHmag}[1]{\ifnum#1=37 %
\hatcurCCtwomassHmagxxxxxA
\else
\ifnum#1=38 %
\hatcurCCtwomassHmagxxxxxB
\else
??????\fi
\fi
}
\newcommand{\hatcurCCtwomassJmag}[1]{\ifnum#1=37 %
\hatcurCCtwomassJmagxxxxxA
\else
\ifnum#1=38 %
\hatcurCCtwomassJmagxxxxxB
\else
??????\fi
\fi
}
\newcommand{\hatcurCCtwomassKmag}[1]{\ifnum#1=37 %
\hatcurCCtwomassKmagxxxxxA
\else
\ifnum#1=38 %
\hatcurCCtwomassKmagxxxxxB
\else
??????\fi
\fi
}
\newcommand{\hatcurCCWfourmag}[1]{\ifnum#1=38 %
\hatcurCCWfourmagxxxxxB
\else
??????\fi
}
\newcommand{\hatcurCCWonemag}[1]{\ifnum#1=37 %
\hatcurCCWonemagxxxxxA
\else
\ifnum#1=38 %
\hatcurCCWonemagxxxxxB
\else
??????\fi
\fi
}
\newcommand{\hatcurCCWthreemag}[1]{\ifnum#1=37 %
\hatcurCCWthreemagxxxxxA
\else
\ifnum#1=38 %
\hatcurCCWthreemagxxxxxB
\else
??????\fi
\fi
}
\newcommand{\hatcurCCWtwomag}[1]{\ifnum#1=37 %
\hatcurCCWtwomagxxxxxA
\else
\ifnum#1=38 %
\hatcurCCWtwomagxxxxxB
\else
??????\fi
\fi
}
\newcommand{\hatcurfield}[1]{\ifnum#1=37 %
\hatcurfieldxxxxxA
\else
\ifnum#1=38 %
\hatcurfieldxxxxxB
\else
??????\fi
\fi
}
\newcommand{\hatcurhtr}[1]{\ifnum#1=37 %
\hatcurhtrxxxxxA
\else
\ifnum#1=38 %
\hatcurhtrxxxxxB
\else
??????\fi
\fi
}
\newcommand{\hatcurISOage}[1]{\ifnum#1=37 %
\hatcurISOagexxxxxA
\else
\ifnum#1=38 %
\hatcurISOagexxxxxB
\else
??????\fi
\fi
}
\newcommand{\hatcurISOlogg}[1]{\ifnum#1=37 %
\hatcurISOloggxxxxxA
\else
\ifnum#1=38 %
\hatcurISOloggxxxxxB
\else
??????\fi
\fi
}
\newcommand{\hatcurISOlum}[1]{\ifnum#1=37 %
\hatcurISOlumxxxxxA
\else
\ifnum#1=38 %
\hatcurISOlumxxxxxB
\else
??????\fi
\fi
}
\newcommand{\hatcurISOlumshort}[1]{\ifnum#1=37 %
\hatcurISOlumshortxxxxxA
\else
\ifnum#1=38 %
\hatcurISOlumshortxxxxxB
\else
??????\fi
\fi
}
\newcommand{\hatcurISOm}[1]{\ifnum#1=37 %
\hatcurISOmxxxxxA
\else
\ifnum#1=38 %
\hatcurISOmxxxxxB
\else
??????\fi
\fi
}
\newcommand{\hatcurISOmlong}[1]{\ifnum#1=37 %
\hatcurISOmlongxxxxxA
\else
\ifnum#1=38 %
\hatcurISOmlongxxxxxB
\else
??????\fi
\fi
}
\newcommand{\hatcurISOmshort}[1]{\ifnum#1=37 %
\hatcurISOmshortxxxxxA
\else
\ifnum#1=38 %
\hatcurISOmshortxxxxxB
\else
??????\fi
\fi
}
\newcommand{\hatcurISOr}[1]{\ifnum#1=37 %
\hatcurISOrxxxxxA
\else
\ifnum#1=38 %
\hatcurISOrxxxxxB
\else
??????\fi
\fi
}
\newcommand{\hatcurISOrho}[1]{\ifnum#1=37 %
\hatcurISOrhoxxxxxA
\else
\ifnum#1=38 %
\hatcurISOrhoxxxxxB
\else
??????\fi
\fi
}
\newcommand{\hatcurISOrholong}[1]{\ifnum#1=37 %
\hatcurISOrholongxxxxxA
\else
\ifnum#1=38 %
\hatcurISOrholongxxxxxB
\else
??????\fi
\fi
}
\newcommand{\hatcurISOrlong}[1]{\ifnum#1=37 %
\hatcurISOrlongxxxxxA
\else
\ifnum#1=38 %
\hatcurISOrlongxxxxxB
\else
??????\fi
\fi
}
\newcommand{\hatcurISOrshort}[1]{\ifnum#1=37 %
\hatcurISOrshortxxxxxA
\else
\ifnum#1=38 %
\hatcurISOrshortxxxxxB
\else
??????\fi
\fi
}
\newcommand{\hatcurISOspec}[1]{\ifnum#1=37 %
\hatcurISOspecxxxxxA
\else
\ifnum#1=38 %
\hatcurISOspecxxxxxB
\else
??????\fi
\fi
}
\newcommand{\hatcurISOteff}[1]{\ifnum#1=37 %
\hatcurISOteffxxxxxA
\else
\ifnum#1=38 %
\hatcurISOteffxxxxxB
\else
??????\fi
\fi
}
\newcommand{\hatcurISOzfeh}[1]{\ifnum#1=37 %
\hatcurISOzfehxxxxxA
\else
\ifnum#1=38 %
\hatcurISOzfehxxxxxB
\else
??????\fi
\fi
}
\newcommand{\hatcurLBiB}[1]{\ifnum#1=38 %
\hatcurLBiBxxxxxB
\else
??????\fi
}
\newcommand{\hatcurLBiC}[1]{\ifnum#1=38 %
\hatcurLBiCxxxxxB
\else
??????\fi
}
\newcommand{\hatcurLBig}[1]{\ifnum#1=37 %
\hatcurLBigxxxxxA
\else
\ifnum#1=38 %
\hatcurLBigxxxxxB
\else
??????\fi
\fi
}
\newcommand{\hatcurLBiH}[1]{\ifnum#1=38 %
\hatcurLBiHxxxxxB
\else
??????\fi
}
\newcommand{\hatcurLBii}[1]{\ifnum#1=37 %
\hatcurLBiixxxxxA
\else
\ifnum#1=38 %
\hatcurLBiixxxxxB
\else
??????\fi
\fi
}
\newcommand{\hatcurLBiI}[1]{\ifnum#1=37 %
\hatcurLBiIxxxxxA
\else
\ifnum#1=38 %
\hatcurLBiIxxxxxB
\else
??????\fi
\fi
}
\newcommand{\hatcurLBiiB}[1]{\ifnum#1=38 %
\hatcurLBiiBxxxxxB
\else
??????\fi
}
\newcommand{\hatcurLBiiC}[1]{\ifnum#1=38 %
\hatcurLBiiCxxxxxB
\else
??????\fi
}
\newcommand{\hatcurLBiig}[1]{\ifnum#1=37 %
\hatcurLBiigxxxxxA
\else
\ifnum#1=38 %
\hatcurLBiigxxxxxB
\else
??????\fi
\fi
}
\newcommand{\hatcurLBiiH}[1]{\ifnum#1=38 %
\hatcurLBiiHxxxxxB
\else
??????\fi
}
\newcommand{\hatcurLBiii}[1]{\ifnum#1=37 %
\hatcurLBiiixxxxxA
\else
\ifnum#1=38 %
\hatcurLBiiixxxxxB
\else
??????\fi
\fi
}
\newcommand{\hatcurLBiiI}[1]{\ifnum#1=37 %
\hatcurLBiiIxxxxxA
\else
\ifnum#1=38 %
\hatcurLBiiIxxxxxB
\else
??????\fi
\fi
}
\newcommand{\hatcurLBiiJ}[1]{\ifnum#1=38 %
\hatcurLBiiJxxxxxB
\else
??????\fi
}
\newcommand{\hatcurLBiiK}[1]{\ifnum#1=38 %
\hatcurLBiiKxxxxxB
\else
??????\fi
}
\newcommand{\hatcurLBiikep}[1]{\ifnum#1=37 %
\hatcurLBiikepxxxxxA
\else
\ifnum#1=38 %
\hatcurLBiikepxxxxxB
\else
??????\fi
\fi
}
\newcommand{\hatcurLBiiM}[1]{\ifnum#1=38 %
\hatcurLBiiMxxxxxB
\else
??????\fi
}
\newcommand{\hatcurLBiir}[1]{\ifnum#1=37 %
\hatcurLBiirxxxxxA
\else
\ifnum#1=38 %
\hatcurLBiirxxxxxB
\else
??????\fi
\fi
}
\newcommand{\hatcurLBiiR}[1]{\ifnum#1=37 %
\hatcurLBiiRxxxxxA
\else
\ifnum#1=38 %
\hatcurLBiiRxxxxxB
\else
??????\fi
\fi
}
\newcommand{\hatcurLBiiT}[1]{\ifnum#1=38 %
\hatcurLBiiTxxxxxB
\else
??????\fi
}
\newcommand{\hatcurLBiiu}[1]{\ifnum#1=38 %
\hatcurLBiiuxxxxxB
\else
??????\fi
}
\newcommand{\hatcurLBiiV}[1]{\ifnum#1=38 %
\hatcurLBiiVxxxxxB
\else
??????\fi
}
\newcommand{\hatcurLBiiz}[1]{\ifnum#1=37 %
\hatcurLBiizxxxxxA
\else
\ifnum#1=38 %
\hatcurLBiizxxxxxB
\else
??????\fi
\fi
}
\newcommand{\hatcurLBiJ}[1]{\ifnum#1=38 %
\hatcurLBiJxxxxxB
\else
??????\fi
}
\newcommand{\hatcurLBiK}[1]{\ifnum#1=38 %
\hatcurLBiKxxxxxB
\else
??????\fi
}
\newcommand{\hatcurLBikep}[1]{\ifnum#1=37 %
\hatcurLBikepxxxxxA
\else
\ifnum#1=38 %
\hatcurLBikepxxxxxB
\else
??????\fi
\fi
}
\newcommand{\hatcurLBiM}[1]{\ifnum#1=38 %
\hatcurLBiMxxxxxB
\else
??????\fi
}
\newcommand{\hatcurLBir}[1]{\ifnum#1=37 %
\hatcurLBirxxxxxA
\else
\ifnum#1=38 %
\hatcurLBirxxxxxB
\else
??????\fi
\fi
}
\newcommand{\hatcurLBiR}[1]{\ifnum#1=37 %
\hatcurLBiRxxxxxA
\else
\ifnum#1=38 %
\hatcurLBiRxxxxxB
\else
??????\fi
\fi
}
\newcommand{\hatcurLBiT}[1]{\ifnum#1=38 %
\hatcurLBiTxxxxxB
\else
??????\fi
}
\newcommand{\hatcurLBiu}[1]{\ifnum#1=38 %
\hatcurLBiuxxxxxB
\else
??????\fi
}
\newcommand{\hatcurLBiV}[1]{\ifnum#1=38 %
\hatcurLBiVxxxxxB
\else
??????\fi
}
\newcommand{\hatcurLBiz}[1]{\ifnum#1=37 %
\hatcurLBizxxxxxA
\else
\ifnum#1=38 %
\hatcurLBizxxxxxB
\else
??????\fi
\fi
}
\newcommand{\hatcurLCbsq}[1]{\ifnum#1=37 %
\hatcurLCbsqxxxxxA
\else
\ifnum#1=38 %
\hatcurLCbsqxxxxxB
\else
??????\fi
\fi
}
\newcommand{\hatcurLCdip}[1]{\ifnum#1=37 %
\hatcurLCdipxxxxxA
\else
\ifnum#1=38 %
\hatcurLCdipxxxxxB
\else
??????\fi
\fi
}
\newcommand{\hatcurLCdur}[1]{\ifnum#1=37 %
\hatcurLCdurxxxxxA
\else
\ifnum#1=38 %
\hatcurLCdurxxxxxB
\else
??????\fi
\fi
}
\newcommand{\hatcurLCdurhr}[1]{\ifnum#1=37 %
\hatcurLCdurhrxxxxxA
\else
\ifnum#1=38 %
\hatcurLCdurhrxxxxxB
\else
??????\fi
\fi
}
\newcommand{\hatcurLCdurhrshort}[1]{\ifnum#1=37 %
\hatcurLCdurhrshortxxxxxA
\else
\ifnum#1=38 %
\hatcurLCdurhrshortxxxxxB
\else
??????\fi
\fi
}
\newcommand{\hatcurLCdurshort}[1]{\ifnum#1=37 %
\hatcurLCdurshortxxxxxA
\else
\ifnum#1=38 %
\hatcurLCdurshortxxxxxB
\else
??????\fi
\fi
}
\newcommand{\hatcurLChatnetm}[1]{\ifnum#1=37 %
\hatcurLChatnetmxxxxxA
\else
??????\fi
}
\newcommand{\hatcurLChatnetmA}[1]{\ifnum#1=38 %
\hatcurLChatnetmAxxxxxB
\else
??????\fi
}
\newcommand{\hatcurLChatnetmB}[1]{\ifnum#1=38 %
\hatcurLChatnetmBxxxxxB
\else
??????\fi
}
\newcommand{\hatcurLCiblend}[1]{\ifnum#1=37 %
\hatcurLCiblendxxxxxA
\else
??????\fi
}
\newcommand{\hatcurLCiblendA}[1]{\ifnum#1=38 %
\hatcurLCiblendAxxxxxB
\else
??????\fi
}
\newcommand{\hatcurLCiblendB}[1]{\ifnum#1=38 %
\hatcurLCiblendBxxxxxB
\else
??????\fi
}
\newcommand{\hatcurLCimp}[1]{\ifnum#1=37 %
\hatcurLCimpxxxxxA
\else
\ifnum#1=38 %
\hatcurLCimpxxxxxB
\else
??????\fi
\fi
}
\newcommand{\hatcurLCingdur}[1]{\ifnum#1=37 %
\hatcurLCingdurxxxxxA
\else
\ifnum#1=38 %
\hatcurLCingdurxxxxxB
\else
??????\fi
\fi
}
\newcommand{\hatcurLCP}[1]{\ifnum#1=37 %
\hatcurLCPxxxxxA
\else
\ifnum#1=38 %
\hatcurLCPxxxxxB
\else
??????\fi
\fi
}
\newcommand{\hatcurLCPprec}[1]{\ifnum#1=37 %
\hatcurLCPprecxxxxxA
\else
\ifnum#1=38 %
\hatcurLCPprecxxxxxB
\else
??????\fi
\fi
}
\newcommand{\hatcurLCPshort}[1]{\ifnum#1=37 %
\hatcurLCPshortxxxxxA
\else
\ifnum#1=38 %
\hatcurLCPshortxxxxxB
\else
??????\fi
\fi
}
\newcommand{\hatcurLCq}[1]{\ifnum#1=37 %
\hatcurLCqxxxxxA
\else
\ifnum#1=38 %
\hatcurLCqxxxxxB
\else
??????\fi
\fi
}
\newcommand{\hatcurLCqshort}[1]{\ifnum#1=37 %
\hatcurLCqshortxxxxxA
\else
\ifnum#1=38 %
\hatcurLCqshortxxxxxB
\else
??????\fi
\fi
}
\newcommand{\hatcurLCrho}[1]{\ifnum#1=37 %
\hatcurLCrhoxxxxxA
\else
\ifnum#1=38 %
\hatcurLCrhoxxxxxB
\else
??????\fi
\fi
}
\newcommand{\hatcurLCrprstar}[1]{\ifnum#1=37 %
\hatcurLCrprstarxxxxxA
\else
\ifnum#1=38 %
\hatcurLCrprstarxxxxxB
\else
??????\fi
\fi
}
\newcommand{\hatcurLCT}[1]{\ifnum#1=37 %
\hatcurLCTxxxxxA
\else
\ifnum#1=38 %
\hatcurLCTxxxxxB
\else
??????\fi
\fi
}
\newcommand{\hatcurLCTA}[1]{\ifnum#1=37 %
\hatcurLCTAxxxxxA
\else
\ifnum#1=38 %
\hatcurLCTAxxxxxB
\else
??????\fi
\fi
}
\newcommand{\hatcurLCTB}[1]{\ifnum#1=37 %
\hatcurLCTBxxxxxA
\else
\ifnum#1=38 %
\hatcurLCTBxxxxxB
\else
??????\fi
\fi
}
\newcommand{\hatcurLCzeta}[1]{\ifnum#1=37 %
\hatcurLCzetaxxxxxA
\else
\ifnum#1=38 %
\hatcurLCzetaxxxxxB
\else
??????\fi
\fi
}
\newcommand{\hatcurPPaequiv}[1]{\ifnum#1=37 %
\hatcurPPaequivxxxxxA
\else
\ifnum#1=38 %
\hatcurPPaequivxxxxxB
\else
??????\fi
\fi
}
\newcommand{\hatcurPPar}[1]{\ifnum#1=37 %
\hatcurPParxxxxxA
\else
\ifnum#1=38 %
\hatcurPParxxxxxB
\else
??????\fi
\fi
}
\newcommand{\hatcurPParel}[1]{\ifnum#1=37 %
\hatcurPParelxxxxxA
\else
\ifnum#1=38 %
\hatcurPParelxxxxxB
\else
??????\fi
\fi
}
\newcommand{\hatcurPPfluxap}[1]{\ifnum#1=37 %
\hatcurPPfluxapxxxxxA
\else
\ifnum#1=38 %
\hatcurPPfluxapxxxxxB
\else
??????\fi
\fi
}
\newcommand{\hatcurPPfluxapdim}[1]{\ifnum#1=37 %
\hatcurPPfluxapdimxxxxxA
\else
\ifnum#1=38 %
\hatcurPPfluxapdimxxxxxB
\else
??????\fi
\fi
}
\newcommand{\hatcurPPfluxavg}[1]{\ifnum#1=37 %
\hatcurPPfluxavgxxxxxA
\else
\ifnum#1=38 %
\hatcurPPfluxavgxxxxxB
\else
??????\fi
\fi
}
\newcommand{\hatcurPPfluxavgdim}[1]{\ifnum#1=37 %
\hatcurPPfluxavgdimxxxxxA
\else
\ifnum#1=38 %
\hatcurPPfluxavgdimxxxxxB
\else
??????\fi
\fi
}
\newcommand{\hatcurPPfluxavglog}[1]{\ifnum#1=37 %
\hatcurPPfluxavglogxxxxxA
\else
\ifnum#1=38 %
\hatcurPPfluxavglogxxxxxB
\else
??????\fi
\fi
}
\newcommand{\hatcurPPfluxperi}[1]{\ifnum#1=37 %
\hatcurPPfluxperixxxxxA
\else
\ifnum#1=38 %
\hatcurPPfluxperixxxxxB
\else
??????\fi
\fi
}
\newcommand{\hatcurPPfluxperidim}[1]{\ifnum#1=37 %
\hatcurPPfluxperidimxxxxxA
\else
\ifnum#1=38 %
\hatcurPPfluxperidimxxxxxB
\else
??????\fi
\fi
}
\newcommand{\hatcurPPg}[1]{\ifnum#1=37 %
\hatcurPPgxxxxxA
\else
\ifnum#1=38 %
\hatcurPPgxxxxxB
\else
??????\fi
\fi
}
\newcommand{\hatcurPPi}[1]{\ifnum#1=37 %
\hatcurPPixxxxxA
\else
\ifnum#1=38 %
\hatcurPPixxxxxB
\else
??????\fi
\fi
}
\newcommand{\hatcurPPlogg}[1]{\ifnum#1=37 %
\hatcurPPloggxxxxxA
\else
\ifnum#1=38 %
\hatcurPPloggxxxxxB
\else
??????\fi
\fi
}
\newcommand{\hatcurPPm}[1]{\ifnum#1=37 %
\hatcurPPmxxxxxA
\else
\ifnum#1=38 %
\hatcurPPmxxxxxB
\else
??????\fi
\fi
}
\newcommand{\hatcurPPme}[1]{\ifnum#1=37 %
\hatcurPPmexxxxxA
\else
\ifnum#1=38 %
\hatcurPPmexxxxxB
\else
??????\fi
\fi
}
\newcommand{\hatcurPPmelong}[1]{\ifnum#1=37 %
\hatcurPPmelongxxxxxA
\else
\ifnum#1=38 %
\hatcurPPmelongxxxxxB
\else
??????\fi
\fi
}
\newcommand{\hatcurPPmeshort}[1]{\ifnum#1=37 %
\hatcurPPmeshortxxxxxA
\else
\ifnum#1=38 %
\hatcurPPmeshortxxxxxB
\else
??????\fi
\fi
}
\newcommand{\hatcurPPmlong}[1]{\ifnum#1=37 %
\hatcurPPmlongxxxxxA
\else
\ifnum#1=38 %
\hatcurPPmlongxxxxxB
\else
??????\fi
\fi
}
\newcommand{\hatcurPPmrcorr}[1]{\ifnum#1=37 %
\hatcurPPmrcorrxxxxxA
\else
\ifnum#1=38 %
\hatcurPPmrcorrxxxxxB
\else
??????\fi
\fi
}
\newcommand{\hatcurPPmshort}[1]{\ifnum#1=37 %
\hatcurPPmshortxxxxxA
\else
\ifnum#1=38 %
\hatcurPPmshortxxxxxB
\else
??????\fi
\fi
}
\newcommand{\hatcurPPperi}[1]{\ifnum#1=37 %
\hatcurPPperixxxxxA
\else
\ifnum#1=38 %
\hatcurPPperixxxxxB
\else
??????\fi
\fi
}
\newcommand{\hatcurPPphiconj}[1]{\ifnum#1=37 %
\hatcurPPphiconjxxxxxA
\else
\ifnum#1=38 %
\hatcurPPphiconjxxxxxB
\else
??????\fi
\fi
}
\newcommand{\hatcurPPr}[1]{\ifnum#1=37 %
\hatcurPPrxxxxxA
\else
\ifnum#1=38 %
\hatcurPPrxxxxxB
\else
??????\fi
\fi
}
\newcommand{\hatcurPPre}[1]{\ifnum#1=37 %
\hatcurPPrexxxxxA
\else
\ifnum#1=38 %
\hatcurPPrexxxxxB
\else
??????\fi
\fi
}
\newcommand{\hatcurPPrelong}[1]{\ifnum#1=37 %
\hatcurPPrelongxxxxxA
\else
\ifnum#1=38 %
\hatcurPPrelongxxxxxB
\else
??????\fi
\fi
}
\newcommand{\hatcurPPreshort}[1]{\ifnum#1=37 %
\hatcurPPreshortxxxxxA
\else
\ifnum#1=38 %
\hatcurPPreshortxxxxxB
\else
??????\fi
\fi
}
\newcommand{\hatcurPPrho}[1]{\ifnum#1=37 %
\hatcurPPrhoxxxxxA
\else
\ifnum#1=38 %
\hatcurPPrhoxxxxxB
\else
??????\fi
\fi
}
\newcommand{\hatcurPPrlong}[1]{\ifnum#1=37 %
\hatcurPPrlongxxxxxA
\else
\ifnum#1=38 %
\hatcurPPrlongxxxxxB
\else
??????\fi
\fi
}
\newcommand{\hatcurPPrshort}[1]{\ifnum#1=37 %
\hatcurPPrshortxxxxxA
\else
\ifnum#1=38 %
\hatcurPPrshortxxxxxB
\else
??????\fi
\fi
}
\newcommand{\hatcurPPtcirc}[1]{\ifnum#1=37 %
\hatcurPPtcircxxxxxA
\else
\ifnum#1=38 %
\hatcurPPtcircxxxxxB
\else
??????\fi
\fi
}
\newcommand{\hatcurPPteff}[1]{\ifnum#1=37 %
\hatcurPPteffxxxxxA
\else
\ifnum#1=38 %
\hatcurPPteffxxxxxB
\else
??????\fi
\fi
}
\newcommand{\hatcurPPtheta}[1]{\ifnum#1=37 %
\hatcurPPthetaxxxxxA
\else
\ifnum#1=38 %
\hatcurPPthetaxxxxxB
\else
??????\fi
\fi
}
\newcommand{\hatcurPPtinfall}[1]{\ifnum#1=37 %
\hatcurPPtinfallxxxxxA
\else
\ifnum#1=38 %
\hatcurPPtinfallxxxxxB
\else
??????\fi
\fi
}
\newcommand{\hatcurRVeccen}[1]{\ifnum#1=37 %
\hatcurRVeccenxxxxxA
\else
\ifnum#1=38 %
\hatcurRVeccenxxxxxB
\else
??????\fi
\fi
}
\newcommand{\hatcurRVeccentwosiglim}[1]{\ifnum#1=37 %
\hatcurRVeccentwosiglimxxxxxA
\else
\ifnum#1=38 %
\hatcurRVeccentwosiglimxxxxxB
\else
??????\fi
\fi
}
\newcommand{\hatcurRVfitrmsA}[1]{\ifnum#1=37 %
\hatcurRVfitrmsAxxxxxA
\else
\ifnum#1=38 %
\hatcurRVfitrmsAxxxxxB
\else
??????\fi
\fi
}
\newcommand{\hatcurRVfitrmsB}[1]{\ifnum#1=37 %
\hatcurRVfitrmsBxxxxxA
\else
\ifnum#1=38 %
\hatcurRVfitrmsBxxxxxB
\else
??????\fi
\fi
}
\newcommand{\hatcurRVfitrmsC}[1]{\ifnum#1=38 %
\hatcurRVfitrmsCxxxxxB
\else
??????\fi
}
\newcommand{\hatcurRVgammaA}[1]{\ifnum#1=37 %
\hatcurRVgammaAxxxxxA
\else
\ifnum#1=38 %
\hatcurRVgammaAxxxxxB
\else
??????\fi
\fi
}
\newcommand{\hatcurRVgammaB}[1]{\ifnum#1=37 %
\hatcurRVgammaBxxxxxA
\else
\ifnum#1=38 %
\hatcurRVgammaBxxxxxB
\else
??????\fi
\fi
}
\newcommand{\hatcurRVgammaC}[1]{\ifnum#1=38 %
\hatcurRVgammaCxxxxxB
\else
??????\fi
}
\newcommand{\hatcurRVh}[1]{\ifnum#1=37 %
\hatcurRVhxxxxxA
\else
\ifnum#1=38 %
\hatcurRVhxxxxxB
\else
??????\fi
\fi
}
\newcommand{\hatcurRVjitterA}[1]{\ifnum#1=37 %
\hatcurRVjitterAxxxxxA
\else
\ifnum#1=38 %
\hatcurRVjitterAxxxxxB
\else
??????\fi
\fi
}
\newcommand{\hatcurRVjitterB}[1]{\ifnum#1=37 %
\hatcurRVjitterBxxxxxA
\else
\ifnum#1=38 %
\hatcurRVjitterBxxxxxB
\else
??????\fi
\fi
}
\newcommand{\hatcurRVjitterC}[1]{\ifnum#1=38 %
\hatcurRVjitterCxxxxxB
\else
??????\fi
}
\newcommand{\hatcurRVjittertwosiglimA}[1]{\ifnum#1=37 %
\hatcurRVjittertwosiglimAxxxxxA
\else
\ifnum#1=38 %
\hatcurRVjittertwosiglimAxxxxxB
\else
??????\fi
\fi
}
\newcommand{\hatcurRVjittertwosiglimB}[1]{\ifnum#1=37 %
\hatcurRVjittertwosiglimBxxxxxA
\else
\ifnum#1=38 %
\hatcurRVjittertwosiglimBxxxxxB
\else
??????\fi
\fi
}
\newcommand{\hatcurRVjittertwosiglimC}[1]{\ifnum#1=38 %
\hatcurRVjittertwosiglimCxxxxxB
\else
??????\fi
}
\newcommand{\hatcurRVk}[1]{\ifnum#1=37 %
\hatcurRVkxxxxxA
\else
\ifnum#1=38 %
\hatcurRVkxxxxxB
\else
??????\fi
\fi
}
\newcommand{\hatcurRVK}[1]{\ifnum#1=37 %
\hatcurRVKxxxxxA
\else
\ifnum#1=38 %
\hatcurRVKxxxxxB
\else
??????\fi
\fi
}
\newcommand{\hatcurRVomega}[1]{\ifnum#1=37 %
\hatcurRVomegaxxxxxA
\else
\ifnum#1=38 %
\hatcurRVomegaxxxxxB
\else
??????\fi
\fi
}
\newcommand{\hatcurRVrh}[1]{\ifnum#1=37 %
\hatcurRVrhxxxxxA
\else
\ifnum#1=38 %
\hatcurRVrhxxxxxB
\else
??????\fi
\fi
}
\newcommand{\hatcurRVrk}[1]{\ifnum#1=37 %
\hatcurRVrkxxxxxA
\else
\ifnum#1=38 %
\hatcurRVrkxxxxxB
\else
??????\fi
\fi
}
\newcommand{\hatcurRVtrone}[1]{\ifnum#1=37 %
\hatcurRVtronexxxxxA
\else
\ifnum#1=38 %
\hatcurRVtronexxxxxB
\else
??????\fi
\fi
}
\newcommand{\hatcurRVtrtwo}[1]{\ifnum#1=37 %
\hatcurRVtrtwoxxxxxA
\else
\ifnum#1=38 %
\hatcurRVtrtwoxxxxxB
\else
??????\fi
\fi
}
\newcommand{\hatcurSMEiilogg}[1]{\ifnum#1=37 %
\hatcurSMEiiloggxxxxxA
\else
\ifnum#1=38 %
\hatcurSMEiiloggxxxxxB
\else
??????\fi
\fi
}
\newcommand{\hatcurSMEiiteff}[1]{\ifnum#1=37 %
\hatcurSMEiiteffxxxxxA
\else
\ifnum#1=38 %
\hatcurSMEiiteffxxxxxB
\else
??????\fi
\fi
}
\newcommand{\hatcurSMEiivmac}[1]{\ifnum#1=38 %
\hatcurSMEiivmacxxxxxB
\else
??????\fi
}
\newcommand{\hatcurSMEiivmic}[1]{\ifnum#1=38 %
\hatcurSMEiivmicxxxxxB
\else
??????\fi
}
\newcommand{\hatcurSMEiivsin}[1]{\ifnum#1=37 %
\hatcurSMEiivsinxxxxxA
\else
\ifnum#1=38 %
\hatcurSMEiivsinxxxxxB
\else
??????\fi
\fi
}
\newcommand{\hatcurSMEiizfeh}[1]{\ifnum#1=37 %
\hatcurSMEiizfehxxxxxA
\else
\ifnum#1=38 %
\hatcurSMEiizfehxxxxxB
\else
??????\fi
\fi
}
\newcommand{\hatcurSMEiizfehshort}[1]{\ifnum#1=37 %
\hatcurSMEiizfehshortxxxxxA
\else
\ifnum#1=38 %
\hatcurSMEiizfehshortxxxxxB
\else
??????\fi
\fi
}
\newcommand{\hatcurSMEilogg}[1]{\ifnum#1=37 %
\hatcurSMEiloggxxxxxA
\else
\ifnum#1=38 %
\hatcurSMEiloggxxxxxB
\else
??????\fi
\fi
}
\newcommand{\hatcurSMEiteff}[1]{\ifnum#1=37 %
\hatcurSMEiteffxxxxxA
\else
\ifnum#1=38 %
\hatcurSMEiteffxxxxxB
\else
??????\fi
\fi
}
\newcommand{\hatcurSMEivmac}[1]{\ifnum#1=37 %
\hatcurSMEivmacxxxxxA
\else
\ifnum#1=38 %
\hatcurSMEivmacxxxxxB
\else
??????\fi
\fi
}
\newcommand{\hatcurSMEivmic}[1]{\ifnum#1=37 %
\hatcurSMEivmicxxxxxA
\else
\ifnum#1=38 %
\hatcurSMEivmicxxxxxB
\else
??????\fi
\fi
}
\newcommand{\hatcurSMEivsin}[1]{\ifnum#1=37 %
\hatcurSMEivsinxxxxxA
\else
\ifnum#1=38 %
\hatcurSMEivsinxxxxxB
\else
??????\fi
\fi
}
\newcommand{\hatcurSMEizfeh}[1]{\ifnum#1=37 %
\hatcurSMEizfehxxxxxA
\else
\ifnum#1=38 %
\hatcurSMEizfehxxxxxB
\else
??????\fi
\fi
}
\newcommand{\hatcurSMEizfehshort}[1]{\ifnum#1=37 %
\hatcurSMEizfehshortxxxxxA
\else
\ifnum#1=38 %
\hatcurSMEizfehshortxxxxxB
\else
??????\fi
\fi
}
\newcommand{\hatcurXAv}[1]{\ifnum#1=37 %
\hatcurXAvxxxxxA
\else
\ifnum#1=38 %
\hatcurXAvxxxxxB
\else
??????\fi
\fi
}
\newcommand{\hatcurXdist}[1]{\ifnum#1=37 %
\hatcurXdistxxxxxA
\else
\ifnum#1=38 %
\hatcurXdistxxxxxB
\else
??????\fi
\fi
}
\newcommand{\hatcurXdistred}[1]{\ifnum#1=37 %
\hatcurXdistredxxxxxA
\else
\ifnum#1=38 %
\hatcurXdistredxxxxxB
\else
??????\fi
\fi
}
\newcommand{\hatcurXEBV}[1]{\ifnum#1=37 %
\hatcurXEBVxxxxxA
\else
\ifnum#1=38 %
\hatcurXEBVxxxxxB
\else
??????\fi
\fi
}
\newcommand{\hatcurXsecdur}[1]{\ifnum#1=37 %
\hatcurXsecdurxxxxxA
\else
\ifnum#1=38 %
\hatcurXsecdurxxxxxB
\else
??????\fi
\fi
}
\newcommand{\hatcurXsecingdur}[1]{\ifnum#1=37 %
\hatcurXsecingdurxxxxxA
\else
\ifnum#1=38 %
\hatcurXsecingdurxxxxxB
\else
??????\fi
\fi
}
\newcommand{\hatcurXsecondary}[1]{\ifnum#1=37 %
\hatcurXsecondaryxxxxxA
\else
\ifnum#1=38 %
\hatcurXsecondaryxxxxxB
\else
??????\fi
\fi
}
\newcommand{\hatcurXsecphase}[1]{\ifnum#1=37 %
\hatcurXsecphasexxxxxA
\else
\ifnum#1=38 %
\hatcurXsecphasexxxxxB
\else
??????\fi
\fi
}

\newcommand{\hatcurhtrhpsxxxxA}{HATS567-001} 
\newcommand{\hatcurLCrprstarhpsxxxxA}{\ensuremath{0.0707\pm0.0018}} 
\newcommand{\hatcurLCbsqhpsxxxxA}{\ensuremath{0.020_{-0.017}^{+0.038}}} 
\newcommand{\hatcurLCimphpsxxxxA}{\ensuremath{0.140_{-0.092}^{+0.100}}} 
\newcommand{\hatcurLCzetahpsxxxxA}{\ensuremath{17.67\pm0.14}}     
\newcommand{\hatcurLCdurhpsxxxxA}{\ensuremath{0.1214\pm0.0010}}   
\newcommand{\hatcurLCdurshorthpsxxxxA}{\ensuremath{0.1214}}       
\newcommand{\hatcurLCingdurhpsxxxxA}{\ensuremath{0.00822\pm0.00030}} 
\newcommand{\hatcurLCPhpsxxxxA}{\ensuremath{4.3315366\pm0.0000041}} 
\newcommand{\hatcurLCPprechpsxxxxA}{\ensuremath{4.3315366}}       
\newcommand{\hatcurLCPshorthpsxxxxA}{\ensuremath{4.3315}}         
\newcommand{\hatcurLCThpsxxxxA}{\ensuremath{2458006.80145\pm0.00050}} 
\newcommand{\hatcurLCTAhpsxxxxA}{\ensuremath{2455646.1140\pm0.0022}} 
\newcommand{\hatcurLCTBhpsxxxxA}{\ensuremath{2458214.71524\pm0.00056}} 
\newcommand{\hatcurLCiblendhpsxxxxA}{\ensuremath{1.000\pm0.063}}  
\newcommand{\hatcurLCrhohpsxxxxA}{\ensuremath{1.762_{-0.098}^{+0.065}}} 
\newcommand{\hatcurISOmhpsxxxxA}{\ensuremath{0.843_{-0.012}^{+0.017}}} 
\newcommand{\hatcurISOmshorthpsxxxxA}{\ensuremath{0.84}}          
\newcommand{\hatcurISOmlonghpsxxxxA}{\ensuremath{0.843_{-0.012}^{+0.017}}} 
\newcommand{\hatcurISOrhpsxxxxA}{\ensuremath{0.877_{-0.012}^{+0.019}}} 
\newcommand{\hatcurISOrshorthpsxxxxA}{\ensuremath{0.88}}          
\newcommand{\hatcurISOrlonghpsxxxxA}{\ensuremath{0.877_{-0.012}^{+0.019}}} 
\newcommand{\hatcurISOlogghpsxxxxA}{\ensuremath{4.478\pm0.017}}   
\newcommand{\hatcurISOlumhpsxxxxA}{\ensuremath{0.555_{-0.028}^{+0.038}}} 
\newcommand{\hatcurISOlumshorthpsxxxxA}{\ensuremath{0.56}}        
\newcommand{\hatcurISOfehhpsxxxxA}{\ensuremath{0.051\pm0.029}}    
\newcommand{\hatcurISOteffhpsxxxxA}{\ensuremath{5326\pm44}}       
\newcommand{\hatcurISOagehpsxxxxA}{\ensuremath{11.46_{-1.45}^{+0.79}}} 
\newcommand{\hatcurISOmBhpsxxxxA}{\ensuremath{0.654\pm0.033}}     
\newcommand{\hatcurISOmshortBhpsxxxxA}{\ensuremath{0.65}}         
\newcommand{\hatcurISOmlongBhpsxxxxA}{\ensuremath{0.654\pm0.033}} 
\newcommand{\hatcurISOrBhpsxxxxA}{\ensuremath{0.654\pm0.032}}     
\newcommand{\hatcurISOrshortBhpsxxxxA}{\ensuremath{0.65}}         
\newcommand{\hatcurISOrlongBhpsxxxxA}{\ensuremath{0.654\pm0.032}} 
\newcommand{\hatcurISOloggBhpsxxxxA}{\ensuremath{4.622\pm0.023}}  
\newcommand{\hatcurISOlumBhpsxxxxA}{\ensuremath{0.120\pm0.023}}   
\newcommand{\hatcurISOlumshortBhpsxxxxA}{\ensuremath{0.12}}       
\newcommand{\hatcurISOfehBhpsxxxxA}{\ensuremath{0.051\pm0.029}}   
\newcommand{\hatcurISOteffBhpsxxxxA}{\ensuremath{4210\pm170}}     
\newcommand{\hatcurISOageBhpsxxxxA}{\ensuremath{11.46_{-1.45}^{+0.79}}} 
\newcommand{\hatcurPPihpsxxxxA}{\ensuremath{89.33\pm0.45}}        
\newcommand{\hatcurPParhpsxxxxA}{\ensuremath{12.05_{-0.23}^{+0.15}}} 
\newcommand{\hatcurPParelhpsxxxxA}{\ensuremath{0.04913_{-0.00023}^{+0.00033}}} 
\newcommand{\hatcurPPrhpsxxxxA}{\ensuremath{0.606\pm0.016}}       
\newcommand{\hatcurPPrshorthpsxxxxA}{\ensuremath{0.61}}           
\newcommand{\hatcurPPrlonghpsxxxxA}{\ensuremath{0.606\pm0.016}}   
\newcommand{\hatcurPPteffhpsxxxxA}{\ensuremath{1085_{-12}^{+16}}} 
\newcommand{\hatcurPPfluxavghpsxxxxA}{\ensuremath{313000000_{-14000000}^{+19000000}}} 
\newcommand{\hatcurPPfluxavgloghpsxxxxA}{\ensuremath{8.495_{-0.020}^{+0.026}}} 
\newcommand{\hatcurXAvhpsxxxxA}{\ensuremath{0.258\pm0.062}}       
\newcommand{\hatcurXdisthpsxxxxA}{\ensuremath{211.1\pm2.5}}       
\newcommand{\hatcurXdistredhpsxxxxA}{\ensuremath{211.1\pm2.5}}    
\newcommand{\hatcurRVKhpsxxxxA}{\ensuremath{13.9\pm5.8}}          
\newcommand{\hatcurPPlogghpsxxxxA}{\ensuremath{2.83\pm0.19}}      
\newcommand{\hatcurPPrhohpsxxxxA}{\ensuremath{0.55\pm0.24}}       
\newcommand{\hatcurPPmtwosiglimhpsxxxxA}{\ensuremath{<0.179}}     
\newcommand{\hatcurPPmhpsxxxxA}{\ensuremath{0.099\pm0.042}}       
\newcommand{\hatcurPPmshorthpsxxxxA}{\ensuremath{0.10}}           
\newcommand{\hatcurPPmlonghpsxxxxA}{\ensuremath{0.099\pm0.042}}   
\newcommand{\hatcurPPthetahpsxxxxA}{\ensuremath{0.0190\pm0.0080}} 

\newcommand{\hatcurhtrhpsxxxxB}{HATS561-006}
\newcommand{\hatcurLCrprstarhpsxxxxB}{\ensuremath{0.0553\pm0.0019}} 
\newcommand{\hatcurLCbsqhpsxxxxB}{\ensuremath{0.113_{-0.090}^{+0.070}}} 
\newcommand{\hatcurLCimphpsxxxxB}{\ensuremath{0.336_{-0.185}^{+0.091}}} 
\newcommand{\hatcurLCzetahpsxxxxB}{\ensuremath{15.97\pm0.21}}     
\newcommand{\hatcurLCdurhpsxxxxB}{\ensuremath{0.1330\pm0.0019}}   
\newcommand{\hatcurLCdurshorthpsxxxxB}{\ensuremath{0.1330}}       
\newcommand{\hatcurLCingdurhpsxxxxB}{\ensuremath{0.00783\pm0.00068}} 
\newcommand{\hatcurLCPhpsxxxxB}{\ensuremath{4.375019\pm0.000016}} 
\newcommand{\hatcurLCPprechpsxxxxB}{\ensuremath{4.3750189}}       
\newcommand{\hatcurLCPshorthpsxxxxB}{\ensuremath{4.3750}}         
\newcommand{\hatcurLCThpsxxxxB}{\ensuremath{2457786.41043\pm0.00076}} 
\newcommand{\hatcurLCTAhpsxxxxB}{\ensuremath{2457012.0321\pm0.0029}} 
\newcommand{\hatcurLCTBhpsxxxxB}{\ensuremath{2457847.66069\pm0.00081}} 
\newcommand{\hatcurLCiblendAhpsxxxxB}{\ensuremath{0.0000\pm0.0010}} 
\newcommand{\hatcurLCiblendBhpsxxxxB}{\ensuremath{0.000\pm0.018}} 
\newcommand{\hatcurLCrhohpsxxxxB}{\ensuremath{1.13_{-0.14}^{+0.19}}} 
\newcommand{\hatcurISOmhpsxxxxB}{\ensuremath{0.935_{-0.051}^{+0.068}}} 
\newcommand{\hatcurISOmshorthpsxxxxB}{\ensuremath{0.94}}          
\newcommand{\hatcurISOmlonghpsxxxxB}{\ensuremath{0.935_{-0.051}^{+0.068}}} 
\newcommand{\hatcurISOrhpsxxxxB}{\ensuremath{1.053\pm0.035}}      
\newcommand{\hatcurISOrshorthpsxxxxB}{\ensuremath{1.05}}          
\newcommand{\hatcurISOrlonghpsxxxxB}{\ensuremath{1.053\pm0.035}}  
\newcommand{\hatcurISOlogghpsxxxxB}{\ensuremath{4.366\pm0.044}}   
\newcommand{\hatcurISOlumhpsxxxxB}{\ensuremath{1.074_{-0.051}^{+0.070}}} 
\newcommand{\hatcurISOlumshorthpsxxxxB}{\ensuremath{1.07}}        
\newcommand{\hatcurISOfehhpsxxxxB}{\ensuremath{-0.01\pm0.11}}     
\newcommand{\hatcurISOteffhpsxxxxB}{\ensuremath{5744\pm36}}       
\newcommand{\hatcurISOagehpsxxxxB}{\ensuremath{9.0_{-3.7}^{+2.7}}} 
\newcommand{\hatcurISOmBhpsxxxxB}{\ensuremath{0.491_{-0.128}^{+0.080}}} 
\newcommand{\hatcurISOmshortBhpsxxxxB}{\ensuremath{0.49}}         
\newcommand{\hatcurISOmlongBhpsxxxxB}{\ensuremath{0.491_{-0.128}^{+0.080}}} 
\newcommand{\hatcurISOrBhpsxxxxB}{\ensuremath{0.487_{-0.123}^{+0.083}}} 
\newcommand{\hatcurISOrshortBhpsxxxxB}{\ensuremath{0.49}}         
\newcommand{\hatcurISOrlongBhpsxxxxB}{\ensuremath{0.487_{-0.123}^{+0.083}}} 
\newcommand{\hatcurISOloggBhpsxxxxB}{\ensuremath{4.75\pm0.10}}    
\newcommand{\hatcurISOlumBhpsxxxxB}{\ensuremath{0.033\pm0.052}}   
\newcommand{\hatcurISOlumshortBhpsxxxxB}{\ensuremath{0.03}}       
\newcommand{\hatcurISOfehBhpsxxxxB}{\ensuremath{-0.01\pm0.11}}    
\newcommand{\hatcurISOteffBhpsxxxxB}{\ensuremath{3520\pm350}}     
\newcommand{\hatcurISOageBhpsxxxxB}{\ensuremath{9.0_{-3.7}^{+2.7}}} 
\newcommand{\hatcurPPihpsxxxxB}{\ensuremath{88.16_{-0.60}^{+1.06}}} 
\newcommand{\hatcurPParhpsxxxxB}{\ensuremath{10.47\pm0.46}}       
\newcommand{\hatcurPParelhpsxxxxB}{\ensuremath{0.05119\pm0.00098}} 
\newcommand{\hatcurPPrhpsxxxxB}{\ensuremath{0.567\pm0.023}}       
\newcommand{\hatcurPPrshorthpsxxxxB}{\ensuremath{0.57}}           
\newcommand{\hatcurPPrlonghpsxxxxB}{\ensuremath{0.567\pm0.023}}   
\newcommand{\hatcurPPteffhpsxxxxB}{\ensuremath{1253\pm25}}        
\newcommand{\hatcurPPfluxavghpsxxxxB}{\ensuremath{556000000\pm46000000}} 
\newcommand{\hatcurPPfluxavgloghpsxxxxB}{\ensuremath{8.745\pm0.035}} 
\newcommand{\hatcurXAvhpsxxxxB}{\ensuremath{0.35_{-0.26}^{+0.53}}} 
\newcommand{\hatcurXdisthpsxxxxB}{\ensuremath{346.1\pm5.6}}       
\newcommand{\hatcurXdistredhpsxxxxB}{\ensuremath{346.1\pm5.6}}    
\newcommand{\hatcurRVKhpsxxxxB}{\ensuremath{9.9\pm2.6}}           
\newcommand{\hatcurPPlogghpsxxxxB}{\ensuremath{2.77\pm0.13}}      
\newcommand{\hatcurPPrhohpsxxxxB}{\ensuremath{0.52\pm0.16}}       
\newcommand{\hatcurPPmtwosiglimhpsxxxxB}{\ensuremath{<0.113}}     
\newcommand{\hatcurPPmhpsxxxxB}{\ensuremath{0.077\pm0.020}}       
\newcommand{\hatcurPPmshorthpsxxxxB}{\ensuremath{0.08}}           
\newcommand{\hatcurPPmlonghpsxxxxB}{\ensuremath{0.077\pm0.020}}   
\newcommand{\hatcurPPthetahpsxxxxB}{\ensuremath{0.0148\pm0.0039}} 

\newcommand{\hatcurhtrhps}[1]{\ifnum#1=37 %
\hatcurhtrhpsxxxxA
\else
\ifnum#1=38 %
\hatcurhtrhpsxxxxB
\else
??????\fi
\fi
}
\newcommand{\hatcurISOageBhps}[1]{\ifnum#1=37 %
\hatcurISOageBhpsxxxxA
\else
\ifnum#1=38 %
\hatcurISOageBhpsxxxxB
\else
??????\fi
\fi
}
\newcommand{\hatcurISOagehps}[1]{\ifnum#1=37 %
\hatcurISOagehpsxxxxA
\else
\ifnum#1=38 %
\hatcurISOagehpsxxxxB
\else
??????\fi
\fi
}
\newcommand{\hatcurISOfehBhps}[1]{\ifnum#1=37 %
\hatcurISOfehBhpsxxxxA
\else
\ifnum#1=38 %
\hatcurISOfehBhpsxxxxB
\else
??????\fi
\fi
}
\newcommand{\hatcurISOfehhps}[1]{\ifnum#1=37 %
\hatcurISOfehhpsxxxxA
\else
\ifnum#1=38 %
\hatcurISOfehhpsxxxxB
\else
??????\fi
\fi
}
\newcommand{\hatcurISOloggBhps}[1]{\ifnum#1=37 %
\hatcurISOloggBhpsxxxxA
\else
\ifnum#1=38 %
\hatcurISOloggBhpsxxxxB
\else
??????\fi
\fi
}
\newcommand{\hatcurISOlogghps}[1]{\ifnum#1=37 %
\hatcurISOlogghpsxxxxA
\else
\ifnum#1=38 %
\hatcurISOlogghpsxxxxB
\else
??????\fi
\fi
}
\newcommand{\hatcurISOlumBhps}[1]{\ifnum#1=37 %
\hatcurISOlumBhpsxxxxA
\else
\ifnum#1=38 %
\hatcurISOlumBhpsxxxxB
\else
??????\fi
\fi
}
\newcommand{\hatcurISOlumhps}[1]{\ifnum#1=37 %
\hatcurISOlumhpsxxxxA
\else
\ifnum#1=38 %
\hatcurISOlumhpsxxxxB
\else
??????\fi
\fi
}
\newcommand{\hatcurISOlumshortBhps}[1]{\ifnum#1=37 %
\hatcurISOlumshortBhpsxxxxA
\else
\ifnum#1=38 %
\hatcurISOlumshortBhpsxxxxB
\else
??????\fi
\fi
}
\newcommand{\hatcurISOlumshorthps}[1]{\ifnum#1=37 %
\hatcurISOlumshorthpsxxxxA
\else
\ifnum#1=38 %
\hatcurISOlumshorthpsxxxxB
\else
??????\fi
\fi
}
\newcommand{\hatcurISOmBhps}[1]{\ifnum#1=37 %
\hatcurISOmBhpsxxxxA
\else
\ifnum#1=38 %
\hatcurISOmBhpsxxxxB
\else
??????\fi
\fi
}
\newcommand{\hatcurISOmhps}[1]{\ifnum#1=37 %
\hatcurISOmhpsxxxxA
\else
\ifnum#1=38 %
\hatcurISOmhpsxxxxB
\else
??????\fi
\fi
}
\newcommand{\hatcurISOmlongBhps}[1]{\ifnum#1=37 %
\hatcurISOmlongBhpsxxxxA
\else
\ifnum#1=38 %
\hatcurISOmlongBhpsxxxxB
\else
??????\fi
\fi
}
\newcommand{\hatcurISOmlonghps}[1]{\ifnum#1=37 %
\hatcurISOmlonghpsxxxxA
\else
\ifnum#1=38 %
\hatcurISOmlonghpsxxxxB
\else
??????\fi
\fi
}
\newcommand{\hatcurISOmshortBhps}[1]{\ifnum#1=37 %
\hatcurISOmshortBhpsxxxxA
\else
\ifnum#1=38 %
\hatcurISOmshortBhpsxxxxB
\else
??????\fi
\fi
}
\newcommand{\hatcurISOmshorthps}[1]{\ifnum#1=37 %
\hatcurISOmshorthpsxxxxA
\else
\ifnum#1=38 %
\hatcurISOmshorthpsxxxxB
\else
??????\fi
\fi
}
\newcommand{\hatcurISOrBhps}[1]{\ifnum#1=37 %
\hatcurISOrBhpsxxxxA
\else
\ifnum#1=38 %
\hatcurISOrBhpsxxxxB
\else
??????\fi
\fi
}
\newcommand{\hatcurISOrhps}[1]{\ifnum#1=37 %
\hatcurISOrhpsxxxxA
\else
\ifnum#1=38 %
\hatcurISOrhpsxxxxB
\else
??????\fi
\fi
}
\newcommand{\hatcurISOrlongBhps}[1]{\ifnum#1=37 %
\hatcurISOrlongBhpsxxxxA
\else
\ifnum#1=38 %
\hatcurISOrlongBhpsxxxxB
\else
??????\fi
\fi
}
\newcommand{\hatcurISOrlonghps}[1]{\ifnum#1=37 %
\hatcurISOrlonghpsxxxxA
\else
\ifnum#1=38 %
\hatcurISOrlonghpsxxxxB
\else
??????\fi
\fi
}
\newcommand{\hatcurISOrshortBhps}[1]{\ifnum#1=37 %
\hatcurISOrshortBhpsxxxxA
\else
\ifnum#1=38 %
\hatcurISOrshortBhpsxxxxB
\else
??????\fi
\fi
}
\newcommand{\hatcurISOrshorthps}[1]{\ifnum#1=37 %
\hatcurISOrshorthpsxxxxA
\else
\ifnum#1=38 %
\hatcurISOrshorthpsxxxxB
\else
??????\fi
\fi
}
\newcommand{\hatcurISOteffBhps}[1]{\ifnum#1=37 %
\hatcurISOteffBhpsxxxxA
\else
\ifnum#1=38 %
\hatcurISOteffBhpsxxxxB
\else
??????\fi
\fi
}
\newcommand{\hatcurISOteffhps}[1]{\ifnum#1=37 %
\hatcurISOteffhpsxxxxA
\else
\ifnum#1=38 %
\hatcurISOteffhpsxxxxB
\else
??????\fi
\fi
}
\newcommand{\hatcurLCbsqhps}[1]{\ifnum#1=37 %
\hatcurLCbsqhpsxxxxA
\else
\ifnum#1=38 %
\hatcurLCbsqhpsxxxxB
\else
??????\fi
\fi
}
\newcommand{\hatcurLCdurhps}[1]{\ifnum#1=37 %
\hatcurLCdurhpsxxxxA
\else
\ifnum#1=38 %
\hatcurLCdurhpsxxxxB
\else
??????\fi
\fi
}
\newcommand{\hatcurLCdurshorthps}[1]{\ifnum#1=37 %
\hatcurLCdurshorthpsxxxxA
\else
\ifnum#1=38 %
\hatcurLCdurshorthpsxxxxB
\else
??????\fi
\fi
}
\newcommand{\hatcurLCiblendAhps}[1]{\ifnum#1=38 %
\hatcurLCiblendAhpsxxxxB
\else
??????\fi
}
\newcommand{\hatcurLCiblendBhps}[1]{\ifnum#1=38 %
\hatcurLCiblendBhpsxxxxB
\else
??????\fi
}
\newcommand{\hatcurLCiblendhps}[1]{\ifnum#1=37 %
\hatcurLCiblendhpsxxxxA
\else
??????\fi
}
\newcommand{\hatcurLCimphps}[1]{\ifnum#1=37 %
\hatcurLCimphpsxxxxA
\else
\ifnum#1=38 %
\hatcurLCimphpsxxxxB
\else
??????\fi
\fi
}
\newcommand{\hatcurLCingdurhps}[1]{\ifnum#1=37 %
\hatcurLCingdurhpsxxxxA
\else
\ifnum#1=38 %
\hatcurLCingdurhpsxxxxB
\else
??????\fi
\fi
}
\newcommand{\hatcurLCPhps}[1]{\ifnum#1=37 %
\hatcurLCPhpsxxxxA
\else
\ifnum#1=38 %
\hatcurLCPhpsxxxxB
\else
??????\fi
\fi
}
\newcommand{\hatcurLCPprechps}[1]{\ifnum#1=37 %
\hatcurLCPprechpsxxxxA
\else
\ifnum#1=38 %
\hatcurLCPprechpsxxxxB
\else
??????\fi
\fi
}
\newcommand{\hatcurLCPshorthps}[1]{\ifnum#1=37 %
\hatcurLCPshorthpsxxxxA
\else
\ifnum#1=38 %
\hatcurLCPshorthpsxxxxB
\else
??????\fi
\fi
}
\newcommand{\hatcurLCrhohps}[1]{\ifnum#1=37 %
\hatcurLCrhohpsxxxxA
\else
\ifnum#1=38 %
\hatcurLCrhohpsxxxxB
\else
??????\fi
\fi
}
\newcommand{\hatcurLCrprstarhps}[1]{\ifnum#1=37 %
\hatcurLCrprstarhpsxxxxA
\else
\ifnum#1=38 %
\hatcurLCrprstarhpsxxxxB
\else
??????\fi
\fi
}
\newcommand{\hatcurLCTAhps}[1]{\ifnum#1=37 %
\hatcurLCTAhpsxxxxA
\else
\ifnum#1=38 %
\hatcurLCTAhpsxxxxB
\else
??????\fi
\fi
}
\newcommand{\hatcurLCTBhps}[1]{\ifnum#1=37 %
\hatcurLCTBhpsxxxxA
\else
\ifnum#1=38 %
\hatcurLCTBhpsxxxxB
\else
??????\fi
\fi
}
\newcommand{\hatcurLCThps}[1]{\ifnum#1=37 %
\hatcurLCThpsxxxxA
\else
\ifnum#1=38 %
\hatcurLCThpsxxxxB
\else
??????\fi
\fi
}
\newcommand{\hatcurLCzetahps}[1]{\ifnum#1=37 %
\hatcurLCzetahpsxxxxA
\else
\ifnum#1=38 %
\hatcurLCzetahpsxxxxB
\else
??????\fi
\fi
}
\newcommand{\hatcurPParelhps}[1]{\ifnum#1=37 %
\hatcurPParelhpsxxxxA
\else
\ifnum#1=38 %
\hatcurPParelhpsxxxxB
\else
??????\fi
\fi
}
\newcommand{\hatcurPParhps}[1]{\ifnum#1=37 %
\hatcurPParhpsxxxxA
\else
\ifnum#1=38 %
\hatcurPParhpsxxxxB
\else
??????\fi
\fi
}
\newcommand{\hatcurPPfluxavghps}[1]{\ifnum#1=37 %
\hatcurPPfluxavghpsxxxxA
\else
\ifnum#1=38 %
\hatcurPPfluxavghpsxxxxB
\else
??????\fi
\fi
}
\newcommand{\hatcurPPfluxavgloghps}[1]{\ifnum#1=37 %
\hatcurPPfluxavgloghpsxxxxA
\else
\ifnum#1=38 %
\hatcurPPfluxavgloghpsxxxxB
\else
??????\fi
\fi
}
\newcommand{\hatcurPPihps}[1]{\ifnum#1=37 %
\hatcurPPihpsxxxxA
\else
\ifnum#1=38 %
\hatcurPPihpsxxxxB
\else
??????\fi
\fi
}
\newcommand{\hatcurPPlogghps}[1]{\ifnum#1=37 %
\hatcurPPlogghpsxxxxA
\else
\ifnum#1=38 %
\hatcurPPlogghpsxxxxB
\else
??????\fi
\fi
}
\newcommand{\hatcurPPmhps}[1]{\ifnum#1=37 %
\hatcurPPmhpsxxxxA
\else
\ifnum#1=38 %
\hatcurPPmhpsxxxxB
\else
??????\fi
\fi
}
\newcommand{\hatcurPPmlonghps}[1]{\ifnum#1=37 %
\hatcurPPmlonghpsxxxxA
\else
\ifnum#1=38 %
\hatcurPPmlonghpsxxxxB
\else
??????\fi
\fi
}
\newcommand{\hatcurPPmshorthps}[1]{\ifnum#1=37 %
\hatcurPPmshorthpsxxxxA
\else
\ifnum#1=38 %
\hatcurPPmshorthpsxxxxB
\else
??????\fi
\fi
}
\newcommand{\hatcurPPmtwosiglimhps}[1]{\ifnum#1=37 %
\hatcurPPmtwosiglimhpsxxxxA
\else
\ifnum#1=38 %
\hatcurPPmtwosiglimhpsxxxxB
\else
??????\fi
\fi
}
\newcommand{\hatcurPPrhohps}[1]{\ifnum#1=37 %
\hatcurPPrhohpsxxxxA
\else
\ifnum#1=38 %
\hatcurPPrhohpsxxxxB
\else
??????\fi
\fi
}
\newcommand{\hatcurPPrhps}[1]{\ifnum#1=37 %
\hatcurPPrhpsxxxxA
\else
\ifnum#1=38 %
\hatcurPPrhpsxxxxB
\else
??????\fi
\fi
}
\newcommand{\hatcurPPrlonghps}[1]{\ifnum#1=37 %
\hatcurPPrlonghpsxxxxA
\else
\ifnum#1=38 %
\hatcurPPrlonghpsxxxxB
\else
??????\fi
\fi
}
\newcommand{\hatcurPPrshorthps}[1]{\ifnum#1=37 %
\hatcurPPrshorthpsxxxxA
\else
\ifnum#1=38 %
\hatcurPPrshorthpsxxxxB
\else
??????\fi
\fi
}
\newcommand{\hatcurPPteffhps}[1]{\ifnum#1=37 %
\hatcurPPteffhpsxxxxA
\else
\ifnum#1=38 %
\hatcurPPteffhpsxxxxB
\else
??????\fi
\fi
}
\newcommand{\hatcurPPthetahps}[1]{\ifnum#1=37 %
\hatcurPPthetahpsxxxxA
\else
\ifnum#1=38 %
\hatcurPPthetahpsxxxxB
\else
??????\fi
\fi
}
\newcommand{\hatcurRVKhps}[1]{\ifnum#1=37 %
\hatcurRVKhpsxxxxA
\else
\ifnum#1=38 %
\hatcurRVKhpsxxxxB
\else
??????\fi
\fi
}
\newcommand{\hatcurXAvhps}[1]{\ifnum#1=37 %
\hatcurXAvhpsxxxxA
\else
\ifnum#1=38 %
\hatcurXAvhpsxxxxB
\else
??????\fi
\fi
}
\newcommand{\hatcurXdisthps}[1]{\ifnum#1=37 %
\hatcurXdisthpsxxxxA
\else
\ifnum#1=38 %
\hatcurXdisthpsxxxxB
\else
??????\fi
\fi
}
\newcommand{\hatcurXdistredhps}[1]{\ifnum#1=37 %
\hatcurXdistredhpsxxxxA
\else
\ifnum#1=38 %
\hatcurXdistredhpsxxxxB
\else
??????\fi
\fi
}

\newcommand{\hatcurhtreccenxxxxxA}{HATS567-001}                      
\newcommand{\hatcurfieldeccenxxxxxA}{\ensuremath{string}}            
\newcommand{\hatcurCCraeccenxxxxxA}{\ensuremath{13^{\mathrm h}19^{\mathrm m}12.4637{\mathrm s}}}                   
\newcommand{\hatcurCCdececcenxxxxxA}{\ensuremath{-22{\arcdeg}59{\arcmin}12.7306{\arcsec}}}                 
\newcommand{\hatcurCCmageccenxxxxxA}{12.266}                         
\newcommand{\hatcurCCtwomasseccenxxxxxA}{2MASS~13191246-2259127}     
\newcommand{\hatcurCCgsceccenxxxxxA}{GSC~6700-00149}                 
\newcommand{\hatcurCCgaiaeccenxxxxxA}{GAIA~6194574671813047424}      
\newcommand{\hatcurCCgaiadrtwoeccenxxxxxA}{GAIA~DR2~6194574671813047424} 
\newcommand{\hatcurCCtassmveccenxxxxxA}{\ensuremath{12.266\pm0.030}} 
\newcommand{\hatcurCCtassmvshorteccenxxxxxA}{\ensuremath{12.3}}      
\newcommand{\hatcurCCtassmBeccenxxxxxA}{\ensuremath{13.222\pm0.060}} 
\newcommand{\hatcurCCtassmBshorteccenxxxxxA}{\ensuremath{13.2}}      
\newcommand{\hatcurCCtassmIeccenxxxxxA}{\ensuremath{nff\pmnff}}      
\newcommand{\hatcurCCtassmIshorteccenxxxxxA}{\ensuremath{0.0}}       
\newcommand{\hatcurCCtassmgeccenxxxxxA}{\ensuremath{12.733\pm0.060}} 
\newcommand{\hatcurCCtassmgshorteccenxxxxxA}{\ensuremath{12.7}}      
\newcommand{\hatcurCCtassmreccenxxxxxA}{\ensuremath{11.906\pm0.030}} 
\newcommand{\hatcurCCtassmrshorteccenxxxxxA}{\ensuremath{11.9}}      
\newcommand{\hatcurCCtassmieccenxxxxxA}{\ensuremath{11.616\pm0.030}} 
\newcommand{\hatcurCCtassmishorteccenxxxxxA}{\ensuremath{11.6}}      
\newcommand{\hatcurCCparallaxeccenxxxxxA}{\ensuremath{4.692\pm0.061}} 
\newcommand{\hatcurCCgaiamGeccenxxxxxA}{\ensuremath{11.99780\pm0.00020}} 
\newcommand{\hatcurCCgaiamBPeccenxxxxxA}{\ensuremath{12.5309\pm0.0023}} 
\newcommand{\hatcurCCgaiamRPeccenxxxxxA}{\ensuremath{11.3387\pm0.0017}} 
\newcommand{\hatcurCCtwomassJmageccenxxxxxA}{\ensuremath{10.528\pm0.024}} 
\newcommand{\hatcurCCtwomassHmageccenxxxxxA}{\ensuremath{10.038\pm0.022}} 
\newcommand{\hatcurCCtwomassKmageccenxxxxxA}{\ensuremath{9.947\pm0.021}} 
\newcommand{\hatcurCCcitJmageccenxxxxxA}{\ensuremath{10.534\pm0.025}} 
\newcommand{\hatcurCCcitHmageccenxxxxxA}{\ensuremath{10.032\pm0.023}} 
\newcommand{\hatcurCCcitKmageccenxxxxxA}{\ensuremath{9.971\pm0.021}} 
\newcommand{\hatcurCCbbJmageccenxxxxxA}{\ensuremath{10.600\pm0.027}} 
\newcommand{\hatcurCCbbHmageccenxxxxxA}{\ensuremath{10.054\pm0.024}} 
\newcommand{\hatcurCCbbKmageccenxxxxxA}{\ensuremath{9.991\pm0.022}}  
\newcommand{\hatcurCCesoJmageccenxxxxxA}{\ensuremath{10.605\pm0.029}} 
\newcommand{\hatcurCCesoHmageccenxxxxxA}{\ensuremath{10.049\pm0.027}} 
\newcommand{\hatcurCCesoKmageccenxxxxxA}{\ensuremath{9.989\pm0.022}} 
\newcommand{\hatcurCCesoJHmageccenxxxxxA}{\ensuremath{0.555\pm0.037}} 
\newcommand{\hatcurCCesoJKmageccenxxxxxA}{\ensuremath{0.617\pm0.036}} 
\newcommand{\hatcurCCesoHKmageccenxxxxxA}{\ensuremath{0.061\pm0.035}} 
\newcommand{\hatcurCCWonemageccenxxxxxA}{\ensuremath{9.866\pm0.022}} 
\newcommand{\hatcurCCWtwomageccenxxxxxA}{\ensuremath{9.942\pm0.021}} 
\newcommand{\hatcurCCWthreemageccenxxxxxA}{\ensuremath{9.896\pm0.047}} 
\newcommand{\hatcurCCWfourmageccenxxxxxA}{\ensuremath{0\pm0}}        
\newcommand{\hatcurLCdipeccenxxxxxA}{\ensuremath{4.6}}               
\newcommand{\hatcurLCrprstareccenxxxxxA}{\ensuremath{0.0671\pm0.0020}} 
\newcommand{\hatcurLCbsqeccenxxxxxA}{\ensuremath{0.251_{-0.085}^{+0.058}}} 
\newcommand{\hatcurLCimpeccenxxxxxA}{\ensuremath{0.501_{-0.094}^{+0.055}}} 
\newcommand{\hatcurLCzetaeccenxxxxxA}{\ensuremath{17.69\pm0.37}}     
\newcommand{\hatcurLCdureccenxxxxxA}{\ensuremath{0.1231\pm0.0026}}   
\newcommand{\hatcurLCdurshorteccenxxxxxA}{\ensuremath{0.1231}}       
\newcommand{\hatcurLCdurhreccenxxxxxA}{\ensuremath{2.955\pm0.063}}   
\newcommand{\hatcurLCdurhrshorteccenxxxxxA}{\ensuremath{2.955}}      
\newcommand{\hatcurLCqeccenxxxxxA}{\ensuremath{0.02840\pm0.00061}}   
\newcommand{\hatcurLCqshorteccenxxxxxA}{\ensuremath{0.028}}          
\newcommand{\hatcurLCingdureccenxxxxxA}{\ensuremath{0.0102\pm0.0011}} 
\newcommand{\hatcurLCPeccenxxxxxA}{\ensuremath{4.3315360\pm0.0000050}} 
\newcommand{\hatcurLCPprececcenxxxxxA}{\ensuremath{4.3315360}}       
\newcommand{\hatcurLCPshorteccenxxxxxA}{\ensuremath{4.3315}}         
\newcommand{\hatcurLCTeccenxxxxxA}{\ensuremath{2457751.23984\pm0.00066}} 
\newcommand{\hatcurLCTAeccenxxxxxA}{\ensuremath{2455646.1133\pm0.0026}} 
\newcommand{\hatcurLCTBeccenxxxxxA}{\ensuremath{2458214.71421\pm0.00080}} 
\newcommand{\hatcurLChatnetmeccenxxxxxA}{\ensuremath{12.038940\pm0.000083}} 
\newcommand{\hatcurLCiblendeccenxxxxxA}{\ensuremath{0.982\pm0.030}}  
\newcommand{\hatcurLCrhoeccenxxxxxA}{\ensuremath{1.553\pm0.070}}     
\newcommand{\hatcurSMEiteffeccenxxxxxA}{\ensuremath{5247\pm50}}      
\newcommand{\hatcurSMEizfeheccenxxxxxA}{\ensuremath{0.040\pm0.030}}  
\newcommand{\hatcurSMEizfehshorteccenxxxxxA}{\ensuremath{0.04}}      
\newcommand{\hatcurSMEiloggeccenxxxxxA}{\ensuremath{4.70\pm0.10}}    
\newcommand{\hatcurSMEivsineccenxxxxxA}{\ensuremath{3.98\pm0.30}}    
\newcommand{\hatcurSMEivmaceccenxxxxxA}{\ensuremath{nff\pmnff}}      
\newcommand{\hatcurSMEivmiceccenxxxxxA}{\ensuremath{nff\pmnff}}      
\newcommand{\hatcurSMEiiteffeccenxxxxxA}{\ensuremath{5188\pm40}}     
\newcommand{\hatcurSMEiizfeheccenxxxxxA}{\ensuremath{0.040\pm0.030}} 
\newcommand{\hatcurSMEiizfehshorteccenxxxxxA}{\ensuremath{0.04}}     
\newcommand{\hatcurSMEiiloggeccenxxxxxA}{\ensuremath{4.371\pm0.018}} 
\newcommand{\hatcurSMEiivsineccenxxxxxA}{\ensuremath{3.38\pm0.30}}   
\newcommand{\hatcurLBizeccenxxxxxA}{\ensuremath{0.3754}}             
\newcommand{\hatcurLBiizeccenxxxxxA}{\ensuremath{0.1769}}            
\newcommand{\hatcurLBiieccenxxxxxA}{\ensuremath{0.4491}}             
\newcommand{\hatcurLBiiieccenxxxxxA}{\ensuremath{0.1683}}            
\newcommand{\hatcurLBiIeccenxxxxxA}{\ensuremath{0.4245}}             
\newcommand{\hatcurLBiiIeccenxxxxxA}{\ensuremath{0.1712}}            
\newcommand{\hatcurLBigeccenxxxxxA}{\ensuremath{0.7785}}             
\newcommand{\hatcurLBiigeccenxxxxxA}{\ensuremath{0.0224}}            
\newcommand{\hatcurLBireccenxxxxxA}{\ensuremath{0.5594}}             
\newcommand{\hatcurLBiireccenxxxxxA}{\ensuremath{0.1497}}            
\newcommand{\hatcurLBiReccenxxxxxA}{\ensuremath{0.5328}}             
\newcommand{\hatcurLBiiReccenxxxxxA}{\ensuremath{0.1491}}            
\newcommand{\hatcurLBikepeccenxxxxxA}{\ensuremath{0.5332}}           
\newcommand{\hatcurLBiikepeccenxxxxxA}{\ensuremath{0.1620}}          
\newcommand{\hatcurISOmeccenxxxxxA}{\ensuremath{0.8713\pm0.0071}}    
\newcommand{\hatcurISOmshorteccenxxxxxA}{\ensuremath{0.87}}          
\newcommand{\hatcurISOmlongeccenxxxxxA}{\ensuremath{0.8713\pm0.0071}} 
\newcommand{\hatcurISOreccenxxxxxA}{\ensuremath{0.925\pm0.016}}      
\newcommand{\hatcurISOrshorteccenxxxxxA}{\ensuremath{0.92}}          
\newcommand{\hatcurISOrlongeccenxxxxxA}{\ensuremath{0.925\pm0.016}}  
\newcommand{\hatcurISOrhoeccenxxxxxA}{\ensuremath{1.553\pm0.070}}    
\newcommand{\hatcurISOrholongeccenxxxxxA}{\ensuremath{1.553\pm0.070}} 
\newcommand{\hatcurISOloggeccenxxxxxA}{\ensuremath{4.446\pm0.012}}   
\newcommand{\hatcurISOlumeccenxxxxxA}{\ensuremath{0.577_{-0.031}^{+0.053}}} 
\newcommand{\hatcurISOlumshorteccenxxxxxA}{\ensuremath{0.58}}        
\newcommand{\hatcurISOteffeccenxxxxxA}{\ensuremath{5239_{-36}^{+71}}} 
\newcommand{\hatcurISOzfeheccenxxxxxA}{\ensuremath{0.232\pm0.039}}   
\newcommand{\hatcurISOageeccenxxxxxA}{\ensuremath{12.21_{-0.46}^{+0.26}}} 
\newcommand{\hatcurISOspececcenxxxxxA}{K}                            
\newcommand{\hatcurRVKeccenxxxxxA}{\ensuremath{2.0\pm1.9}}           
\newcommand{\hatcurRVrkeccenxxxxxA}{\ensuremath{-0.33\pm0.14}}       
\newcommand{\hatcurRVrheccenxxxxxA}{\ensuremath{-0.241_{-0.063}^{+0.110}}} 
\newcommand{\hatcurRVkeccenxxxxxA}{\ensuremath{-0.140\pm0.084}}      
\newcommand{\hatcurRVheccenxxxxxA}{\ensuremath{-0.102\pm0.059}}      
\newcommand{\hatcurRVtroneeccenxxxxxA}{\ensuremath{0.4551\pm0.0086}} 
\newcommand{\hatcurRVtrtwoeccenxxxxxA}{\ensuremath{0\pm0}}           
\newcommand{\hatcurRVgammaAeccenxxxxxA}{\ensuremath{6417\pm0}}       
\newcommand{\hatcurRVjitterAeccenxxxxxA}{\ensuremath{15\pm14}}       
\newcommand{\hatcurRVjittertwosiglimAeccenxxxxxA}{\ensuremath{<43.1}} 
\newcommand{\hatcurRVfitrmsAeccenxxxxxA}{\ensuremath{0.0}}           
\newcommand{\hatcurRVgammaBeccenxxxxxA}{\ensuremath{-893\pm0}}       
\newcommand{\hatcurRVjitterBeccenxxxxxA}{\ensuremath{72\pm13}}       
\newcommand{\hatcurRVjittertwosiglimBeccenxxxxxA}{\ensuremath{<92.5}} 
\newcommand{\hatcurRVfitrmsBeccenxxxxxA}{\ensuremath{0.0}}           
\newcommand{\hatcurRVecceneccenxxxxxA}{\ensuremath{0.177\pm0.099}}   
\newcommand{\hatcurRVeccentwosiglimeccenxxxxxA}{\ensuremath{<0.345}} 
\newcommand{\hatcurRVomegaeccenxxxxxA}{\ensuremath{215\pm43}}        
\newcommand{\hatcurPPieccenxxxxxA}{\ensuremath{87.69_{-0.23}^{+0.36}}} 
\newcommand{\hatcurPPgeccenxxxxxA}{\ensuremath{0.95_{-0.64}^{+1.01}}} 
\newcommand{\hatcurPPloggeccenxxxxxA}{\ensuremath{1.98_{-0.49}^{+0.32}}} 
\newcommand{\hatcurPPareccenxxxxxA}{\ensuremath{11.55\pm0.17}}       
\newcommand{\hatcurPPareleccenxxxxxA}{\ensuremath{0.04967\pm0.00013}} 
\newcommand{\hatcurPPrhoeccenxxxxxA}{\ensuremath{0.079_{-0.054}^{+0.084}}} 
\newcommand{\hatcurPPmeccenxxxxxA}{\ensuremath{0.0139_{-0.0094}^{+0.0150}}} 
\newcommand{\hatcurPPmshorteccenxxxxxA}{\ensuremath{0.01}}           
\newcommand{\hatcurPPmlongeccenxxxxxA}{\ensuremath{0.0139_{-0.0094}^{+0.0150}}} 
\newcommand{\hatcurPPmeeccenxxxxxA}{\ensuremath{4.4_{-3.0}^{+4.8}}}  
\newcommand{\hatcurPPmeshorteccenxxxxxA}{\ensuremath{4.4}}           
\newcommand{\hatcurPPmelongeccenxxxxxA}{\ensuremath{4.4_{-3.0}^{+4.8}}} 
\newcommand{\hatcurPPreccenxxxxxA}{\ensuremath{0.604\pm0.018}}       
\newcommand{\hatcurPPrshorteccenxxxxxA}{\ensuremath{0.60}}           
\newcommand{\hatcurPPrlongeccenxxxxxA}{\ensuremath{0.604\pm0.018}}   
\newcommand{\hatcurPPreeccenxxxxxA}{\ensuremath{6.77\pm0.20}}        
\newcommand{\hatcurPPreshorteccenxxxxxA}{\ensuremath{6.8}}           
\newcommand{\hatcurPPrelongeccenxxxxxA}{\ensuremath{6.77\pm0.20}}    
\newcommand{\hatcurPPmrcorreccenxxxxxA}{\ensuremath{-0.07}}          
\newcommand{\hatcurPPteffeccenxxxxxA}{\ensuremath{1096_{-15}^{+24}}} 
\newcommand{\hatcurPPthetaeccenxxxxxA}{\ensuremath{0.0027_{-0.0018}^{+0.0029}}} 
\newcommand{\hatcurPPfluxperieccenxxxxxA}{\ensuremath{4.8_{-1.1}^{+1.5}}} 
\newcommand{\hatcurPPfluxperidimeccenxxxxxA}{\ensuremath{8}}         
\newcommand{\hatcurPPfluxapeccenxxxxxA}{\ensuremath{2.33\pm0.42}}    
\newcommand{\hatcurPPfluxapdimeccenxxxxxA}{\ensuremath{8}}           
\newcommand{\hatcurPPfluxavgeccenxxxxxA}{\ensuremath{3.25_{-0.17}^{+0.29}}} 
\newcommand{\hatcurPPfluxavgdimeccenxxxxxA}{\ensuremath{8}}          
\newcommand{\hatcurPPfluxavglogeccenxxxxxA}{\ensuremath{8.512_{-0.023}^{+0.037}}} 
\newcommand{\hatcurXsecphaseeccenxxxxxA}{\ensuremath{0.411\pm0.054}} 
\newcommand{\hatcurXsecondaryeccenxxxxxA}{\ensuremath{2457753.02\pm0.23}} 
\newcommand{\hatcurXsecdureccenxxxxxA}{\ensuremath{0.1054\pm0.0081}} 
\newcommand{\hatcurXsecingdureccenxxxxxA}{\ensuremath{0.00782\pm0.00050}} 
\newcommand{\hatcurPPphiconjeccenxxxxxA}{\ensuremath{-0.29\pm0.11}}  
\newcommand{\hatcurPPperieccenxxxxxA}{\ensuremath{2457752.48\pm0.47}} 
\newcommand{\hatcurPPaequiveccenxxxxxA}{\ensuremath{0.0654_{-0.0028}^{+0.0017}}} 
\newcommand{\hatcurPPtcirceccenxxxxxA}{\ensuremath{210_{-150}^{+240}}} 
\newcommand{\hatcurPPtinfalleccenxxxxxA}{\ensuremath{540000_{-290000}^{+1090000}}} 
\newcommand{\hatcurXdisteccenxxxxxA}{\ensuremath{194.1\pm3.2}}       
\newcommand{\hatcurXAveccenxxxxxA}{\ensuremath{0.326_{-0.060}^{+0.101}}} 
\newcommand{\hatcurXdistredeccenxxxxxA}{\ensuremath{194.1\pm3.2}}    
\newcommand{\hatcurXEBVeccenxxxxxA}{\ensuremath{0.105_{-0.019}^{+0.033}}} 
\newcommand{\hatcurCCpmraeccenxxxxxA}{\ensuremath{-21.78\pm0.11}}    
\newcommand{\hatcurCCpmdececcenxxxxxA}{\ensuremath{6.15\pm0.11}}     
\newcommand{\hatcurCCpmeccenxxxxxA}{\ensuremath{22.64\pm0.15}}       

\newcommand{\hatcurhtreccenxxxxxB}{HATS561-006}                      
\newcommand{\hatcurfieldeccenxxxxxB}{\ensuremath{string}}            
\newcommand{\hatcurCCraeccenxxxxxB}{\ensuremath{10^{\mathrm h}17^{\mathrm m}05.0796{\mathrm s}}}                   
\newcommand{\hatcurCCdececcenxxxxxB}{\ensuremath{-25{\arcdeg}16{\arcmin}34.5568{\arcsec}}}                 
\newcommand{\hatcurCCmageccenxxxxxB}{12.411}                         
\newcommand{\hatcurCCtwomasseccenxxxxxB}{2MASS~10170509-2516345}     
\newcommand{\hatcurCCgsceccenxxxxxB}{GSC~6622-00794}                 
\newcommand{\hatcurCCgaiaeccenxxxxxB}{GAIA~5472386847387498496}      
\newcommand{\hatcurCCgaiadrtwoeccenxxxxxB}{GAIA~DR2~5472386851683941376} 
\newcommand{\hatcurCCtassmveccenxxxxxB}{\ensuremath{12.411\pm0.030}} 
\newcommand{\hatcurCCtassmvshorteccenxxxxxB}{\ensuremath{12.4}}      
\newcommand{\hatcurCCtassmBeccenxxxxxB}{\ensuremath{13.22\pm0.11}}   
\newcommand{\hatcurCCtassmBshorteccenxxxxxB}{\ensuremath{13.2}}      
\newcommand{\hatcurCCtassmIeccenxxxxxB}{\ensuremath{nff\pmnff}}      
\newcommand{\hatcurCCtassmIshorteccenxxxxxB}{\ensuremath{0.0}}       
\newcommand{\hatcurCCtassmgeccenxxxxxB}{\ensuremath{12.780\pm0.037}} 
\newcommand{\hatcurCCtassmgshorteccenxxxxxB}{\ensuremath{12.8}}      
\newcommand{\hatcurCCtassmreccenxxxxxB}{\ensuremath{12.220\pm0.057}} 
\newcommand{\hatcurCCtassmrshorteccenxxxxxB}{\ensuremath{12.2}}      
\newcommand{\hatcurCCtassmieccenxxxxxB}{\ensuremath{12.26\pm0.19}}   
\newcommand{\hatcurCCtassmishorteccenxxxxxB}{\ensuremath{12.3}}      
\newcommand{\hatcurCCparallaxeccenxxxxxB}{\ensuremath{2.883\pm0.043}} 
\newcommand{\hatcurCCgaiamGeccenxxxxxB}{\ensuremath{12.27810\pm0.00020}} 
\newcommand{\hatcurCCgaiamBPeccenxxxxxB}{\ensuremath{12.6494\pm0.0012}} 
\newcommand{\hatcurCCgaiamRPeccenxxxxxB}{\ensuremath{11.76070\pm0.00060}} 
\newcommand{\hatcurCCtwomassJmageccenxxxxxB}{\ensuremath{11.184\pm0.026}} 
\newcommand{\hatcurCCtwomassHmageccenxxxxxB}{\ensuremath{10.850\pm0.024}} 
\newcommand{\hatcurCCtwomassKmageccenxxxxxB}{\ensuremath{10.768\pm0.024}} 
\newcommand{\hatcurCCcitJmageccenxxxxxB}{\ensuremath{11.198\pm0.026}} 
\newcommand{\hatcurCCcitHmageccenxxxxxB}{\ensuremath{10.845\pm0.025}} 
\newcommand{\hatcurCCcitKmageccenxxxxxB}{\ensuremath{10.792\pm0.024}} 
\newcommand{\hatcurCCbbJmageccenxxxxxB}{\ensuremath{11.252\pm0.028}} 
\newcommand{\hatcurCCbbHmageccenxxxxxB}{\ensuremath{10.866\pm0.025}} 
\newcommand{\hatcurCCbbKmageccenxxxxxB}{\ensuremath{10.812\pm0.024}} 
\newcommand{\hatcurCCesoJmageccenxxxxxB}{\ensuremath{11.254\pm0.029}} 
\newcommand{\hatcurCCesoHmageccenxxxxxB}{\ensuremath{10.862\pm0.027}} 
\newcommand{\hatcurCCesoKmageccenxxxxxB}{\ensuremath{10.811\pm0.025}} 
\newcommand{\hatcurCCesoJHmageccenxxxxxB}{\ensuremath{0.392\pm0.038}} 
\newcommand{\hatcurCCesoJKmageccenxxxxxB}{\ensuremath{0.443\pm0.038}} 
\newcommand{\hatcurCCesoHKmageccenxxxxxB}{\ensuremath{0.050\pm0.010}} 
\newcommand{\hatcurCCWonemageccenxxxxxB}{\ensuremath{10.714\pm0.023}} 
\newcommand{\hatcurCCWtwomageccenxxxxxB}{\ensuremath{10.783\pm0.022}} 
\newcommand{\hatcurCCWthreemageccenxxxxxB}{\ensuremath{10.736\pm0.091}} 
\newcommand{\hatcurCCWfourmageccenxxxxxB}{\ensuremath{0\pm0}}        
\newcommand{\hatcurLCdipeccenxxxxxB}{\ensuremath{4.1}}               
\newcommand{\hatcurLCrprstareccenxxxxxB}{\ensuremath{0.0573\pm0.0012}} 
\newcommand{\hatcurLCbsqeccenxxxxxB}{\ensuremath{0.227_{-0.033}^{+0.037}}} 
\newcommand{\hatcurLCimpeccenxxxxxB}{\ensuremath{0.476_{-0.036}^{+0.037}}} 
\newcommand{\hatcurLCzetaeccenxxxxxB}{\ensuremath{16.08\pm0.22}}     
\newcommand{\hatcurLCdureccenxxxxxB}{\ensuremath{0.1334\pm0.0017}}   
\newcommand{\hatcurLCdurshorteccenxxxxxB}{\ensuremath{0.1334}}       
\newcommand{\hatcurLCdurhreccenxxxxxB}{\ensuremath{3.203\pm0.041}}   
\newcommand{\hatcurLCdurhrshorteccenxxxxxB}{\ensuremath{3.203}}      
\newcommand{\hatcurLCqeccenxxxxxB}{\ensuremath{0.03050\pm0.00039}}   
\newcommand{\hatcurLCqshorteccenxxxxxB}{\ensuremath{0.030}}          
\newcommand{\hatcurLCingdureccenxxxxxB}{\ensuremath{0.00925\pm0.00045}} 
\newcommand{\hatcurLCPeccenxxxxxB}{\ensuremath{4.375021\pm0.000011}} 
\newcommand{\hatcurLCPprececcenxxxxxB}{\ensuremath{4.3750211}}       
\newcommand{\hatcurLCPshorteccenxxxxxB}{\ensuremath{4.3750}}         
\newcommand{\hatcurLCTeccenxxxxxB}{\ensuremath{2457786.41057\pm0.00064}} 
\newcommand{\hatcurLCTAeccenxxxxxB}{\ensuremath{2457012.0319\pm0.0022}} 
\newcommand{\hatcurLCTBeccenxxxxxB}{\ensuremath{2457847.66086\pm0.00062}} 
\newcommand{\hatcurLChatnetmAeccenxxxxxB}{\ensuremath{12.354730\pm0.000036}} 
\newcommand{\hatcurLCiblendAeccenxxxxxB}{\ensuremath{0.960\pm0.037}} 
\newcommand{\hatcurLChatnetmBeccenxxxxxB}{\ensuremath{12.35471\pm0.00016}} 
\newcommand{\hatcurLCiblendBeccenxxxxxB}{\ensuremath{0.822\pm0.078}} 
\newcommand{\hatcurLCrhoeccenxxxxxB}{\ensuremath{0.940\pm0.045}}     
\newcommand{\hatcurSMEiteffeccenxxxxxB}{\ensuremath{5740\pm50}}      
\newcommand{\hatcurSMEizfeheccenxxxxxB}{\ensuremath{0.060\pm0.026}}  
\newcommand{\hatcurSMEizfehshorteccenxxxxxB}{\ensuremath{0.06}}      
\newcommand{\hatcurSMEiloggeccenxxxxxB}{\ensuremath{4.550\pm0.095}}  
\newcommand{\hatcurSMEivsineccenxxxxxB}{\ensuremath{3.10\pm0.27}}    
\newcommand{\hatcurSMEivmaceccenxxxxxB}{\ensuremath{3.934\pm0.076}}  
\newcommand{\hatcurSMEivmiceccenxxxxxB}{\ensuremath{1.059\pm0.028}}  
\newcommand{\hatcurSMEiiteffeccenxxxxxB}{\ensuremath{5696\pm50}}     
\newcommand{\hatcurSMEiizfeheccenxxxxxB}{\ensuremath{0.02\pm0.32}}   
\newcommand{\hatcurSMEiizfehshorteccenxxxxxB}{\ensuremath{0.02}}     
\newcommand{\hatcurSMEiiloggeccenxxxxxB}{\ensuremath{4.296\pm0.013}} 
\newcommand{\hatcurSMEiivsineccenxxxxxB}{\ensuremath{3.19\pm0.17}}   
\newcommand{\hatcurSMEiivmaceccenxxxxxB}{\ensuremath{3.76\pm0.10}}   
\newcommand{\hatcurSMEiivmiceccenxxxxxB}{\ensuremath{0.999\pm0.035}} 
\newcommand{\hatcurLBiBeccenxxxxxB}{\ensuremath{0.7272}}             
\newcommand{\hatcurLBiiBeccenxxxxxB}{\ensuremath{0.1002}}            
\newcommand{\hatcurLBiVeccenxxxxxB}{\ensuremath{0.5631}}             
\newcommand{\hatcurLBiiVeccenxxxxxB}{\ensuremath{0.1695}}            
\newcommand{\hatcurLBiReccenxxxxxB}{\ensuremath{0.4660}}             
\newcommand{\hatcurLBiiReccenxxxxxB}{\ensuremath{0.1829}}            
\newcommand{\hatcurLBiIeccenxxxxxB}{\ensuremath{0.3693}}             
\newcommand{\hatcurLBiiIeccenxxxxxB}{\ensuremath{0.1915}}            
\newcommand{\hatcurLBiueccenxxxxxB}{\ensuremath{0.8957}}             
\newcommand{\hatcurLBiiueccenxxxxxB}{\ensuremath{-0.0545}}           
\newcommand{\hatcurLBigeccenxxxxxB}{\ensuremath{0.6706}}             
\newcommand{\hatcurLBiigeccenxxxxxB}{\ensuremath{0.1178}}            
\newcommand{\hatcurLBireccenxxxxxB}{\ensuremath{0.20\pm0.11}}        
\newcommand{\hatcurLBiireccenxxxxxB}{\ensuremath{0.35\pm0.16}}       
\newcommand{\hatcurLBiieccenxxxxxB}{\ensuremath{0.35\pm0.13}}        
\newcommand{\hatcurLBiiieccenxxxxxB}{\ensuremath{0.31\pm0.15}}       
\newcommand{\hatcurLBizeccenxxxxxB}{\ensuremath{0.3244}}             
\newcommand{\hatcurLBiizeccenxxxxxB}{\ensuremath{0.1946}}            
\newcommand{\hatcurLBiJeccenxxxxxB}{\ensuremath{0.2130}}             
\newcommand{\hatcurLBiiJeccenxxxxxB}{\ensuremath{0.2265}}            
\newcommand{\hatcurLBiHeccenxxxxxB}{\ensuremath{0.1233}}             
\newcommand{\hatcurLBiiHeccenxxxxxB}{\ensuremath{0.2628}}            
\newcommand{\hatcurLBiKeccenxxxxxB}{\ensuremath{0.1173}}             
\newcommand{\hatcurLBiiKeccenxxxxxB}{\ensuremath{0.2079}}            
\newcommand{\hatcurLBiTeccenxxxxxB}{\ensuremath{0.32\pm0.14}}        
\newcommand{\hatcurLBiiTeccenxxxxxB}{\ensuremath{0.29\pm0.17}}       
\newcommand{\hatcurLBikepeccenxxxxxB}{\ensuremath{0.4878}}           
\newcommand{\hatcurLBiikepeccenxxxxxB}{\ensuremath{0.1842}}          
\newcommand{\hatcurLBiCeccenxxxxxB}{\ensuremath{0.4784}}             
\newcommand{\hatcurLBiiCeccenxxxxxB}{\ensuremath{0.1823}}            
\newcommand{\hatcurLBiMeccenxxxxxB}{\ensuremath{0.5727}}             
\newcommand{\hatcurLBiiMeccenxxxxxB}{\ensuremath{0.1573}}            
\newcommand{\hatcurLBiSoneeccenxxxxxB}{\ensuremath{0.0978}}            
\newcommand{\hatcurLBiiSoneeccenxxxxxB}{\ensuremath{0.1240}}           
\newcommand{\hatcurLBiStwoeccenxxxxxB}{\ensuremath{0.0819}}            
\newcommand{\hatcurLBiiStwoeccenxxxxxB}{\ensuremath{0.1058}}           
\newcommand{\hatcurLBiSthreeeccenxxxxxB}{\ensuremath{0.0689}}            
\newcommand{\hatcurLBiiSthreeeccenxxxxxB}{\ensuremath{0.0866}}           
\newcommand{\hatcurLBiSfoureccenxxxxxB}{\ensuremath{0.0532}}            
\newcommand{\hatcurLBiiSfoureccenxxxxxB}{\ensuremath{0.0685}}           
\newcommand{\hatcurISOmeccenxxxxxB}{\ensuremath{0.891\pm0.011}}      
\newcommand{\hatcurISOmshorteccenxxxxxB}{\ensuremath{0.89}}          
\newcommand{\hatcurISOmlongeccenxxxxxB}{\ensuremath{0.891\pm0.011}}  
\newcommand{\hatcurISOreccenxxxxxB}{\ensuremath{1.102\pm0.016}}      
\newcommand{\hatcurISOrshorteccenxxxxxB}{\ensuremath{1.10}}          
\newcommand{\hatcurISOrlongeccenxxxxxB}{\ensuremath{1.102\pm0.016}}  
\newcommand{\hatcurISOrhoeccenxxxxxB}{\ensuremath{0.940\pm0.045}}    
\newcommand{\hatcurISOrholongeccenxxxxxB}{\ensuremath{0.940\pm0.045}} 
\newcommand{\hatcurISOloggeccenxxxxxB}{\ensuremath{4.304\pm0.014}}   
\newcommand{\hatcurISOlumeccenxxxxxB}{\ensuremath{1.165\pm0.045}}    
\newcommand{\hatcurISOlumshorteccenxxxxxB}{\ensuremath{1.16}}        
\newcommand{\hatcurISOteffeccenxxxxxB}{\ensuremath{5720_{-22}^{+30}}} 
\newcommand{\hatcurISOzfeheccenxxxxxB}{\ensuremath{-0.090\pm0.036}}  
\newcommand{\hatcurISOageeccenxxxxxB}{\ensuremath{11.94_{-0.63}^{+0.47}}} 
\newcommand{\hatcurISOspececcenxxxxxB}{G}                            
\newcommand{\hatcurRVKeccenxxxxxB}{\ensuremath{9.8\pm1.4}}           
\newcommand{\hatcurRVrkeccenxxxxxB}{\ensuremath{-0.11\pm0.14}}       
\newcommand{\hatcurRVrheccenxxxxxB}{\ensuremath{-0.009\pm0.075}}     
\newcommand{\hatcurRVkeccenxxxxxB}{\ensuremath{-0.014_{-0.047}^{+0.017}}} 
\newcommand{\hatcurRVheccenxxxxxB}{\ensuremath{-0.001_{-0.014}^{+0.011}}} 
\newcommand{\hatcurRVtroneeccenxxxxxB}{\ensuremath{0\pm0}}           
\newcommand{\hatcurRVtrtwoeccenxxxxxB}{\ensuremath{0\pm0}}           
\newcommand{\hatcurRVgammaAeccenxxxxxB}{\ensuremath{4133.3\pm4.3}}   
\newcommand{\hatcurRVjitterAeccenxxxxxB}{\ensuremath{14.2\pm5.4}}    
\newcommand{\hatcurRVjittertwosiglimAeccenxxxxxB}{\ensuremath{<24.3}} 
\newcommand{\hatcurRVfitrmsAeccenxxxxxB}{\ensuremath{0.0}}           
\newcommand{\hatcurRVgammaBeccenxxxxxB}{\ensuremath{4143.6\pm1.2}}   
\newcommand{\hatcurRVjitterBeccenxxxxxB}{\ensuremath{0.04\pm0.49}}   
\newcommand{\hatcurRVjittertwosiglimBeccenxxxxxB}{\ensuremath{<1.2}} 
\newcommand{\hatcurRVfitrmsBeccenxxxxxB}{\ensuremath{0.0}}           
\newcommand{\hatcurRVgammaCeccenxxxxxB}{\ensuremath{-3.6\pm1.4}}     
\newcommand{\hatcurRVjitterCeccenxxxxxB}{\ensuremath{0.04\pm0.65}}   
\newcommand{\hatcurRVjittertwosiglimCeccenxxxxxB}{\ensuremath{<1.5}} 
\newcommand{\hatcurRVfitrmsCeccenxxxxxB}{\ensuremath{0.0}}           
\newcommand{\hatcurRVecceneccenxxxxxB}{\ensuremath{0.023\pm0.041}}   
\newcommand{\hatcurRVeccentwosiglimeccenxxxxxB}{\ensuremath{<0.122}} 
\newcommand{\hatcurRVomegaeccenxxxxxB}{\ensuremath{183\pm76}}        
\newcommand{\hatcurPPieccenxxxxxB}{\ensuremath{87.22\pm0.24}}        
\newcommand{\hatcurPPgeccenxxxxxB}{\ensuremath{4.75\pm0.78}}         
\newcommand{\hatcurPPloggeccenxxxxxB}{\ensuremath{2.677\pm0.071}}    
\newcommand{\hatcurPPareccenxxxxxB}{\ensuremath{9.83\pm0.16}}        
\newcommand{\hatcurPPareleccenxxxxxB}{\ensuremath{0.05038\pm0.00021}} 
\newcommand{\hatcurPPrhoeccenxxxxxB}{\ensuremath{0.385\pm0.068}}     
\newcommand{\hatcurPPmeccenxxxxxB}{\ensuremath{0.073\pm0.011}}       
\newcommand{\hatcurPPmshorteccenxxxxxB}{\ensuremath{0.07}}           
\newcommand{\hatcurPPmlongeccenxxxxxB}{\ensuremath{0.073\pm0.011}}   
\newcommand{\hatcurPPmeeccenxxxxxB}{\ensuremath{23.1\pm3.4}}         
\newcommand{\hatcurPPmeshorteccenxxxxxB}{\ensuremath{23.1}}          
\newcommand{\hatcurPPmelongeccenxxxxxB}{\ensuremath{23.1\pm3.4}}     
\newcommand{\hatcurPPreccenxxxxxB}{\ensuremath{0.615\pm0.015}}       
\newcommand{\hatcurPPrshorteccenxxxxxB}{\ensuremath{0.61}}           
\newcommand{\hatcurPPrlongeccenxxxxxB}{\ensuremath{0.615\pm0.015}}   
\newcommand{\hatcurPPreeccenxxxxxB}{\ensuremath{6.89\pm0.17}}        
\newcommand{\hatcurPPreshorteccenxxxxxB}{\ensuremath{6.9}}           
\newcommand{\hatcurPPrelongeccenxxxxxB}{\ensuremath{6.89\pm0.17}}    
\newcommand{\hatcurPPmrcorreccenxxxxxB}{\ensuremath{-0.09}}          
\newcommand{\hatcurPPteffeccenxxxxxB}{\ensuremath{1291\pm13}}        
\newcommand{\hatcurPPthetaeccenxxxxxB}{\ensuremath{0.0133\pm0.0020}} 
\newcommand{\hatcurPPfluxperieccenxxxxxB}{\ensuremath{6.61_{-0.42}^{+0.67}}} 
\newcommand{\hatcurPPfluxperidimeccenxxxxxB}{\ensuremath{8}}         
\newcommand{\hatcurPPfluxapeccenxxxxxB}{\ensuremath{5.91_{-0.47}^{+0.31}}} 
\newcommand{\hatcurPPfluxapdimeccenxxxxxB}{\ensuremath{8}}           
\newcommand{\hatcurPPfluxavgeccenxxxxxB}{\ensuremath{6.26\pm0.25}}   
\newcommand{\hatcurPPfluxavgdimeccenxxxxxB}{\ensuremath{8}}          
\newcommand{\hatcurPPfluxavglogeccenxxxxxB}{\ensuremath{8.796\pm0.018}} 
\newcommand{\hatcurXsecphaseeccenxxxxxB}{\ensuremath{0.491\pm0.028}} 
\newcommand{\hatcurXsecondaryeccenxxxxxB}{\ensuremath{2457788.56\pm0.12}} 
\newcommand{\hatcurXsecdureccenxxxxxB}{\ensuremath{0.1333\pm0.0030}} 
\newcommand{\hatcurXsecingdureccenxxxxxB}{\ensuremath{0.00918\pm0.00037}} 
\newcommand{\hatcurPPphiconjeccenxxxxxB}{\ensuremath{-0.208_{-0.090}^{+0.348}}} 
\newcommand{\hatcurPPperieccenxxxxxB}{\ensuremath{2457787.3\pm1.1}}  
\newcommand{\hatcurPPaequiveccenxxxxxB}{\ensuremath{0.04670\pm0.00095}} 
\newcommand{\hatcurPPtcirceccenxxxxxB}{\ensuremath{1360_{-250}^{+360}}} 
\newcommand{\hatcurPPtinfalleccenxxxxxB}{\ensuremath{47100_{-6700}^{+8800}}} 
\newcommand{\hatcurXdisteccenxxxxxB}{\ensuremath{346.8\pm4.9}}       
\newcommand{\hatcurXAveccenxxxxxB}{\ensuremath{0.109\pm0.025}}       
\newcommand{\hatcurXdistredeccenxxxxxB}{\ensuremath{346.8\pm4.9}}    
\newcommand{\hatcurXEBVeccenxxxxxB}{\ensuremath{0.0350_{-0.0060}^{+0.0090}}} 
\newcommand{\hatcurCCpmraeccenxxxxxB}{\ensuremath{-21.752\pm0.066}}  
\newcommand{\hatcurCCpmdececcenxxxxxB}{\ensuremath{-7.540\pm0.070}}  
\newcommand{\hatcurCCpmeccenxxxxxB}{\ensuremath{23.022\pm0.096}}     

\newcommand{\hatcurCCbbHmageccen}[1]{\ifnum#1=37 %
\hatcurCCbbHmageccenxxxxxA
\else
\ifnum#1=38 %
\hatcurCCbbHmageccenxxxxxB
\else
??????\fi
\fi
}
\newcommand{\hatcurCCbbJmageccen}[1]{\ifnum#1=37 %
\hatcurCCbbJmageccenxxxxxA
\else
\ifnum#1=38 %
\hatcurCCbbJmageccenxxxxxB
\else
??????\fi
\fi
}
\newcommand{\hatcurCCbbKmageccen}[1]{\ifnum#1=37 %
\hatcurCCbbKmageccenxxxxxA
\else
\ifnum#1=38 %
\hatcurCCbbKmageccenxxxxxB
\else
??????\fi
\fi
}
\newcommand{\hatcurCCcitHmageccen}[1]{\ifnum#1=37 %
\hatcurCCcitHmageccenxxxxxA
\else
\ifnum#1=38 %
\hatcurCCcitHmageccenxxxxxB
\else
??????\fi
\fi
}
\newcommand{\hatcurCCcitJmageccen}[1]{\ifnum#1=37 %
\hatcurCCcitJmageccenxxxxxA
\else
\ifnum#1=38 %
\hatcurCCcitJmageccenxxxxxB
\else
??????\fi
\fi
}
\newcommand{\hatcurCCcitKmageccen}[1]{\ifnum#1=37 %
\hatcurCCcitKmageccenxxxxxA
\else
\ifnum#1=38 %
\hatcurCCcitKmageccenxxxxxB
\else
??????\fi
\fi
}
\newcommand{\hatcurCCdececcen}[1]{\ifnum#1=37 %
\hatcurCCdececcenxxxxxA
\else
\ifnum#1=38 %
\hatcurCCdececcenxxxxxB
\else
??????\fi
\fi
}
\newcommand{\hatcurCCesoHKmageccen}[1]{\ifnum#1=37 %
\hatcurCCesoHKmageccenxxxxxA
\else
\ifnum#1=38 %
\hatcurCCesoHKmageccenxxxxxB
\else
??????\fi
\fi
}
\newcommand{\hatcurCCesoHmageccen}[1]{\ifnum#1=37 %
\hatcurCCesoHmageccenxxxxxA
\else
\ifnum#1=38 %
\hatcurCCesoHmageccenxxxxxB
\else
??????\fi
\fi
}
\newcommand{\hatcurCCesoJHmageccen}[1]{\ifnum#1=37 %
\hatcurCCesoJHmageccenxxxxxA
\else
\ifnum#1=38 %
\hatcurCCesoJHmageccenxxxxxB
\else
??????\fi
\fi
}
\newcommand{\hatcurCCesoJKmageccen}[1]{\ifnum#1=37 %
\hatcurCCesoJKmageccenxxxxxA
\else
\ifnum#1=38 %
\hatcurCCesoJKmageccenxxxxxB
\else
??????\fi
\fi
}
\newcommand{\hatcurCCesoJmageccen}[1]{\ifnum#1=37 %
\hatcurCCesoJmageccenxxxxxA
\else
\ifnum#1=38 %
\hatcurCCesoJmageccenxxxxxB
\else
??????\fi
\fi
}
\newcommand{\hatcurCCesoKmageccen}[1]{\ifnum#1=37 %
\hatcurCCesoKmageccenxxxxxA
\else
\ifnum#1=38 %
\hatcurCCesoKmageccenxxxxxB
\else
??????\fi
\fi
}
\newcommand{\hatcurCCgaiadrtwoeccen}[1]{\ifnum#1=37 %
\hatcurCCgaiadrtwoeccenxxxxxA
\else
\ifnum#1=38 %
\hatcurCCgaiadrtwoeccenxxxxxB
\else
??????\fi
\fi
}
\newcommand{\hatcurCCgaiaeccen}[1]{\ifnum#1=37 %
\hatcurCCgaiaeccenxxxxxA
\else
\ifnum#1=38 %
\hatcurCCgaiaeccenxxxxxB
\else
??????\fi
\fi
}
\newcommand{\hatcurCCgaiamBPeccen}[1]{\ifnum#1=37 %
\hatcurCCgaiamBPeccenxxxxxA
\else
\ifnum#1=38 %
\hatcurCCgaiamBPeccenxxxxxB
\else
??????\fi
\fi
}
\newcommand{\hatcurCCgaiamGeccen}[1]{\ifnum#1=37 %
\hatcurCCgaiamGeccenxxxxxA
\else
\ifnum#1=38 %
\hatcurCCgaiamGeccenxxxxxB
\else
??????\fi
\fi
}
\newcommand{\hatcurCCgaiamRPeccen}[1]{\ifnum#1=37 %
\hatcurCCgaiamRPeccenxxxxxA
\else
\ifnum#1=38 %
\hatcurCCgaiamRPeccenxxxxxB
\else
??????\fi
\fi
}
\newcommand{\hatcurCCgsceccen}[1]{\ifnum#1=37 %
\hatcurCCgsceccenxxxxxA
\else
\ifnum#1=38 %
\hatcurCCgsceccenxxxxxB
\else
??????\fi
\fi
}
\newcommand{\hatcurCCmageccen}[1]{\ifnum#1=37 %
\hatcurCCmageccenxxxxxA
\else
\ifnum#1=38 %
\hatcurCCmageccenxxxxxB
\else
??????\fi
\fi
}
\newcommand{\hatcurCCparallaxeccen}[1]{\ifnum#1=37 %
\hatcurCCparallaxeccenxxxxxA
\else
\ifnum#1=38 %
\hatcurCCparallaxeccenxxxxxB
\else
??????\fi
\fi
}
\newcommand{\hatcurCCpmdececcen}[1]{\ifnum#1=37 %
\hatcurCCpmdececcenxxxxxA
\else
\ifnum#1=38 %
\hatcurCCpmdececcenxxxxxB
\else
??????\fi
\fi
}
\newcommand{\hatcurCCpmeccen}[1]{\ifnum#1=37 %
\hatcurCCpmeccenxxxxxA
\else
\ifnum#1=38 %
\hatcurCCpmeccenxxxxxB
\else
??????\fi
\fi
}
\newcommand{\hatcurCCpmraeccen}[1]{\ifnum#1=37 %
\hatcurCCpmraeccenxxxxxA
\else
\ifnum#1=38 %
\hatcurCCpmraeccenxxxxxB
\else
??????\fi
\fi
}
\newcommand{\hatcurCCraeccen}[1]{\ifnum#1=37 %
\hatcurCCraeccenxxxxxA
\else
\ifnum#1=38 %
\hatcurCCraeccenxxxxxB
\else
??????\fi
\fi
}
\newcommand{\hatcurCCtassmBeccen}[1]{\ifnum#1=37 %
\hatcurCCtassmBeccenxxxxxA
\else
\ifnum#1=38 %
\hatcurCCtassmBeccenxxxxxB
\else
??????\fi
\fi
}
\newcommand{\hatcurCCtassmBshorteccen}[1]{\ifnum#1=37 %
\hatcurCCtassmBshorteccenxxxxxA
\else
\ifnum#1=38 %
\hatcurCCtassmBshorteccenxxxxxB
\else
??????\fi
\fi
}
\newcommand{\hatcurCCtassmgeccen}[1]{\ifnum#1=37 %
\hatcurCCtassmgeccenxxxxxA
\else
\ifnum#1=38 %
\hatcurCCtassmgeccenxxxxxB
\else
??????\fi
\fi
}
\newcommand{\hatcurCCtassmgshorteccen}[1]{\ifnum#1=37 %
\hatcurCCtassmgshorteccenxxxxxA
\else
\ifnum#1=38 %
\hatcurCCtassmgshorteccenxxxxxB
\else
??????\fi
\fi
}
\newcommand{\hatcurCCtassmieccen}[1]{\ifnum#1=37 %
\hatcurCCtassmieccenxxxxxA
\else
\ifnum#1=38 %
\hatcurCCtassmieccenxxxxxB
\else
??????\fi
\fi
}
\newcommand{\hatcurCCtassmIeccen}[1]{\ifnum#1=37 %
\hatcurCCtassmIeccenxxxxxA
\else
\ifnum#1=38 %
\hatcurCCtassmIeccenxxxxxB
\else
??????\fi
\fi
}
\newcommand{\hatcurCCtassmishorteccen}[1]{\ifnum#1=37 %
\hatcurCCtassmishorteccenxxxxxA
\else
\ifnum#1=38 %
\hatcurCCtassmishorteccenxxxxxB
\else
??????\fi
\fi
}
\newcommand{\hatcurCCtassmIshorteccen}[1]{\ifnum#1=37 %
\hatcurCCtassmIshorteccenxxxxxA
\else
\ifnum#1=38 %
\hatcurCCtassmIshorteccenxxxxxB
\else
??????\fi
\fi
}
\newcommand{\hatcurCCtassmreccen}[1]{\ifnum#1=37 %
\hatcurCCtassmreccenxxxxxA
\else
\ifnum#1=38 %
\hatcurCCtassmreccenxxxxxB
\else
??????\fi
\fi
}
\newcommand{\hatcurCCtassmrshorteccen}[1]{\ifnum#1=37 %
\hatcurCCtassmrshorteccenxxxxxA
\else
\ifnum#1=38 %
\hatcurCCtassmrshorteccenxxxxxB
\else
??????\fi
\fi
}
\newcommand{\hatcurCCtassmveccen}[1]{\ifnum#1=37 %
\hatcurCCtassmveccenxxxxxA
\else
\ifnum#1=38 %
\hatcurCCtassmveccenxxxxxB
\else
??????\fi
\fi
}
\newcommand{\hatcurCCtassmvshorteccen}[1]{\ifnum#1=37 %
\hatcurCCtassmvshorteccenxxxxxA
\else
\ifnum#1=38 %
\hatcurCCtassmvshorteccenxxxxxB
\else
??????\fi
\fi
}
\newcommand{\hatcurCCtwomasseccen}[1]{\ifnum#1=37 %
\hatcurCCtwomasseccenxxxxxA
\else
\ifnum#1=38 %
\hatcurCCtwomasseccenxxxxxB
\else
??????\fi
\fi
}
\newcommand{\hatcurCCtwomassHmageccen}[1]{\ifnum#1=37 %
\hatcurCCtwomassHmageccenxxxxxA
\else
\ifnum#1=38 %
\hatcurCCtwomassHmageccenxxxxxB
\else
??????\fi
\fi
}
\newcommand{\hatcurCCtwomassJmageccen}[1]{\ifnum#1=37 %
\hatcurCCtwomassJmageccenxxxxxA
\else
\ifnum#1=38 %
\hatcurCCtwomassJmageccenxxxxxB
\else
??????\fi
\fi
}
\newcommand{\hatcurCCtwomassKmageccen}[1]{\ifnum#1=37 %
\hatcurCCtwomassKmageccenxxxxxA
\else
\ifnum#1=38 %
\hatcurCCtwomassKmageccenxxxxxB
\else
??????\fi
\fi
}
\newcommand{\hatcurCCWfourmageccen}[1]{\ifnum#1=37 %
\hatcurCCWfourmageccenxxxxxA
\else
\ifnum#1=38 %
\hatcurCCWfourmageccenxxxxxB
\else
??????\fi
\fi
}
\newcommand{\hatcurCCWonemageccen}[1]{\ifnum#1=37 %
\hatcurCCWonemageccenxxxxxA
\else
\ifnum#1=38 %
\hatcurCCWonemageccenxxxxxB
\else
??????\fi
\fi
}
\newcommand{\hatcurCCWthreemageccen}[1]{\ifnum#1=37 %
\hatcurCCWthreemageccenxxxxxA
\else
\ifnum#1=38 %
\hatcurCCWthreemageccenxxxxxB
\else
??????\fi
\fi
}
\newcommand{\hatcurCCWtwomageccen}[1]{\ifnum#1=37 %
\hatcurCCWtwomageccenxxxxxA
\else
\ifnum#1=38 %
\hatcurCCWtwomageccenxxxxxB
\else
??????\fi
\fi
}
\newcommand{\hatcurfieldeccen}[1]{\ifnum#1=37 %
\hatcurfieldeccenxxxxxA
\else
\ifnum#1=38 %
\hatcurfieldeccenxxxxxB
\else
??????\fi
\fi
}
\newcommand{\hatcurhtreccen}[1]{\ifnum#1=37 %
\hatcurhtreccenxxxxxA
\else
\ifnum#1=38 %
\hatcurhtreccenxxxxxB
\else
??????\fi
\fi
}
\newcommand{\hatcurISOageeccen}[1]{\ifnum#1=37 %
\hatcurISOageeccenxxxxxA
\else
\ifnum#1=38 %
\hatcurISOageeccenxxxxxB
\else
??????\fi
\fi
}
\newcommand{\hatcurISOloggeccen}[1]{\ifnum#1=37 %
\hatcurISOloggeccenxxxxxA
\else
\ifnum#1=38 %
\hatcurISOloggeccenxxxxxB
\else
??????\fi
\fi
}
\newcommand{\hatcurISOlumeccen}[1]{\ifnum#1=37 %
\hatcurISOlumeccenxxxxxA
\else
\ifnum#1=38 %
\hatcurISOlumeccenxxxxxB
\else
??????\fi
\fi
}
\newcommand{\hatcurISOlumshorteccen}[1]{\ifnum#1=37 %
\hatcurISOlumshorteccenxxxxxA
\else
\ifnum#1=38 %
\hatcurISOlumshorteccenxxxxxB
\else
??????\fi
\fi
}
\newcommand{\hatcurISOmeccen}[1]{\ifnum#1=37 %
\hatcurISOmeccenxxxxxA
\else
\ifnum#1=38 %
\hatcurISOmeccenxxxxxB
\else
??????\fi
\fi
}
\newcommand{\hatcurISOmlongeccen}[1]{\ifnum#1=37 %
\hatcurISOmlongeccenxxxxxA
\else
\ifnum#1=38 %
\hatcurISOmlongeccenxxxxxB
\else
??????\fi
\fi
}
\newcommand{\hatcurISOmshorteccen}[1]{\ifnum#1=37 %
\hatcurISOmshorteccenxxxxxA
\else
\ifnum#1=38 %
\hatcurISOmshorteccenxxxxxB
\else
??????\fi
\fi
}
\newcommand{\hatcurISOreccen}[1]{\ifnum#1=37 %
\hatcurISOreccenxxxxxA
\else
\ifnum#1=38 %
\hatcurISOreccenxxxxxB
\else
??????\fi
\fi
}
\newcommand{\hatcurISOrhoeccen}[1]{\ifnum#1=37 %
\hatcurISOrhoeccenxxxxxA
\else
\ifnum#1=38 %
\hatcurISOrhoeccenxxxxxB
\else
??????\fi
\fi
}
\newcommand{\hatcurISOrholongeccen}[1]{\ifnum#1=37 %
\hatcurISOrholongeccenxxxxxA
\else
\ifnum#1=38 %
\hatcurISOrholongeccenxxxxxB
\else
??????\fi
\fi
}
\newcommand{\hatcurISOrlongeccen}[1]{\ifnum#1=37 %
\hatcurISOrlongeccenxxxxxA
\else
\ifnum#1=38 %
\hatcurISOrlongeccenxxxxxB
\else
??????\fi
\fi
}
\newcommand{\hatcurISOrshorteccen}[1]{\ifnum#1=37 %
\hatcurISOrshorteccenxxxxxA
\else
\ifnum#1=38 %
\hatcurISOrshorteccenxxxxxB
\else
??????\fi
\fi
}
\newcommand{\hatcurISOspececcen}[1]{\ifnum#1=37 %
\hatcurISOspececcenxxxxxA
\else
\ifnum#1=38 %
\hatcurISOspececcenxxxxxB
\else
??????\fi
\fi
}
\newcommand{\hatcurISOteffeccen}[1]{\ifnum#1=37 %
\hatcurISOteffeccenxxxxxA
\else
\ifnum#1=38 %
\hatcurISOteffeccenxxxxxB
\else
??????\fi
\fi
}
\newcommand{\hatcurISOzfeheccen}[1]{\ifnum#1=37 %
\hatcurISOzfeheccenxxxxxA
\else
\ifnum#1=38 %
\hatcurISOzfeheccenxxxxxB
\else
??????\fi
\fi
}
\newcommand{\hatcurLBiBeccen}[1]{\ifnum#1=38 %
\hatcurLBiBeccenxxxxxB
\else
??????\fi
}
\newcommand{\hatcurLBiCeccen}[1]{\ifnum#1=38 %
\hatcurLBiCeccenxxxxxB
\else
??????\fi
}
\newcommand{\hatcurLBigeccen}[1]{\ifnum#1=37 %
\hatcurLBigeccenxxxxxA
\else
\ifnum#1=38 %
\hatcurLBigeccenxxxxxB
\else
??????\fi
\fi
}
\newcommand{\hatcurLBiHeccen}[1]{\ifnum#1=38 %
\hatcurLBiHeccenxxxxxB
\else
??????\fi
}
\newcommand{\hatcurLBiiBeccen}[1]{\ifnum#1=38 %
\hatcurLBiiBeccenxxxxxB
\else
??????\fi
}
\newcommand{\hatcurLBiiCeccen}[1]{\ifnum#1=38 %
\hatcurLBiiCeccenxxxxxB
\else
??????\fi
}
\newcommand{\hatcurLBiieccen}[1]{\ifnum#1=37 %
\hatcurLBiieccenxxxxxA
\else
\ifnum#1=38 %
\hatcurLBiieccenxxxxxB
\else
??????\fi
\fi
}
\newcommand{\hatcurLBiIeccen}[1]{\ifnum#1=37 %
\hatcurLBiIeccenxxxxxA
\else
\ifnum#1=38 %
\hatcurLBiIeccenxxxxxB
\else
??????\fi
\fi
}
\newcommand{\hatcurLBiigeccen}[1]{\ifnum#1=37 %
\hatcurLBiigeccenxxxxxA
\else
\ifnum#1=38 %
\hatcurLBiigeccenxxxxxB
\else
??????\fi
\fi
}
\newcommand{\hatcurLBiiHeccen}[1]{\ifnum#1=38 %
\hatcurLBiiHeccenxxxxxB
\else
??????\fi
}
\newcommand{\hatcurLBiiieccen}[1]{\ifnum#1=37 %
\hatcurLBiiieccenxxxxxA
\else
\ifnum#1=38 %
\hatcurLBiiieccenxxxxxB
\else
??????\fi
\fi
}
\newcommand{\hatcurLBiiIeccen}[1]{\ifnum#1=37 %
\hatcurLBiiIeccenxxxxxA
\else
\ifnum#1=38 %
\hatcurLBiiIeccenxxxxxB
\else
??????\fi
\fi
}
\newcommand{\hatcurLBiiJeccen}[1]{\ifnum#1=38 %
\hatcurLBiiJeccenxxxxxB
\else
??????\fi
}
\newcommand{\hatcurLBiiKeccen}[1]{\ifnum#1=38 %
\hatcurLBiiKeccenxxxxxB
\else
??????\fi
}
\newcommand{\hatcurLBiikepeccen}[1]{\ifnum#1=37 %
\hatcurLBiikepeccenxxxxxA
\else
\ifnum#1=38 %
\hatcurLBiikepeccenxxxxxB
\else
??????\fi
\fi
}
\newcommand{\hatcurLBiiMeccen}[1]{\ifnum#1=38 %
\hatcurLBiiMeccenxxxxxB
\else
??????\fi
}
\newcommand{\hatcurLBiireccen}[1]{\ifnum#1=37 %
\hatcurLBiireccenxxxxxA
\else
\ifnum#1=38 %
\hatcurLBiireccenxxxxxB
\else
??????\fi
\fi
}
\newcommand{\hatcurLBiiReccen}[1]{\ifnum#1=37 %
\hatcurLBiiReccenxxxxxA
\else
\ifnum#1=38 %
\hatcurLBiiReccenxxxxxB
\else
??????\fi
\fi
}
\newcommand{\hatcurLBiiSfoureccen}[1]{\ifnum#1=38 %
\hatcurLBiiSfoureccenxxxxxB
\else
??????\fi
}
\newcommand{\hatcurLBiiSoneeccen}[1]{\ifnum#1=38 %
\hatcurLBiiSoneeccenxxxxxB
\else
??????\fi
}
\newcommand{\hatcurLBiiSthreeeccen}[1]{\ifnum#1=38 %
\hatcurLBiiSthreeeccenxxxxxB
\else
??????\fi
}
\newcommand{\hatcurLBiiStwoeccen}[1]{\ifnum#1=38 %
\hatcurLBiiStwoeccenxxxxxB
\else
??????\fi
}
\newcommand{\hatcurLBiiTeccen}[1]{\ifnum#1=38 %
\hatcurLBiiTeccenxxxxxB
\else
??????\fi
}
\newcommand{\hatcurLBiiueccen}[1]{\ifnum#1=38 %
\hatcurLBiiueccenxxxxxB
\else
??????\fi
}
\newcommand{\hatcurLBiiVeccen}[1]{\ifnum#1=38 %
\hatcurLBiiVeccenxxxxxB
\else
??????\fi
}
\newcommand{\hatcurLBiizeccen}[1]{\ifnum#1=37 %
\hatcurLBiizeccenxxxxxA
\else
\ifnum#1=38 %
\hatcurLBiizeccenxxxxxB
\else
??????\fi
\fi
}
\newcommand{\hatcurLBiJeccen}[1]{\ifnum#1=38 %
\hatcurLBiJeccenxxxxxB
\else
??????\fi
}
\newcommand{\hatcurLBiKeccen}[1]{\ifnum#1=38 %
\hatcurLBiKeccenxxxxxB
\else
??????\fi
}
\newcommand{\hatcurLBikepeccen}[1]{\ifnum#1=37 %
\hatcurLBikepeccenxxxxxA
\else
\ifnum#1=38 %
\hatcurLBikepeccenxxxxxB
\else
??????\fi
\fi
}
\newcommand{\hatcurLBiMeccen}[1]{\ifnum#1=38 %
\hatcurLBiMeccenxxxxxB
\else
??????\fi
}
\newcommand{\hatcurLBireccen}[1]{\ifnum#1=37 %
\hatcurLBireccenxxxxxA
\else
\ifnum#1=38 %
\hatcurLBireccenxxxxxB
\else
??????\fi
\fi
}
\newcommand{\hatcurLBiReccen}[1]{\ifnum#1=37 %
\hatcurLBiReccenxxxxxA
\else
\ifnum#1=38 %
\hatcurLBiReccenxxxxxB
\else
??????\fi
\fi
}
\newcommand{\hatcurLBiSfoureccen}[1]{\ifnum#1=38 %
\hatcurLBiSfoureccenxxxxxB
\else
??????\fi
}
\newcommand{\hatcurLBiSoneeccen}[1]{\ifnum#1=38 %
\hatcurLBiSoneeccenxxxxxB
\else
??????\fi
}
\newcommand{\hatcurLBiSthreeeccen}[1]{\ifnum#1=38 %
\hatcurLBiSthreeeccenxxxxxB
\else
??????\fi
}
\newcommand{\hatcurLBiStwoeccen}[1]{\ifnum#1=38 %
\hatcurLBiStwoeccenxxxxxB
\else
??????\fi
}
\newcommand{\hatcurLBiTeccen}[1]{\ifnum#1=38 %
\hatcurLBiTeccenxxxxxB
\else
??????\fi
}
\newcommand{\hatcurLBiueccen}[1]{\ifnum#1=38 %
\hatcurLBiueccenxxxxxB
\else
??????\fi
}
\newcommand{\hatcurLBiVeccen}[1]{\ifnum#1=38 %
\hatcurLBiVeccenxxxxxB
\else
??????\fi
}
\newcommand{\hatcurLBizeccen}[1]{\ifnum#1=37 %
\hatcurLBizeccenxxxxxA
\else
\ifnum#1=38 %
\hatcurLBizeccenxxxxxB
\else
??????\fi
\fi
}
\newcommand{\hatcurLCbsqeccen}[1]{\ifnum#1=37 %
\hatcurLCbsqeccenxxxxxA
\else
\ifnum#1=38 %
\hatcurLCbsqeccenxxxxxB
\else
??????\fi
\fi
}
\newcommand{\hatcurLCdipeccen}[1]{\ifnum#1=37 %
\hatcurLCdipeccenxxxxxA
\else
\ifnum#1=38 %
\hatcurLCdipeccenxxxxxB
\else
??????\fi
\fi
}
\newcommand{\hatcurLCdureccen}[1]{\ifnum#1=37 %
\hatcurLCdureccenxxxxxA
\else
\ifnum#1=38 %
\hatcurLCdureccenxxxxxB
\else
??????\fi
\fi
}
\newcommand{\hatcurLCdurhreccen}[1]{\ifnum#1=37 %
\hatcurLCdurhreccenxxxxxA
\else
\ifnum#1=38 %
\hatcurLCdurhreccenxxxxxB
\else
??????\fi
\fi
}
\newcommand{\hatcurLCdurhrshorteccen}[1]{\ifnum#1=37 %
\hatcurLCdurhrshorteccenxxxxxA
\else
\ifnum#1=38 %
\hatcurLCdurhrshorteccenxxxxxB
\else
??????\fi
\fi
}
\newcommand{\hatcurLCdurshorteccen}[1]{\ifnum#1=37 %
\hatcurLCdurshorteccenxxxxxA
\else
\ifnum#1=38 %
\hatcurLCdurshorteccenxxxxxB
\else
??????\fi
\fi
}
\newcommand{\hatcurLChatnetmAeccen}[1]{\ifnum#1=38 %
\hatcurLChatnetmAeccenxxxxxB
\else
??????\fi
}
\newcommand{\hatcurLChatnetmBeccen}[1]{\ifnum#1=38 %
\hatcurLChatnetmBeccenxxxxxB
\else
??????\fi
}
\newcommand{\hatcurLChatnetmeccen}[1]{\ifnum#1=37 %
\hatcurLChatnetmeccenxxxxxA
\else
??????\fi
}
\newcommand{\hatcurLCiblendAeccen}[1]{\ifnum#1=38 %
\hatcurLCiblendAeccenxxxxxB
\else
??????\fi
}
\newcommand{\hatcurLCiblendBeccen}[1]{\ifnum#1=38 %
\hatcurLCiblendBeccenxxxxxB
\else
??????\fi
}
\newcommand{\hatcurLCiblendeccen}[1]{\ifnum#1=37 %
\hatcurLCiblendeccenxxxxxA
\else
??????\fi
}
\newcommand{\hatcurLCimpeccen}[1]{\ifnum#1=37 %
\hatcurLCimpeccenxxxxxA
\else
\ifnum#1=38 %
\hatcurLCimpeccenxxxxxB
\else
??????\fi
\fi
}
\newcommand{\hatcurLCingdureccen}[1]{\ifnum#1=37 %
\hatcurLCingdureccenxxxxxA
\else
\ifnum#1=38 %
\hatcurLCingdureccenxxxxxB
\else
??????\fi
\fi
}
\newcommand{\hatcurLCPeccen}[1]{\ifnum#1=37 %
\hatcurLCPeccenxxxxxA
\else
\ifnum#1=38 %
\hatcurLCPeccenxxxxxB
\else
??????\fi
\fi
}
\newcommand{\hatcurLCPprececcen}[1]{\ifnum#1=37 %
\hatcurLCPprececcenxxxxxA
\else
\ifnum#1=38 %
\hatcurLCPprececcenxxxxxB
\else
??????\fi
\fi
}
\newcommand{\hatcurLCPshorteccen}[1]{\ifnum#1=37 %
\hatcurLCPshorteccenxxxxxA
\else
\ifnum#1=38 %
\hatcurLCPshorteccenxxxxxB
\else
??????\fi
\fi
}
\newcommand{\hatcurLCqeccen}[1]{\ifnum#1=37 %
\hatcurLCqeccenxxxxxA
\else
\ifnum#1=38 %
\hatcurLCqeccenxxxxxB
\else
??????\fi
\fi
}
\newcommand{\hatcurLCqshorteccen}[1]{\ifnum#1=37 %
\hatcurLCqshorteccenxxxxxA
\else
\ifnum#1=38 %
\hatcurLCqshorteccenxxxxxB
\else
??????\fi
\fi
}
\newcommand{\hatcurLCrhoeccen}[1]{\ifnum#1=37 %
\hatcurLCrhoeccenxxxxxA
\else
\ifnum#1=38 %
\hatcurLCrhoeccenxxxxxB
\else
??????\fi
\fi
}
\newcommand{\hatcurLCrprstareccen}[1]{\ifnum#1=37 %
\hatcurLCrprstareccenxxxxxA
\else
\ifnum#1=38 %
\hatcurLCrprstareccenxxxxxB
\else
??????\fi
\fi
}
\newcommand{\hatcurLCTAeccen}[1]{\ifnum#1=37 %
\hatcurLCTAeccenxxxxxA
\else
\ifnum#1=38 %
\hatcurLCTAeccenxxxxxB
\else
??????\fi
\fi
}
\newcommand{\hatcurLCTBeccen}[1]{\ifnum#1=37 %
\hatcurLCTBeccenxxxxxA
\else
\ifnum#1=38 %
\hatcurLCTBeccenxxxxxB
\else
??????\fi
\fi
}
\newcommand{\hatcurLCTeccen}[1]{\ifnum#1=37 %
\hatcurLCTeccenxxxxxA
\else
\ifnum#1=38 %
\hatcurLCTeccenxxxxxB
\else
??????\fi
\fi
}
\newcommand{\hatcurLCzetaeccen}[1]{\ifnum#1=37 %
\hatcurLCzetaeccenxxxxxA
\else
\ifnum#1=38 %
\hatcurLCzetaeccenxxxxxB
\else
??????\fi
\fi
}
\newcommand{\hatcurPPaequiveccen}[1]{\ifnum#1=37 %
\hatcurPPaequiveccenxxxxxA
\else
\ifnum#1=38 %
\hatcurPPaequiveccenxxxxxB
\else
??????\fi
\fi
}
\newcommand{\hatcurPPareccen}[1]{\ifnum#1=37 %
\hatcurPPareccenxxxxxA
\else
\ifnum#1=38 %
\hatcurPPareccenxxxxxB
\else
??????\fi
\fi
}
\newcommand{\hatcurPPareleccen}[1]{\ifnum#1=37 %
\hatcurPPareleccenxxxxxA
\else
\ifnum#1=38 %
\hatcurPPareleccenxxxxxB
\else
??????\fi
\fi
}
\newcommand{\hatcurPPfluxapdimeccen}[1]{\ifnum#1=37 %
\hatcurPPfluxapdimeccenxxxxxA
\else
\ifnum#1=38 %
\hatcurPPfluxapdimeccenxxxxxB
\else
??????\fi
\fi
}
\newcommand{\hatcurPPfluxapeccen}[1]{\ifnum#1=37 %
\hatcurPPfluxapeccenxxxxxA
\else
\ifnum#1=38 %
\hatcurPPfluxapeccenxxxxxB
\else
??????\fi
\fi
}
\newcommand{\hatcurPPfluxavgdimeccen}[1]{\ifnum#1=37 %
\hatcurPPfluxavgdimeccenxxxxxA
\else
\ifnum#1=38 %
\hatcurPPfluxavgdimeccenxxxxxB
\else
??????\fi
\fi
}
\newcommand{\hatcurPPfluxavgeccen}[1]{\ifnum#1=37 %
\hatcurPPfluxavgeccenxxxxxA
\else
\ifnum#1=38 %
\hatcurPPfluxavgeccenxxxxxB
\else
??????\fi
\fi
}
\newcommand{\hatcurPPfluxavglogeccen}[1]{\ifnum#1=37 %
\hatcurPPfluxavglogeccenxxxxxA
\else
\ifnum#1=38 %
\hatcurPPfluxavglogeccenxxxxxB
\else
??????\fi
\fi
}
\newcommand{\hatcurPPfluxperidimeccen}[1]{\ifnum#1=37 %
\hatcurPPfluxperidimeccenxxxxxA
\else
\ifnum#1=38 %
\hatcurPPfluxperidimeccenxxxxxB
\else
??????\fi
\fi
}
\newcommand{\hatcurPPfluxperieccen}[1]{\ifnum#1=37 %
\hatcurPPfluxperieccenxxxxxA
\else
\ifnum#1=38 %
\hatcurPPfluxperieccenxxxxxB
\else
??????\fi
\fi
}
\newcommand{\hatcurPPgeccen}[1]{\ifnum#1=37 %
\hatcurPPgeccenxxxxxA
\else
\ifnum#1=38 %
\hatcurPPgeccenxxxxxB
\else
??????\fi
\fi
}
\newcommand{\hatcurPPieccen}[1]{\ifnum#1=37 %
\hatcurPPieccenxxxxxA
\else
\ifnum#1=38 %
\hatcurPPieccenxxxxxB
\else
??????\fi
\fi
}
\newcommand{\hatcurPPloggeccen}[1]{\ifnum#1=37 %
\hatcurPPloggeccenxxxxxA
\else
\ifnum#1=38 %
\hatcurPPloggeccenxxxxxB
\else
??????\fi
\fi
}
\newcommand{\hatcurPPmeccen}[1]{\ifnum#1=37 %
\hatcurPPmeccenxxxxxA
\else
\ifnum#1=38 %
\hatcurPPmeccenxxxxxB
\else
??????\fi
\fi
}
\newcommand{\hatcurPPmeeccen}[1]{\ifnum#1=37 %
\hatcurPPmeeccenxxxxxA
\else
\ifnum#1=38 %
\hatcurPPmeeccenxxxxxB
\else
??????\fi
\fi
}
\newcommand{\hatcurPPmelongeccen}[1]{\ifnum#1=37 %
\hatcurPPmelongeccenxxxxxA
\else
\ifnum#1=38 %
\hatcurPPmelongeccenxxxxxB
\else
??????\fi
\fi
}
\newcommand{\hatcurPPmeshorteccen}[1]{\ifnum#1=37 %
\hatcurPPmeshorteccenxxxxxA
\else
\ifnum#1=38 %
\hatcurPPmeshorteccenxxxxxB
\else
??????\fi
\fi
}
\newcommand{\hatcurPPmlongeccen}[1]{\ifnum#1=37 %
\hatcurPPmlongeccenxxxxxA
\else
\ifnum#1=38 %
\hatcurPPmlongeccenxxxxxB
\else
??????\fi
\fi
}
\newcommand{\hatcurPPmrcorreccen}[1]{\ifnum#1=37 %
\hatcurPPmrcorreccenxxxxxA
\else
\ifnum#1=38 %
\hatcurPPmrcorreccenxxxxxB
\else
??????\fi
\fi
}
\newcommand{\hatcurPPmshorteccen}[1]{\ifnum#1=37 %
\hatcurPPmshorteccenxxxxxA
\else
\ifnum#1=38 %
\hatcurPPmshorteccenxxxxxB
\else
??????\fi
\fi
}
\newcommand{\hatcurPPperieccen}[1]{\ifnum#1=37 %
\hatcurPPperieccenxxxxxA
\else
\ifnum#1=38 %
\hatcurPPperieccenxxxxxB
\else
??????\fi
\fi
}
\newcommand{\hatcurPPphiconjeccen}[1]{\ifnum#1=37 %
\hatcurPPphiconjeccenxxxxxA
\else
\ifnum#1=38 %
\hatcurPPphiconjeccenxxxxxB
\else
??????\fi
\fi
}
\newcommand{\hatcurPPreccen}[1]{\ifnum#1=37 %
\hatcurPPreccenxxxxxA
\else
\ifnum#1=38 %
\hatcurPPreccenxxxxxB
\else
??????\fi
\fi
}
\newcommand{\hatcurPPreeccen}[1]{\ifnum#1=37 %
\hatcurPPreeccenxxxxxA
\else
\ifnum#1=38 %
\hatcurPPreeccenxxxxxB
\else
??????\fi
\fi
}
\newcommand{\hatcurPPrelongeccen}[1]{\ifnum#1=37 %
\hatcurPPrelongeccenxxxxxA
\else
\ifnum#1=38 %
\hatcurPPrelongeccenxxxxxB
\else
??????\fi
\fi
}
\newcommand{\hatcurPPreshorteccen}[1]{\ifnum#1=37 %
\hatcurPPreshorteccenxxxxxA
\else
\ifnum#1=38 %
\hatcurPPreshorteccenxxxxxB
\else
??????\fi
\fi
}
\newcommand{\hatcurPPrhoeccen}[1]{\ifnum#1=37 %
\hatcurPPrhoeccenxxxxxA
\else
\ifnum#1=38 %
\hatcurPPrhoeccenxxxxxB
\else
??????\fi
\fi
}
\newcommand{\hatcurPPrlongeccen}[1]{\ifnum#1=37 %
\hatcurPPrlongeccenxxxxxA
\else
\ifnum#1=38 %
\hatcurPPrlongeccenxxxxxB
\else
??????\fi
\fi
}
\newcommand{\hatcurPPrshorteccen}[1]{\ifnum#1=37 %
\hatcurPPrshorteccenxxxxxA
\else
\ifnum#1=38 %
\hatcurPPrshorteccenxxxxxB
\else
??????\fi
\fi
}
\newcommand{\hatcurPPtcirceccen}[1]{\ifnum#1=37 %
\hatcurPPtcirceccenxxxxxA
\else
\ifnum#1=38 %
\hatcurPPtcirceccenxxxxxB
\else
??????\fi
\fi
}
\newcommand{\hatcurPPteffeccen}[1]{\ifnum#1=37 %
\hatcurPPteffeccenxxxxxA
\else
\ifnum#1=38 %
\hatcurPPteffeccenxxxxxB
\else
??????\fi
\fi
}
\newcommand{\hatcurPPthetaeccen}[1]{\ifnum#1=37 %
\hatcurPPthetaeccenxxxxxA
\else
\ifnum#1=38 %
\hatcurPPthetaeccenxxxxxB
\else
??????\fi
\fi
}
\newcommand{\hatcurPPtinfalleccen}[1]{\ifnum#1=37 %
\hatcurPPtinfalleccenxxxxxA
\else
\ifnum#1=38 %
\hatcurPPtinfalleccenxxxxxB
\else
??????\fi
\fi
}
\newcommand{\hatcurRVecceneccen}[1]{\ifnum#1=37 %
\hatcurRVecceneccenxxxxxA
\else
\ifnum#1=38 %
\hatcurRVecceneccenxxxxxB
\else
??????\fi
\fi
}
\newcommand{\hatcurRVeccentwosiglimeccen}[1]{\ifnum#1=37 %
\hatcurRVeccentwosiglimeccenxxxxxA
\else
\ifnum#1=38 %
\hatcurRVeccentwosiglimeccenxxxxxB
\else
??????\fi
\fi
}
\newcommand{\hatcurRVfitrmsAeccen}[1]{\ifnum#1=37 %
\hatcurRVfitrmsAeccenxxxxxA
\else
\ifnum#1=38 %
\hatcurRVfitrmsAeccenxxxxxB
\else
??????\fi
\fi
}
\newcommand{\hatcurRVfitrmsBeccen}[1]{\ifnum#1=37 %
\hatcurRVfitrmsBeccenxxxxxA
\else
\ifnum#1=38 %
\hatcurRVfitrmsBeccenxxxxxB
\else
??????\fi
\fi
}
\newcommand{\hatcurRVfitrmsCeccen}[1]{\ifnum#1=38 %
\hatcurRVfitrmsCeccenxxxxxB
\else
??????\fi
}
\newcommand{\hatcurRVgammaAeccen}[1]{\ifnum#1=37 %
\hatcurRVgammaAeccenxxxxxA
\else
\ifnum#1=38 %
\hatcurRVgammaAeccenxxxxxB
\else
??????\fi
\fi
}
\newcommand{\hatcurRVgammaBeccen}[1]{\ifnum#1=37 %
\hatcurRVgammaBeccenxxxxxA
\else
\ifnum#1=38 %
\hatcurRVgammaBeccenxxxxxB
\else
??????\fi
\fi
}
\newcommand{\hatcurRVgammaCeccen}[1]{\ifnum#1=38 %
\hatcurRVgammaCeccenxxxxxB
\else
??????\fi
}
\newcommand{\hatcurRVheccen}[1]{\ifnum#1=37 %
\hatcurRVheccenxxxxxA
\else
\ifnum#1=38 %
\hatcurRVheccenxxxxxB
\else
??????\fi
\fi
}
\newcommand{\hatcurRVjitterAeccen}[1]{\ifnum#1=37 %
\hatcurRVjitterAeccenxxxxxA
\else
\ifnum#1=38 %
\hatcurRVjitterAeccenxxxxxB
\else
??????\fi
\fi
}
\newcommand{\hatcurRVjitterBeccen}[1]{\ifnum#1=37 %
\hatcurRVjitterBeccenxxxxxA
\else
\ifnum#1=38 %
\hatcurRVjitterBeccenxxxxxB
\else
??????\fi
\fi
}
\newcommand{\hatcurRVjitterCeccen}[1]{\ifnum#1=38 %
\hatcurRVjitterCeccenxxxxxB
\else
??????\fi
}
\newcommand{\hatcurRVjittertwosiglimAeccen}[1]{\ifnum#1=37 %
\hatcurRVjittertwosiglimAeccenxxxxxA
\else
\ifnum#1=38 %
\hatcurRVjittertwosiglimAeccenxxxxxB
\else
??????\fi
\fi
}
\newcommand{\hatcurRVjittertwosiglimBeccen}[1]{\ifnum#1=37 %
\hatcurRVjittertwosiglimBeccenxxxxxA
\else
\ifnum#1=38 %
\hatcurRVjittertwosiglimBeccenxxxxxB
\else
??????\fi
\fi
}
\newcommand{\hatcurRVjittertwosiglimCeccen}[1]{\ifnum#1=38 %
\hatcurRVjittertwosiglimCeccenxxxxxB
\else
??????\fi
}
\newcommand{\hatcurRVkeccen}[1]{\ifnum#1=37 %
\hatcurRVkeccenxxxxxA
\else
\ifnum#1=38 %
\hatcurRVkeccenxxxxxB
\else
??????\fi
\fi
}
\newcommand{\hatcurRVKeccen}[1]{\ifnum#1=37 %
\hatcurRVKeccenxxxxxA
\else
\ifnum#1=38 %
\hatcurRVKeccenxxxxxB
\else
??????\fi
\fi
}
\newcommand{\hatcurRVomegaeccen}[1]{\ifnum#1=37 %
\hatcurRVomegaeccenxxxxxA
\else
\ifnum#1=38 %
\hatcurRVomegaeccenxxxxxB
\else
??????\fi
\fi
}
\newcommand{\hatcurRVrheccen}[1]{\ifnum#1=37 %
\hatcurRVrheccenxxxxxA
\else
\ifnum#1=38 %
\hatcurRVrheccenxxxxxB
\else
??????\fi
\fi
}
\newcommand{\hatcurRVrkeccen}[1]{\ifnum#1=37 %
\hatcurRVrkeccenxxxxxA
\else
\ifnum#1=38 %
\hatcurRVrkeccenxxxxxB
\else
??????\fi
\fi
}
\newcommand{\hatcurRVtroneeccen}[1]{\ifnum#1=37 %
\hatcurRVtroneeccenxxxxxA
\else
\ifnum#1=38 %
\hatcurRVtroneeccenxxxxxB
\else
??????\fi
\fi
}
\newcommand{\hatcurRVtrtwoeccen}[1]{\ifnum#1=37 %
\hatcurRVtrtwoeccenxxxxxA
\else
\ifnum#1=38 %
\hatcurRVtrtwoeccenxxxxxB
\else
??????\fi
\fi
}
\newcommand{\hatcurSMEiiloggeccen}[1]{\ifnum#1=37 %
\hatcurSMEiiloggeccenxxxxxA
\else
\ifnum#1=38 %
\hatcurSMEiiloggeccenxxxxxB
\else
??????\fi
\fi
}
\newcommand{\hatcurSMEiiteffeccen}[1]{\ifnum#1=37 %
\hatcurSMEiiteffeccenxxxxxA
\else
\ifnum#1=38 %
\hatcurSMEiiteffeccenxxxxxB
\else
??????\fi
\fi
}
\newcommand{\hatcurSMEiivmaceccen}[1]{\ifnum#1=38 %
\hatcurSMEiivmaceccenxxxxxB
\else
??????\fi
}
\newcommand{\hatcurSMEiivmiceccen}[1]{\ifnum#1=38 %
\hatcurSMEiivmiceccenxxxxxB
\else
??????\fi
}
\newcommand{\hatcurSMEiivsineccen}[1]{\ifnum#1=37 %
\hatcurSMEiivsineccenxxxxxA
\else
\ifnum#1=38 %
\hatcurSMEiivsineccenxxxxxB
\else
??????\fi
\fi
}
\newcommand{\hatcurSMEiizfeheccen}[1]{\ifnum#1=37 %
\hatcurSMEiizfeheccenxxxxxA
\else
\ifnum#1=38 %
\hatcurSMEiizfeheccenxxxxxB
\else
??????\fi
\fi
}
\newcommand{\hatcurSMEiizfehshorteccen}[1]{\ifnum#1=37 %
\hatcurSMEiizfehshorteccenxxxxxA
\else
\ifnum#1=38 %
\hatcurSMEiizfehshorteccenxxxxxB
\else
??????\fi
\fi
}
\newcommand{\hatcurSMEiloggeccen}[1]{\ifnum#1=37 %
\hatcurSMEiloggeccenxxxxxA
\else
\ifnum#1=38 %
\hatcurSMEiloggeccenxxxxxB
\else
??????\fi
\fi
}
\newcommand{\hatcurSMEiteffeccen}[1]{\ifnum#1=37 %
\hatcurSMEiteffeccenxxxxxA
\else
\ifnum#1=38 %
\hatcurSMEiteffeccenxxxxxB
\else
??????\fi
\fi
}
\newcommand{\hatcurSMEivmaceccen}[1]{\ifnum#1=37 %
\hatcurSMEivmaceccenxxxxxA
\else
\ifnum#1=38 %
\hatcurSMEivmaceccenxxxxxB
\else
??????\fi
\fi
}
\newcommand{\hatcurSMEivmiceccen}[1]{\ifnum#1=37 %
\hatcurSMEivmiceccenxxxxxA
\else
\ifnum#1=38 %
\hatcurSMEivmiceccenxxxxxB
\else
??????\fi
\fi
}
\newcommand{\hatcurSMEivsineccen}[1]{\ifnum#1=37 %
\hatcurSMEivsineccenxxxxxA
\else
\ifnum#1=38 %
\hatcurSMEivsineccenxxxxxB
\else
??????\fi
\fi
}
\newcommand{\hatcurSMEizfeheccen}[1]{\ifnum#1=37 %
\hatcurSMEizfeheccenxxxxxA
\else
\ifnum#1=38 %
\hatcurSMEizfeheccenxxxxxB
\else
??????\fi
\fi
}
\newcommand{\hatcurSMEizfehshorteccen}[1]{\ifnum#1=37 %
\hatcurSMEizfehshorteccenxxxxxA
\else
\ifnum#1=38 %
\hatcurSMEizfehshorteccenxxxxxB
\else
??????\fi
\fi
}
\newcommand{\hatcurXAveccen}[1]{\ifnum#1=37 %
\hatcurXAveccenxxxxxA
\else
\ifnum#1=38 %
\hatcurXAveccenxxxxxB
\else
??????\fi
\fi
}
\newcommand{\hatcurXdisteccen}[1]{\ifnum#1=37 %
\hatcurXdisteccenxxxxxA
\else
\ifnum#1=38 %
\hatcurXdisteccenxxxxxB
\else
??????\fi
\fi
}
\newcommand{\hatcurXdistredeccen}[1]{\ifnum#1=37 %
\hatcurXdistredeccenxxxxxA
\else
\ifnum#1=38 %
\hatcurXdistredeccenxxxxxB
\else
??????\fi
\fi
}
\newcommand{\hatcurXEBVeccen}[1]{\ifnum#1=37 %
\hatcurXEBVeccenxxxxxA
\else
\ifnum#1=38 %
\hatcurXEBVeccenxxxxxB
\else
??????\fi
\fi
}
\newcommand{\hatcurXsecdureccen}[1]{\ifnum#1=37 %
\hatcurXsecdureccenxxxxxA
\else
\ifnum#1=38 %
\hatcurXsecdureccenxxxxxB
\else
??????\fi
\fi
}
\newcommand{\hatcurXsecingdureccen}[1]{\ifnum#1=37 %
\hatcurXsecingdureccenxxxxxA
\else
\ifnum#1=38 %
\hatcurXsecingdureccenxxxxxB
\else
??????\fi
\fi
}
\newcommand{\hatcurXsecondaryeccen}[1]{\ifnum#1=37 %
\hatcurXsecondaryeccenxxxxxA
\else
\ifnum#1=38 %
\hatcurXsecondaryeccenxxxxxB
\else
??????\fi
\fi
}
\newcommand{\hatcurXsecphaseeccen}[1]{\ifnum#1=37 %
\hatcurXsecphaseeccenxxxxxA
\else
\ifnum#1=38 %
\hatcurXsecphaseeccenxxxxxB
\else
??????\fi
\fi
}

\newcommand{\hatcurxxxxA}{HATS-37}
\newcommand{\hatcurAxxxxA}{HATS-37A}
\newcommand{\hatcurBxxxxA}{HATS-37B}
\newcommand{\hatcurbxxxxA}{HATS-37Ab}
\newcommand{\hatcurcxxxxA}{HATS-37Ac}

\newcommand{\hatcurplanetnumxxxxA}{37}

\newcommand{\hatcurCCtwomassshortxxxxA}{13191246-2259127}
\newcommand{\hatcurCCgaiadrtwoshortxxxxA}{6194574671813047424}

\newcommand{\hatcurRVgammaabsxxxxA}{\hatcurRVgamma{\hatcurplanetnumxxxxA}}                           

\newcommand{\hatcurRVgammarelxxxxA}{\hatcurRVgamma{\hatcurplanetnumxxxxA}}                           

\newcommand{\hatcurCCtassvixxxxA}{\ensuremath{NULL\pm NULL}}                  

\newcommand{\hatcurSMEversionxxxxA}{i}                                       

\newcommand{\hatcurisoshortxxxxA}{YY}
\newcommand{\hatcurisofullxxxxA}{Yonsei-Yale (YY)}
\newcommand{\hatcurisocitexxxxA}{yi:2001}

\newcommand{\hatcurlumindxxxxA}{\arstar}

\newcommand{\hatcurjhkfilsetxxxxA}{MASS}

\newcommand{\hatcurSMEteffxxxxA}{\ifthenelse{\equal{\hatcurSMEversionxxxxA}{i}}{\hatcurSMEiteff{\hatcurplanetnumxxxxA}}{\hatcurSMEiiteff{\hatcurplanetnumxxxxA}}}
\newcommand{\hatcurSMEzfehxxxxA}{\ifthenelse{\equal{\hatcurSMEversionxxxxA}{i}}{\hatcurSMEizfeh{\hatcurplanetnumxxxxA}}{\hatcurSMEiizfeh{\hatcurplanetnumxxxxA}}}
\newcommand{\hatcurSMEzfehshortxxxxA}{\ifthenelse{\equal{\hatcurSMEversionxxxxA}{i}}{\hatcurSMEizfehshort{\hatcurplanetnumxxxxA}}{\hatcurSMEiizfehshort{\hatcurplanetnumxxxxA}}}
\newcommand{\hatcurSMEloggxxxxA}{\ifthenelse{\equal{\hatcurSMEversionxxxxA}{i}}{\hatcurSMEilogg{\hatcurplanetnumxxxxA}}{\hatcurSMEiilogg{\hatcurplanetnumxxxxA}}}
\newcommand{\hatcurSMEvsinxxxxA}{\ifthenelse{\equal{\hatcurSMEversionxxxxA}{i}}{\hatcurSMEivsin{\hatcurplanetnumxxxxA}}{\hatcurSMEiivsin{\hatcurplanetnumxxxxA}}}
\newcommand{\hatcurSMEvmacxxxxA}{\ifthenelse{\equal{\hatcurSMEversionxxxxA}{i}}{\hatcurSMEivmac{\hatcurplanetnumxxxxA}}{\hatcurSMEiivmac{\hatcurplanetnumxxxxA}}}
\newcommand{\hatcurSMEvmicxxxxA}{\ifthenelse{\equal{\hatcurSMEversionxxxxA}{i}}{\hatcurSMEivmic{\hatcurplanetnumxxxxA}}{\hatcurSMEiivmic{\hatcurplanetnumxxxxA}}}

\newcommand{\hatcurxxxxB}{HATS-38}
\newcommand{\hatcurbxxxxB}{HATS-38b}
\newcommand{\hatcurcxxxxB}{HATS-38c}

\newcommand{\hatcurplanetnumxxxxB}{38}

\newcommand{\hatcurCCtwomassshortxxxxB}{10170509-2516345}
\newcommand{\hatcurCCgaiadrtwoshortxxxxB}{5472386851683941376}

\newcommand{\hatcurRVgammaabsxxxxB}{\hatcurRVgamma{\hatcurplanetnumxxxxB}}                           

\newcommand{\hatcurRVgammarelxxxxB}{\hatcurRVgamma{\hatcurplanetnumxxxxB}}                           

\newcommand{\hatcurCCtassvixxxxB}{\ensuremath{NULL\pm NULL}}                  

\newcommand{\hatcurSMEversionxxxxB}{i}                                       

\newcommand{\hatcurisoshortxxxxB}{YY}
\newcommand{\hatcurisofullxxxxB}{Yonsei-Yale (YY)}
\newcommand{\hatcurisocitexxxxB}{yi:2001}

\newcommand{\hatcurlumindxxxxB}{\arstar}

\newcommand{\hatcurjhkfilsetxxxxB}{MASS}

\newcommand{\hatcurSMEteffxxxxB}{\ifthenelse{\equal{\hatcurSMEversionxxxxB}{i}}{\hatcurSMEiteff{\hatcurplanetnumxxxxB}}{\hatcurSMEiiteff{\hatcurplanetnumxxxxB}}}
\newcommand{\hatcurSMEzfehxxxxB}{\ifthenelse{\equal{\hatcurSMEversionxxxxB}{i}}{\hatcurSMEizfeh{\hatcurplanetnumxxxxB}}{\hatcurSMEiizfeh{\hatcurplanetnumxxxxB}}}
\newcommand{\hatcurSMEzfehshortxxxxB}{\ifthenelse{\equal{\hatcurSMEversionxxxxB}{i}}{\hatcurSMEizfehshort{\hatcurplanetnumxxxxB}}{\hatcurSMEiizfehshort{\hatcurplanetnumxxxxB}}}
\newcommand{\hatcurSMEloggxxxxB}{\ifthenelse{\equal{\hatcurSMEversionxxxxB}{i}}{\hatcurSMEilogg{\hatcurplanetnumxxxxB}}{\hatcurSMEiilogg{\hatcurplanetnumxxxxB}}}
\newcommand{\hatcurSMEvsinxxxxB}{\ifthenelse{\equal{\hatcurSMEversionxxxxB}{i}}{\hatcurSMEivsin{\hatcurplanetnumxxxxB}}{\hatcurSMEiivsin{\hatcurplanetnumxxxxB}}}
\newcommand{\hatcurSMEvmacxxxxB}{\ifthenelse{\equal{\hatcurSMEversionxxxxB}{i}}{\hatcurSMEivmac{\hatcurplanetnumxxxxB}}{\hatcurSMEiivmac{\hatcurplanetnumxxxxB}}}
\newcommand{\hatcurSMEvmicxxxxB}{\ifthenelse{\equal{\hatcurSMEversionxxxxB}{i}}{\hatcurSMEivmic{\hatcurplanetnumxxxxB}}{\hatcurSMEiivmic{\hatcurplanetnumxxxxB}}}

\newcommand{\hatcur}[1]{\ifnum#1=37 %
\hatcurxxxxA
\else
\ifnum#1=38 %
\hatcurxxxxB
\else
??????\fi
\fi
}
\newcommand{\hatcurA}[1]{\ifnum#1=37 %
\hatcurAxxxxA
\else
??????\fi
}
\newcommand{\hatcurb}[1]{\ifnum#1=37 %
\hatcurbxxxxA
\else
\ifnum#1=38 %
\hatcurbxxxxB
\else
??????\fi
\fi
}
\newcommand{\hatcurB}[1]{\ifnum#1=37 %
\hatcurBxxxxA
\else
??????\fi
}
\newcommand{\hatcurc}[1]{\ifnum#1=37 %
\hatcurcxxxxA
\else
\ifnum#1=38 %
\hatcurcxxxxB
\else
??????\fi
\fi
}
\newcommand{\hatcurCCgaiadrtwoshort}[1]{\ifnum#1=37 %
\hatcurCCgaiadrtwoshortxxxxA
\else
\ifnum#1=38 %
\hatcurCCgaiadrtwoshortxxxxB
\else
??????\fi
\fi
}
\newcommand{\hatcurCCtassvi}[1]{\ifnum#1=37 %
\hatcurCCtassvixxxxA
\else
\ifnum#1=38 %
\hatcurCCtassvixxxxB
\else
??????\fi
\fi
}
\newcommand{\hatcurCCtwomassshort}[1]{\ifnum#1=37 %
\hatcurCCtwomassshortxxxxA
\else
\ifnum#1=38 %
\hatcurCCtwomassshortxxxxB
\else
??????\fi
\fi
}
\newcommand{\hatcurisocite}[1]{\ifnum#1=37 %
\hatcurisocitexxxxA
\else
\ifnum#1=38 %
\hatcurisocitexxxxB
\else
??????\fi
\fi
}
\newcommand{\hatcurisofull}[1]{\ifnum#1=37 %
\hatcurisofullxxxxA
\else
\ifnum#1=38 %
\hatcurisofullxxxxB
\else
??????\fi
\fi
}
\newcommand{\hatcurisoshort}[1]{\ifnum#1=37 %
\hatcurisoshortxxxxA
\else
\ifnum#1=38 %
\hatcurisoshortxxxxB
\else
??????\fi
\fi
}
\newcommand{\hatcurjhkfilset}[1]{\ifnum#1=37 %
\hatcurjhkfilsetxxxxA
\else
\ifnum#1=38 %
\hatcurjhkfilsetxxxxB
\else
??????\fi
\fi
}
\newcommand{\hatcurlumind}[1]{\ifnum#1=37 %
\hatcurlumindxxxxA
\else
\ifnum#1=38 %
\hatcurlumindxxxxB
\else
??????\fi
\fi
}
\newcommand{\hatcurplanetnum}[1]{\ifnum#1=37 %
\hatcurplanetnumxxxxA
\else
\ifnum#1=38 %
\hatcurplanetnumxxxxB
\else
??????\fi
\fi
}
\newcommand{\hatcurRVgammaabs}[1]{\ifnum#1=37 %
\hatcurRVgammaabsxxxxA
\else
\ifnum#1=38 %
\hatcurRVgammaabsxxxxB
\else
??????\fi
\fi
}
\newcommand{\hatcurRVgammarel}[1]{\ifnum#1=37 %
\hatcurRVgammarelxxxxA
\else
\ifnum#1=38 %
\hatcurRVgammarelxxxxB
\else
??????\fi
\fi
}
\newcommand{\hatcurSMElogg}[1]{\ifnum#1=37 %
\hatcurSMEloggxxxxA
\else
\ifnum#1=38 %
\hatcurSMEloggxxxxB
\else
??????\fi
\fi
}
\newcommand{\hatcurSMEteff}[1]{\ifnum#1=37 %
\hatcurSMEteffxxxxA
\else
\ifnum#1=38 %
\hatcurSMEteffxxxxB
\else
??????\fi
\fi
}
\newcommand{\hatcurSMEversion}[1]{\ifnum#1=37 %
\hatcurSMEversionxxxxA
\else
\ifnum#1=38 %
\hatcurSMEversionxxxxB
\else
??????\fi
\fi
}
\newcommand{\hatcurSMEvmac}[1]{\ifnum#1=37 %
\hatcurSMEvmacxxxxA
\else
\ifnum#1=38 %
\hatcurSMEvmacxxxxB
\else
??????\fi
\fi
}
\newcommand{\hatcurSMEvmic}[1]{\ifnum#1=37 %
\hatcurSMEvmicxxxxA
\else
\ifnum#1=38 %
\hatcurSMEvmicxxxxB
\else
??????\fi
\fi
}
\newcommand{\hatcurSMEvsin}[1]{\ifnum#1=37 %
\hatcurSMEvsinxxxxA
\else
\ifnum#1=38 %
\hatcurSMEvsinxxxxB
\else
??????\fi
\fi
}
\newcommand{\hatcurSMEzfeh}[1]{\ifnum#1=37 %
\hatcurSMEzfehxxxxA
\else
\ifnum#1=38 %
\hatcurSMEzfehxxxxB
\else
??????\fi
\fi
}
\newcommand{\hatcurSMEzfehshort}[1]{\ifnum#1=37 %
\hatcurSMEzfehshortxxxxA
\else
\ifnum#1=38 %
\hatcurSMEzfehshortxxxxB
\else
??????\fi
\fi
}

\newcounter{planetcounter}

\newboolean{emulateapj}
\setboolean{emulateapj}{true}

\newboolean{rvtablelong}
\setboolean{rvtablelong}{true}

\newboolean{astroph}
\setboolean{astroph}{true}

\shortauthors{Jord\'an et al.}
\shorttitle{\hatcur{37}A\lowercase{b} and \hatcur{38}\lowercase{b}}
\ifthenelse{\boolean{emulateapj}}{
    
}{
    
}

\begin{document}

\title{
\hatcur{37}A\lowercase{b} and \hatcur{38}\lowercase{b}: Two Transiting Hot Neptunes in the Desert\footnote{The HATSouth network is operated by a collaboration consisting of
Princeton University (PU), the Max Planck Institute f\"ur Astronomie
(MPIA), the Australian National University (ANU), and Universidad Adolfo Ibá\~nez (UAI).  T
he station at Las Campanas
Observatory (LCO) of the Carnegie Institution for Science is operated by PU in
conjunction with UAI, the station at the High Energy Spectroscopic
Survey (H.E.S.S.) site is operated in conjunction with MPIA, and the
station at Siding Spring Observatory (SSO) is operated jointly with
ANU.
 Based in
 part on observations made with the MPG~2.2\,m Telescope at the ESO
 Observatory in La Silla.
Based on observations collected at the European Southern Observatory under ESO programmes 094.C-0428(A), 095.C-0367(A), 097.C-0571(A), 098.C-0292(A), 099.C-0374(A), 0100.C-0406(A), 0100.C-0406(B).
 This paper includes data gathered with the 6.5 meter Magellan Telescopes at Las Campanas Observatory, Chile.
 Based in part on observations made with the Anglo-Australian Telescope operated by the Australian Astronomical Observatory.
}
}

\correspondingauthor{Andr\'es Jord\'an}
\email{andres.jordan@uai.cl}

\author[0000-0002-5389-3944]{A.\ Jord\'an}
\affil{Facultad de Ingenier\'ia y Ciencias, Universidad Adolfo Ib\'a\~nez, Av.\ Diagonal las Torres 2640, Pe\~nalol\'en, Santiago, Chile}
\affil{Millenium Institute of Astrophysics,  Santiago, Chile}

\author[0000-0001-7204-6727]{G.~\'A.\ Bakos}
\affil{Department of Astrophysical Sciences, Princeton University, NJ 08544, USA}
\affil{MTA Distinguished Guest Fellow, Konkoly Observatory, Hungary}

\author[0000-0001-6023-1335]{D.\ Bayliss}
\affil{Department of Physics, University of Warwick, Coventry CV4 7AL, UK}

\author[0000-0002-9832-9271]{J.\ Bento}
\affil{Research School of Astronomy and Astrophysics, Australian National University, Canberra, ACT 2611, Australia}

\author[0000-0002-0628-0088]{W.\ Bhatti}
\affil{Department of Astrophysical Sciences, Princeton University, NJ 08544, USA}

\author[0000-0002-9158-7315]{R.\ Brahm}
\affil{Facultad de Ingeniería y Ciencias, Universidad Adolfo Ib\'a\~nez, Av.\ Diagonal las Torres 2640, Pe\~nalol\'en, Santiago, Chile}
\affil{Millenium Institute of Astrophysics,  Santiago, Chile}

\author{Z.\ Csubry}
\affil{Department of Astrophysical Sciences, Princeton University, NJ 08544, USA}

\author[0000-0001-9513-1449]{N.\ Espinoza}
\affil{Space Telescope Science Institute, 3700 San Martin Drive, Baltimore, MD 21218, USA}

\author[0000-0001-8732-6166]{J.~D.\ Hartman}
\affil{Department of Astrophysical Sciences, Princeton University, NJ 08544, USA}

\author{Th.\ Henning}
\affil{Max Planck Institute for Astronomy, K{\"{o}}nigstuhl 17, 69117 - Heidelberg, Germany}

\author[0000-0002-9428-8732]{L.\ Mancini}
\affil{Department of Physics, University of Rome Tor Vergata, Via della
Ricerca Scientifica 1, I-00133 - Roma, Italy}
\affil{Max Planck Institute for Astronomy, K{\"{o}}nigstuhl 17, 69117 - Heidelberg, Germany}
\affil{INAF - Astrophysical Observatory of Turin, Via Osservatorio 20, I-10025 - Pino Torinese, Italy}

\author[0000-0003-4464-1371]{K.\ Penev}
\affil{Department of Physics, University of Texas at Dallas, Richardson, TX 75080, USA}

\author[0000-0003-2935-7196]{M.\ Rabus}
\affil{Las Cumbres Observatory Global Telescope Network, Santa Barbara, CA 93117, USA}
\affil{Department of Physics, University of California, Santa Barbara, CA 93106-9530, USA}

\author[0000-0001-8128-3126]{P.\ Sarkis}
\affil{Max Planck Institute for Astronomy, K{\"{o}}nigstuhl 17, 69117 - Heidelberg, Germany}

\author[0000-0001-7070-3842]{V.\ Suc}
\affil{Facultad de Ingeniería y Ciencias, Universidad Adolfo Ib\'a\~nez, Av.\ Diagonal las Torres 2640, Pe\~nalol\'en, Santiago, Chile}

\author[0000-0002-0455-9384]{M.\ de Val-Borro}
\affil{Astrochemistry Laboratory, Goddard Space Flight Center, NASA, 8800 Greenbelt Rd, Greenbelt, MD 20771, USA}

\author[0000-0002-4891-3517]{G.\ Zhou}
\affil{Harvard-Smithsonian Center for Astrophysics, 60 Garden St., Cambridge, MA 02138, USA}

\author[0000-0003-1305-3761]{R.~P.\ Butler}
\affil{Earth \& Planets Laboratory, Carnegie Institution for Science, Washington DC 20015, USA}

\author{J.\ Teske}
\affil{The Observatories of the Carnegie Institution for Science, 813 Santa Barbara St, Pasadena, CA 91101, USA}

\author{J.\ Crane}
\affil{The Observatories of the Carnegie Institution for Science, 813 Santa Barbara St, Pasadena, CA 91101, USA}

\author{S.\ Shectman}
\affil{The Observatories of the Carnegie Institution for Science, 813 Santa Barbara St, Pasadena, CA 91101, USA}

\author[0000-0001-5603-6895]{T.~G.\ Tan}
\affil{Perth Exoplanet Survey Telescope, Perth, Australia}

\author{I.\ Thompson}
\affil{The Observatories of the Carnegie Institution for Science, 813 Santa Barbara St, Pasadena, CA 91101, USA}

\author[0000-0001-6135-3086]{J.~J.\ Wallace}
\affil{Department of Astrophysical Sciences, Princeton University, NJ 08544, USA}

\author{J.\ L\'az\'ar}
\affil{Hungarian Astronomical Association, 1451 Budapest, Hungary}

\author{I.\ Papp}
\affil{Hungarian Astronomical Association, 1451 Budapest, Hungary}

\author{P.\ S\'ari}
\affil{Hungarian Astronomical Association, 1451 Budapest, Hungary}


\begin{abstract}


\setcounter{footnote}{10}
We report the discovery of two transiting Neptunes by the HATSouth survey. The planet \hatcurb{37} has a mass of \hatcurPPmhps{37}\,\mjup\ ($31.5 \pm 13.4\, M_\oplus$) and a radius of \hatcurPPrhps{37}\,\rjup, and is on a $P = \hatcurLCPshorthps{37}$\,days orbit around a $V = \hatcurCCtassmv{37}$\,mag, $\hatcurISOmhps{37}$\,\msun\ star with a radius of $\hatcurISOrhps{37}$\,\rsun. We also present evidence that the star \hatcur{37}A has an unresolved stellar companion \hatcur{37}B, with a photometrically estimated mass of \hatcurISOmBhps{37}\,\msun. The planet \hatcurb{38} has a mass of \hatcurPPm{38}\,\mjup\ ($23.5 \pm 3.5\, M_\oplus$) and a radius of \hatcurPPr{38}\,\rjup, and is on a $P = \hatcurLCPshort{38}$\,days orbit around a $V = \hatcurCCtassmv{38}$\,mag, $\hatcurISOm{38}$\,\msun\ star with a radius of $\hatcurISOr{38}$\,\rsun. Both systems appear to be old, with isochrone-based ages of \hatcurISOagehps{37}\,Gyr, and \hatcurISOage{38}\,Gyr, respectively. Both \hatcurb{37} and \hatcurb{38} lie in the Neptune desert and are thus examples of a population with a low occurrence rate. They are also among the lowest mass planets found from ground-based wide-field surveys to date.
\setcounter{footnote}{0}
\end{abstract}

\keywords{
    planetary systems ---
    stars: individual (
\setcounter{planetcounter}{1}
\hatcur{37},
\hatcurCCgsc{37}\loopcommanoperiod
\setcounter{planetcounter}{2}
\hatcur{38},
\hatcurCCgsc{38}\loopcommanoperiod
\setcounter{planetcounter}{3}
)
    techniques: spectroscopic, photometric
}


\section{Introduction}
\label{sec:introduction}

In the past two decades the population of known transiting exoplanets has grown at an accelerating pace, with the \textit{Kepler} satellite \citep{borucki:2010} dominating the overall number of discoveries. The distribution of the discoveries is far from homogeneous in terms of the planetary parameters, both due to observational biases and variations in the intrinsic occurrence of planets as a function of their physical parameters and those of their host stars. An example of an observational bias is the paucity of known transiting exoplanets with periods $P\gtrsim10$ days, a region of parameter space that the ground-based survey HATSouth \citep{bakos:2013:hatsouth} was designed to target, and that is currently being explored efficiently by the Transiting Exoplanet Survey Satellite mission \citep[\emph{TESS},][]{ricker:2015}. An example of intrinsically low occurrence rates is the so-called \emph{Neptune desert}, a term coined by \citet{mazeh:2016} to describe a wedge in the period-mass or period-radius diagram where close-in ($P\lesssim 5$ d) planets with radii similar to Neptune are very rare, and essentially non-existent for $P\lesssim 3$ d \citep[see also][]{szabo:2011,beauge:2013}.

In order to uncover more planetary systems in sparsely populated regions such as the Neptune desert it pays to survey to fainter magnitudes than what \emph{TESS} is optimized for. Ground based wide-field surveys currently operating such as HATSouth  or NGTS \citep{wheatley:2018:ngts} can provide that complement to \emph{TESS} by uncovering an additional number of intrinsically rare systems. Indeed, one of the most extreme systems in the Neptune desert was recently uncovered by the NGTS \citep[NGTS4-b,][]{west:2019}. The reason for the existence of the desert is still a subject of active investigation. The physical processes thought to be relevant are photoevaporation and the tidal disruption barrier for gas giants after high-eccentricity migration \citep[see][and references therein]{owen:2018}.

In this paper we report the discovery by the HATSouth survey of two transiting Neptunes in the desert. They both have similar values of radii and periods, and fairly similar masses. We thus contribute two more systems to the sparsely populated Neptune desert. The paper is structured as follows. In \S~\ref{sec:obs} we describe the observational data which were used to perform the modeling of the system as described in \S~\ref{sec:analysis}. The results are discussed in \S~\ref{sec:discussion}.

\section{Observations}
\label{sec:obs}

Figures~\ref{fig:hats37} and \ref{fig:hats38} show the
observations collected for \hatcur{37} and \hatcur{38},
respectively. Each figure shows the HATSouth light curve used to
detect the transits, the ground-based follow-up transit light curves,
the high-precision radial velocities (RVs) and spectral line bisector spans (BSs), and the
catalog broad-band photometry, including parallax corrections from
Gaia~DR2, used in characterizing the host stars. Below we describe the observations of these objects that were collected by our team.

\subsection{Photometric detection}
\label{sec:detection}

Both of the systems presented here were initially detected as transiting
planet candidates based on observations by the HATSouth network. The
operations of the network are described in
\citet{bakos:2013:hatsouth}, while our methods for reducing the data
to trend-filtered light curves \citep[filtered using the method
  of][]{kovacs:2005:TFA} and identifying transiting planet signals
\citep[using the Box-fitting Least Squares or BLS method;][]{kovacs:2002:BLS}
are described in \citet{penev:2013:hats1}. The HATSouth observations
of each system are summarized in \reftabl{photobs}, while the light
curve data are made available in
\reftabl{phfu}.

We also searched the light curves for other periodic signals using the
Generalized Lomb-Scargle method \citep[GLS;][]{zechmeister:2009}, and
for additional transit signals by applying a second iteration of
BLS. Both of these searches were performed on the residual light
curves after subtracting the best-fit primary transit models. No
additional periodic signals are detected for \hatcur{37}. For
\hatcur{38} we detect a periodic signal at a period of $P =
21.52$\,day, semi-amplitude of $0.43$\,mmag, and a false alarm
probability, determined via bootstrap simulations, of $10^{-6.3}$. We
do not detect any additional transit signals in its light curve. The
periodic signal detected for \hatcur{38} may correspond to the
photometric rotation period of this $\teff=\hatcurSMEteff{38}$\,K star. The
star has $\vsini = \hatcurSMEvsin{38}$\,\kms, which gives an upper
limit of $18.7 \pm 1.7$\,days on the equatorial rotation period. The
photometric period of $21.52$\,days is 1.7$\sigma$ larger than this
upper limit, but a larger value is possible if the rotation
axis has $\sin i \approx 1$ and the spots are at a latitude that is
rotating more slowly than the equator.

%
%
\ifthenelse{\boolean{emulateapj}}{
    \begin{figure*}[!ht]
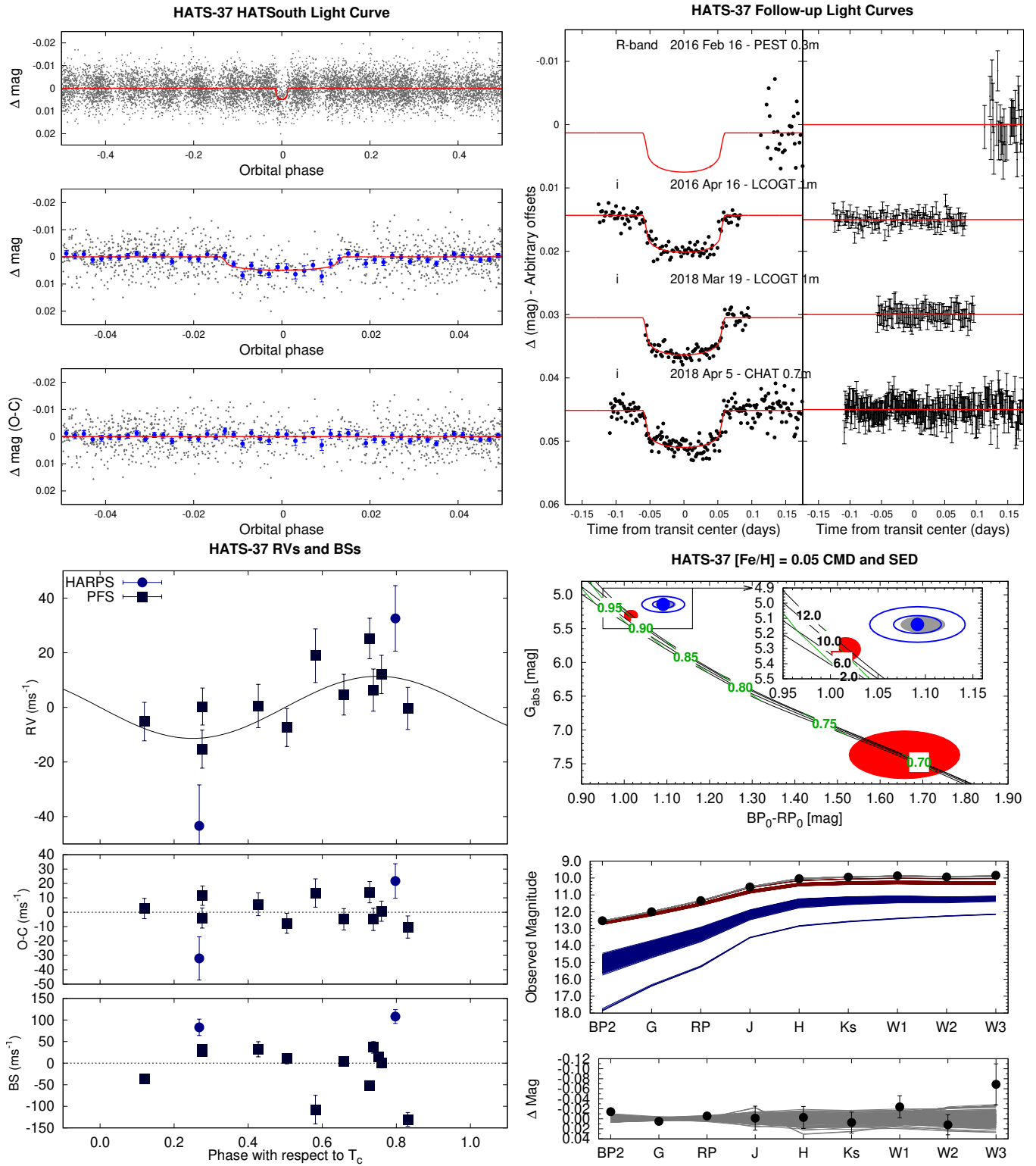

}{
    \begin{figure}[!ht]
}
 {
 \centering
 \leavevmode
 \includegraphics[width={1.0\linewidth}]{\hatcurhtr{37}-banner-hps-eps-converted-to.pdf}
}
 {
 \centering
 \leavevmode
 \includegraphics[width={0.5\linewidth}]{\hatcurhtr{37}-hs-hps-eps-converted-to.pdf}%
 \hfil
 \includegraphics[width={0.5\linewidth}]{\hatcurhtr{37}-lc-hps-eps-converted-to.pdf}%
 }
 {
 \centering
 \leavevmode
 \includegraphics[width={0.5\linewidth}]{\hatcurhtr{37}-rv-SSP-eps-converted-to.pdf}%
 \hfil
 \includegraphics[width={0.5\linewidth}]{\hatcurhtr{37}-iso-bprp-gabs-SED-hps-eps-converted-to.pdf}%
 }
\caption{
    Observations used to confirm the transiting planet system \hatcur{37}. {\em Top Left:} Phase-folded unbinned HATSouth light curve. The
    top panel shows the full light curve, the middle panel shows
    the light curve zoomed-in on the transit, and the bottom panel shows the residuals from the best-fit model zoomed-in on the transit. The solid lines show the
    model fits to the light curves. The dark filled circles show the light curves binned in phase with a bin
    size of 0.002. (Caption continued on next page).
\label{fig:hats37}
}
\ifthenelse{\boolean{emulateapj}}{
    \end{figure*}
}{
    \end{figure}
}

%
%
\addtocounter{figure}{-1}
\ifthenelse{\boolean{emulateapj}}{
    \begin{figure*}[!ht]
}{
    \begin{figure}[!ht]
}
\caption{
    (Caption continued from previous page)
{\em Top Right:} Unbinned follow-up transit light curves
    corrected for instrumental trends fitted
    simultaneously with the transit model, which is overplotted.
    The dates, filters and instruments used are
    indicated.
    The
    residuals are shown on the right-hand-side in the same order
    as the original light curves.
    The error bars represent the
    photon and background shot noise, plus the readout noise. Note that these uncertainties are scaled up in the fitting procedure to achieve a reduced $\chi^2$ of unity, but the uncertainties shown in the plot have not been scaled.
{\em Bottom Left:}
High-precision RVs phased with respect to the mid-transit-time. The instruments used are labelled in the plot.
The top panel shows the phased measurements together with the best-fit model.
The center-of-mass velocity has been subtracted. Both the observations and the model have also had a linear trend in time subtracted (Fig.~\ref{fig:hats37rvjd}). In this case the model has not been corrected for dilution from the unresolved stellar component \hatcurB{37}. We find that the dilution corrected orbit has a semi-amplitude that is $\sim 20$\% larger than what is shown here. The second panel shows the velocity $O\!-\!C$ residuals.
The error bars include the estimated jitter.
The third panel shows the bisector spans.
{\em Bottom Right:} Color-magnitude diagram (CMD) and spectral energy distribution (SED). The top panel shows the absolute $G$ magnitude vs.\ the de-reddened $BP - RP$ color compared to
  theoretical isochrones (black lines) and stellar evolution tracks
  (green lines) from the PARSEC models interpolated at
  the best-estimate value for the metallicity of the host. The age
  of each isochrone is listed in black in Gyr, while the mass of each
  evolution track is listed in green in solar mass units. The filled
  blue circles show the measured reddening- and distance-corrected
  values from Gaia DR2, while the blue lines indicate
  the $1\sigma$ and $2\sigma$ confidence regions, including the
  estimated systematic errors in the photometry. Here we model the system as a binary star with a planet transiting one component. The $1\sigma$ posterior distributions for the primary star \hatcurA{37} and secondary star \hatcurB{37} are shown as the red ellipses. The gray ellipse shows the $1\sigma$ posterior distribution for the combined photometry of the system. The inset shows a zoomed-in view around the primary star and the combined photometry. The middle panel shows the SED as measured via broadband photometry through the listed filters. Here we plot the observed magnitudes without correcting for distance or extinction. Overplotted are 200 model SEDs randomly selected from the MCMC posterior distribution produced through the global analysis (gray lines). The SEDs of the primary and secondary components are also shown with red and blue lines, respectively. Note that the solution has two modes. The first mode consists of a primary star with mass $\hatcurISOmshorthps{37}$\,\msun, and a secondary star with mass $\hatcurISOmshortBhps{37}$\,\msun. The second mode consists of a primary star with mass $0.88$\,\msun, and a fainter secondary star with mass $0.48$\,\msun. The first mode is $\sim 35$ times more likely based on its representation in the posterior distribution.
The model makes use of the predicted absolute magnitudes in each bandpass from the PARSEC isochrones, the distance to the system (constrained largely via Gaia DR2) and extinction (constrained from the SED with a prior coming from the {\sc mwdust} 3D Galactic extinction model).
The bottom panel shows the $O\!-\!C$ residuals from the best-fit model SED.
\label{fig:hats37:labcontinue}}
\ifthenelse{\boolean{emulateapj}}{
    \end{figure*}
}{
    \end{figure}
}

%
%
\ifthenelse{\boolean{emulateapj}}{
    \begin{figure*}[!ht]
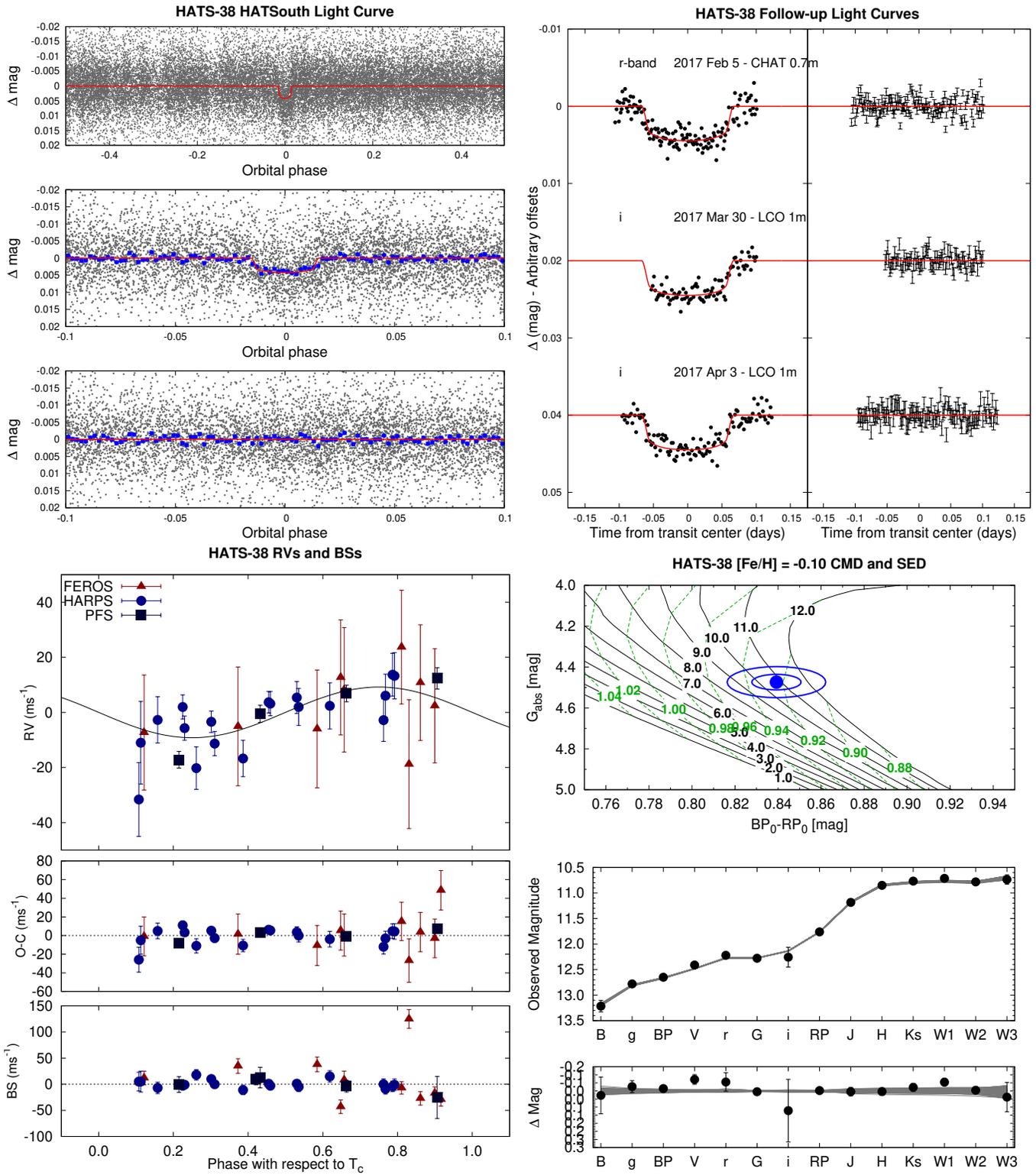

}{
    \begin{figure}[!ht]
}
 {
 \centering
 \leavevmode
 \includegraphics[width={1.0\linewidth}]{\hatcurhtr{38}-banner-eps-converted-to.pdf}
}
 {
 \centering
 \leavevmode
 \includegraphics[width={0.5\linewidth}]{\hatcurhtr{38}-hs-eps-converted-to.pdf}%
 \hfil
 \includegraphics[width={0.5\linewidth}]{\hatcurhtr{38}-lc-eps-converted-to.pdf}%
 }
 {
 \centering
 \leavevmode
 \includegraphics[width={0.5\linewidth}]{\hatcurhtr{38}-rv-eps-converted-to.pdf}%
 \hfil
 \includegraphics[width={0.5\linewidth}]{\hatcurhtr{38}-iso-bprp-gabs-isofeh-SED-eps-converted-to.pdf}%
 }
\caption{
    Same as Figure~\ref{fig:hats38}, here we show the observations of \hatcur{38} when modelled as a single star with a transiting planet.
\label{fig:hats38}
}
\ifthenelse{\boolean{emulateapj}}{
    \end{figure*}
}{
    \end{figure}
}

\begin{figure*}[!ht]
 {
 \centering
 \includegraphics[width={1.0\linewidth}]{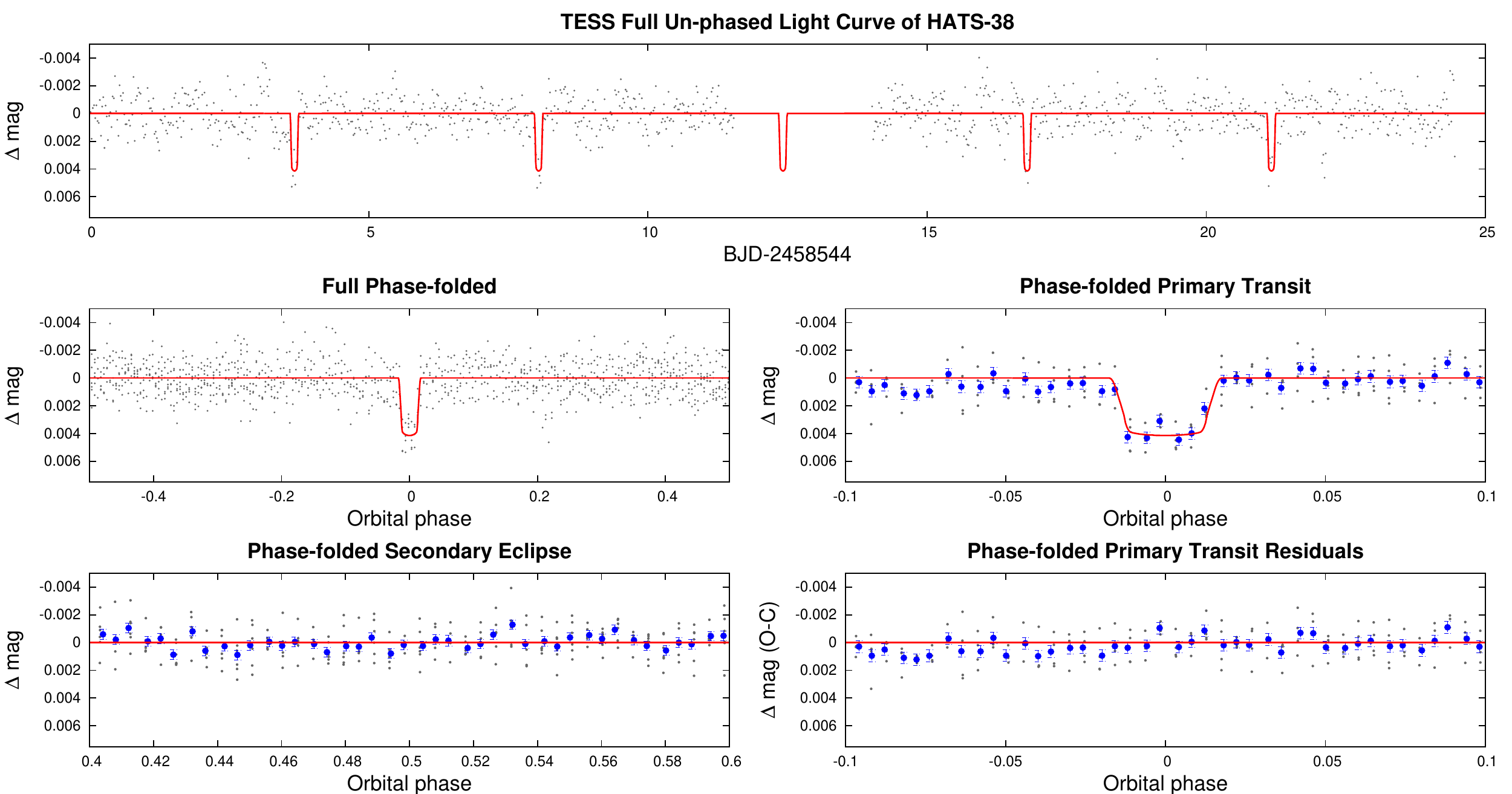}
}
\caption{
    {\em Top}: Unbinned \emph{TESS} observations of \hatcur{38} plotted against time
    simultaneously with the transit model, which is overplotted.
    {\em Middle}: Phase-folded unbinned \emph{TESS} light curve. The
    left panel shows the full light curve, the right panel shows
    the light curve zoomed-in on the transit. The solid lines show the
    model fits to the light curves. The blue filled circles in the middle right panel show
    the light curve binned in phase with a bin
    size of 0.002.
    {\em Bottom}: Phase-folded \emph{TESS} light curve around the predicted time of secondary eclipse (left panel) and residuals with respect to the transit model shown in the middle right panel. The black dots show the unbinned data, while the blue filled circle show values binned in phase with a bin
    size of 0.002.
\label{fig:hats38TESS}
}
\end{figure*}

\begin{figure}[!ht]
 {
 \centering
 \includegraphics[width={1.0\linewidth}]{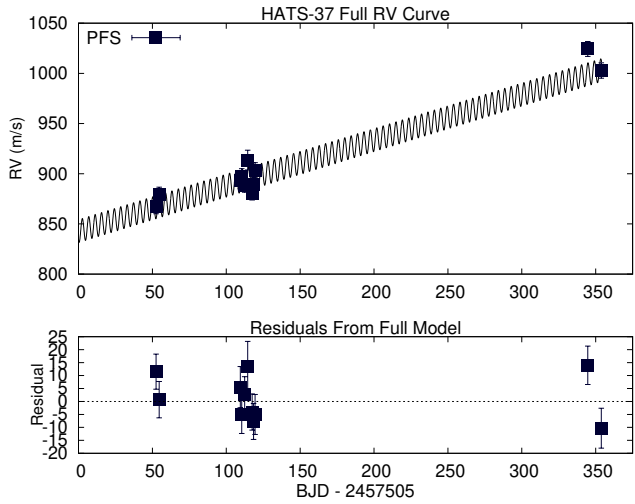}
}
\caption{
    {\em Top:} RV observations of \hatcur{37} plotted against time. The solid line shows the best-fit model including a linear trend and the Keplerian orbital variation of the host star due to the planet \hatcurb{37}. As in Figure~\ref{fig:hats37}, the RV model plotted here is not corrected for dilution from the unresolved stellar component \hatcurB{37}. The corrected semi-amplitude of the orbit is $\sim 20$\% larger than what is shown. {\em Bottom:} RV residuals from the best-fit model plotted against time.
\label{fig:hats37rvjd}
}
\end{figure}


\clearpage

\startlongtable
\ifthenelse{\boolean{emulateapj}}{
    \begin{deluxetable*}{llrrrr}
}{
    \begin{deluxetable}{llrrrr}
}
\tablewidth{0pc}
\tabletypesize{\scriptsize}
\tablecaption{
    Summary of photometric observations
    \label{tab:photobs}
}
\tablehead{
    \multicolumn{1}{c}{Instrument/Field\tablenotemark{a}} &
    \multicolumn{1}{c}{Date(s)} &
    \multicolumn{1}{c}{\# Images} &
    \multicolumn{1}{c}{Cadence\tablenotemark{b}} &
    \multicolumn{1}{c}{Filter} &
    \multicolumn{1}{c}{Precision\tablenotemark{c}} \\
    \multicolumn{1}{c}{} &
    \multicolumn{1}{c}{} &
    \multicolumn{1}{c}{} &
    \multicolumn{1}{c}{(sec)} &
    \multicolumn{1}{c}{} &
    \multicolumn{1}{c}{(mmag)}
}
\startdata
\sidehead{\textbf{\hatcur{37}}}
~~~~HS-1/G567.1 & 2011 Mar--2011 Aug & 4975 & 294 & $r$ & 5.2 \\
~~~~HS-3/G567.1 & 2011 Jul--2011 Aug & 735 & 297 & $r$ & 5.7 \\
~~~~HS-5/G567.1 & 2011 Mar--2011 Aug & 3217 & 291 & $r$ & 5.0 \\
~~~~PEST~0.3\,m & 2016 Feb 16 & 113 & 132 & $R_{C}$ & 2.8 \\
~~~~Swope~1\,m/e2v & 2017 Apr 04 & 161 & 104 & $i$ & 1.6 \\
~~~~LCO~1\,m/sinistro & 2016 Apr 16 & 108 & 159 & $i^{\prime}$ & 1.0 \\
~~~~LCO~1\,m/sinistro & 2018 Mar 19 & 82 & 163 & $i^{\prime}$ & 0.8 \\
~~~~CHAT~0.7\,m & 2018 Apr 05 & 217 & 113 & $i$ & 1.4 \\
\sidehead{\textbf{\hatcur{38}}}
~~~~HS-1/G561.1 & 2014 Dec--2015 Jul & 4892 & 319 & $r$ & 6.7 \\
~~~~HS-2/G561.1 & 2014 Dec--2015 Jul & 5718 & 349 & $r$ & 4.7 \\
~~~~HS-3/G561.1 & 2014 Dec--2015 Jul & 3691 & 353 & $r$ & 5.1 \\
~~~~HS-4/G561.1 & 2014 Dec--2015 Jul & 2862 & 352 & $r$ & 6.9 \\
~~~~HS-5/G561.1 & 2014 Dec--2015 Jul & 2959 & 356 & $r$ & 5.7 \\
~~~~HS-6/G561.1 & 2014 Dec--2015 Jul & 3058 & 342 & $r$ & 6.9 \\
~~~~HS-1/G561.1.focus & 2014 Dec--2015 Jul & 2026 & 1122 & $r$ & 14.5 \\
~~~~HS-2/G561.1.focus & 2014 Dec--2015 Jul & 2134 & 1204 & $r$ & 13.4 \\
~~~~HS-3/G561.1.focus & 2014 Dec--2015 Jul & 1217 & 1227 & $r$ & 14.1 \\
~~~~HS-4/G561.1.focus & 2014 Dec--2015 Jul & 977 & 1221 & $r$ & 15.5 \\
~~~~HS-5/G561.1.focus & 2014 Dec--2015 Jul & 1190 & 1232 & $r$ & 15.0 \\
~~~~HS-6/G561.1.focus & 2014 Dec--2015 Jul & 1174 & 1206 & $r$ & 15.7 \\
~~~~CHAT~0.7\,m & 2017 Feb 05 & 146 & 112 & $r$ & 1.1 \\
~~~~LCO~1\,m/sinistro & 2017 Mar 30 & 83 & 161 & $i^{\prime}$ & 0.9 \\
~~~~LCO~1\,m/sinistro & 2017 Apr 03 & 118 & 160 & $i^{\prime}$ & 1.0 \\
\enddata \tablenotetext{a}{ For HATSouth data we list the HATSouth
  unit, CCD and field name from which the observations are taken. HS-1
  and -2 are located at Las Campanas Observatory in Chile, HS-3 and -4
  are located at the H.E.S.S. site in Namibia, and HS-5 and -6 are
  located at Siding Spring Observatory in Australia. Each unit has 4
  CCDs. Each field corresponds to one of 838 fixed pointings used to
  cover the full 4$\pi$ celestial sphere. All data from a given
  HATSouth field and CCD number are reduced together, while detrending
  through External Parameter Decorrelation (EPD) is done independently
  for each unique unit+CCD+field combination. For \hatcur{38} we also
  derived light curves from short (30\,s) focus frames that were taken
  by the HATSouth instruments every $\sim 20$\,minutes. The Swope~1\,m light curve for \hatcur{37} covered a predicted secondary eclipse event.
}
\tablenotetext{b}{ The median time between consecutive images rounded
  to the nearest second. Due to factors such as weather, the
  day--night cycle, guiding and focus corrections the cadence is only
  approximately uniform over short timescales.  } \tablenotetext{c}{
  The RMS of the residuals from the best-fit model.  }
\ifthenelse{\boolean{emulateapj}}{
    \end{deluxetable*}
}{
    \end{deluxetable}
}

\subsection{Spectroscopic Observations}
\label{sec:obsspec}

The spectroscopic observations carried out to confirm and characterize
both of the transiting planet systems are summarized in
\reftabl{specobs}. The facilities used include FEROS on the MPG~2.2\,m
\citep{kaufer:1998},
Coralie on the Euler~1.2\,m \citep{queloz:2001}, HARPS on the ESO~3.6\,m \citep{mayor:2003}, WiFeS on the ANU~2.3\,m \citep{dopita:2007} and PFS on the
Magellan~6.5\,m \citep{crane:2006,crane:2008,crane:2010}.

The FEROS, Coralie, and HARPS observations were reduced to
wavelength-calibrated spectra and high-precision RV and Bisector Span
(BS) measurements using the CERES pipeline
\citep{brahm:2017:ceres}.

The WiFeS observations of \hatcur{37}, which were used for
reconnaissance, were reduced following
\citet{bayliss:2013:hats3}. We obtained a single
spectrum at resolution $R \equiv \Delta\,\lambda\,/\,\lambda \approx
3000$ from which we estimated the effective temperature, \logg\ and
\feh\ of the star. Three observations at $R \approx 7000$ were
also obtained to search for any large amplitude radial velocity
variations at the $\sim 4$\,\kms\ level, which would indicate a
stellar mass companion.

The PFS observations of both \hatcur{37} and \hatcur{38} include observations through
an I$_{2}$ cell, and observations without the cell used to
construct a spectral template. The observations were reduced to
spectra and used to determine high precision relative RV measurements
following \citet{butler:1996}. Spectral line bisector spans and their uncertainties were
measured as described by \citet{jordan:2014:hats4} and \citet{brahm:2017:ceres}.

We also used the HARPS and I$_{2}$-free PFS observations
to determine high-precision stellar atmospheric parameters, including
the effective temperature \teffstar, surface gravity \logg, metallicity
\feh, and \vsini\ via the ZASPE package \citep{brahm:2017:zaspe}. For \hatcur{37} we used the PFS observations to perform this analysis, while for \hatcur{38} this analysis was performed on the HARPS observations.

The high-precision RV and BS measurements are given in \reftabl{rvs} for both systems at the end of the paper.

\ifthenelse{\boolean{emulateapj}}{
    \begin{deluxetable*}{llrrrrr}
}{
    \begin{deluxetable}{llrrrrrrrr}
}
\tablewidth{0pc}
\tabletypesize{\scriptsize}
\tablecaption{
    Summary of spectroscopy observations.
    \label{tab:specobs}
}
\tablehead{
    \multicolumn{1}{c}{Instrument}          &
    \multicolumn{1}{c}{UT Date(s)}             &
    \multicolumn{1}{c}{\# Spec.}   &
    \multicolumn{1}{c}{Res.}          &
    \multicolumn{1}{c}{S/N Range\tablenotemark{a}}           &
    \multicolumn{1}{c}{$\gamma_{\rm RV}$\tablenotemark{b}} &
    \multicolumn{1}{c}{RV Precision\tablenotemark{c}} \\
    &
    &
    &
    \multicolumn{1}{c}{$\Delta \lambda$/$\lambda$/1000} &
    &
    \multicolumn{1}{c}{(\kms)}              &
    \multicolumn{1}{c}{(\ms)}
}
\startdata
\noalign{\vskip -3pt}
\sidehead{\textbf{\hatcur{37}}}\\
\noalign{\vskip -13pt}
ANU~2.3\,m/WiFeS & 2014 Feb 20 & 1 & 3 & 35 & $\cdots$ & $\cdots$ \\
ANU~2.3\,m/WiFeS & 2014 Feb 20--23 & 3 & 7 & 38--72 & 8.2 & 4000 \\
Euler~1.2\,m/Coralie & 2014 Mar--2016 Jun & 6 & 60 & 20--29 & 7.05 & 149 \\
ESO~3.6\,m/HARPS & 2016 Feb 27--29 & 2 & 115 & 19--22 & 6.417 & 38 \\
Magellan~6.5\,m/PFS+I$_{2}$ & 2016 Jun--2017 Apr & 11 & 76 & $\cdots$ & $\cdots$ & 8.7 \\
Magellan~6.5\,m/PFS & 2016 Jun 20 & 1 & 76 & $\cdots$ & $\cdots$ & $\cdots$ \\
\noalign{\vskip -3pt}
\sidehead{\textbf{\hatcur{38}}}\\
\noalign{\vskip -13pt}
Euler~1.2\,m/Coralie & 2016 Nov 16--18 & 3 & 60 & 17--20 & 4.143 & 54 \\
ESO~3.6\,m/HARPS & 2016 Nov--2017 May & 18 & 115 & 17--47 & 4.144 & 9.2 \\
MPG~2.2\,m/FEROS & 2016 Dec--2017 Mar & 10 & 48 & 36--67 & 4.130 & 19 \\
Magellan~6.5\,m/PFS+I$_{2}$ & 2017 Apr 5--8 & 4 & 76 & $\cdots$ & $\cdots$ & 5.7 \\
Magellan~6.5\,m/PFS & 2017 Apr 19 & 1 & 76 & $\cdots$ & $\cdots$ & $\cdots$ \\
\enddata
\tablenotetext{a}{
    S/N per resolution element near 5180\,\AA. This was not measured for all of the instruments.
}
\tablenotetext{b}{
    For high-precision RV observations included in the orbit determination this is the zero-point RV from the best-fit orbit. For other instruments it is the mean value. We only provide this quantity when applicable.
}
\tablenotetext{c}{
    For high-precision RV observations included in the orbit
    determination this is the scatter in the RV residuals from the
    best-fit orbit (which may include astrophysical jitter), for other
    instruments this is either an estimate of the precision (not
    including jitter), or the measured standard deviation.  We only provide this quantity when applicable.
}
\ifthenelse{\boolean{emulateapj}}{
    \end{deluxetable*}
}{
    \end{deluxetable}
}


\subsection{Photometric follow-up observations}
\label{sec:phot}

Follow-up higher-precision ground-based photometric transit
observations were obtained for both systems, as summarized in
Table~\ref{tab:photobs}. The facilities used for this purpose include:
the Chilean-Hungarian Automated Telescope (CHAT)~0.7\,m telescope at
Las Campanas Observatory, Chile \citep{jordan:2018}; 1\,m telescopes
from the Las Cumbres Observatory (LCO) network,
\citep{brown:2013:lcogt}; the 0.3\,m Perth Exoplanet Survey Telescope
in Australia (PEST)\footnote{\url{http://pestobservatory.com/}}; and
the Swope 1\,m telescope at Las Campanas Observatory in Chile.

Our methods for carrying out the observations with these
facilities and reducing the data to light curves have been described in
our previous papers
\citep{penev:2013:hats1,mohlerfischer:2013:hats2,bayliss:2013:hats3,jordan:2014:hats4,hartman:2015:hats6,rabus:2016:hats11hats12,hartman:2019:hats6069}.

The time-series photometry data are available in Table~\ref{tab:phfu},
and are plotted for each object in
Figures~\ref{fig:hats37} and \ref{fig:hats38}.

%
%
\ifthenelse{\boolean{emulateapj}}{
    \begin{deluxetable*}{llrrrrl}
}{
    \begin{deluxetable}{llrrrrl}
}
\tablewidth{0pc}
\tablecaption{
    Light curve data for \hatcur{37} and \hatcur{38}\label{tab:phfu}.
}
\tablehead{
    \colhead{Object\tablenotemark{a}} &
    \colhead{BJD\tablenotemark{b}} &
    \colhead{Mag\tablenotemark{c}} &
    \colhead{\ensuremath{\sigma_{\rm Mag}}} &
    \colhead{Mag(orig)\tablenotemark{d}} &
    \colhead{Filter} &
    \colhead{Instrument} \\
    \colhead{} &
    \colhead{} &
    \colhead{} &
    \colhead{} &
    \colhead{} &
    \colhead{} &
    \colhead{}
}
\startdata
HATS-37 & $ 2455765.23265 $ & $   0.00383 $ & $   0.00267 $ & $ \cdots $ & $ r$ &         HS\\
HATS-37 & $ 2455691.59689 $ & $  -0.00538 $ & $   0.00294 $ & $ \cdots $ & $ r$ &         HS\\
HATS-37 & $ 2455678.60275 $ & $   0.00353 $ & $   0.00337 $ & $ \cdots $ & $ r$ &         HS\\
HATS-37 & $ 2455747.90786 $ & $   0.00160 $ & $   0.00295 $ & $ \cdots $ & $ r$ &         HS\\
HATS-37 & $ 2455682.93494 $ & $  -0.00208 $ & $   0.00250 $ & $ \cdots $ & $ r$ &         HS\\
HATS-37 & $ 2455665.61003 $ & $  -0.00583 $ & $   0.00305 $ & $ \cdots $ & $ r$ &         HS\\
HATS-37 & $ 2455708.92562 $ & $   0.00583 $ & $   0.00249 $ & $ \cdots $ & $ r$ &         HS\\
HATS-37 & $ 2455691.60029 $ & $   0.00317 $ & $   0.00300 $ & $ \cdots $ & $ r$ &         HS\\
HATS-37 & $ 2455652.61745 $ & $  -0.00718 $ & $   0.00291 $ & $ \cdots $ & $ r$ &         HS\\
HATS-37 & $ 2455747.91127 $ & $   0.00013 $ & $   0.00262 $ & $ \cdots $ & $ r$ &         HS\\

\enddata
\tablenotetext{a}{
    Either \hatcur{37} or \hatcur{38}.
}
\tablenotetext{b}{
    Barycentric Julian Date is computed directly from the UTC time
    without correction for leap seconds.
}
\tablenotetext{c}{
    The out-of-transit level has been subtracted. For observations
    made with the HATSouth instruments (identified by ``HS'' in the
    ``Instrument'' column) these magnitudes have been corrected for
    trends using the EPD and TFA procedures applied {\em prior} to
    fitting the transit model. This procedure may lead to an
    artificial dilution in the transit depths. The blend factors for
    the HATSouth light curves are listed in
    Table~\ref{tab:planetparam}. For
    observations made with follow-up instruments (anything other than
    ``HS'' in the ``Instrument'' column), the magnitudes have been
    corrected for a quadratic trend in time, and for variations
    correlated with up to three PSF shape parameters, fit simultaneously
    with the transit.
}
\tablenotetext{d}{
    Raw magnitude values without correction for the quadratic trend in
    time, or for trends correlated with the seeing. These are only
    reported for the follow-up observations.
}
\tablecomments{
    This table is available in a machine-readable form in the online
    journal.  A portion is shown here for guidance regarding its form
    and content.
}
\ifthenelse{\boolean{emulateapj}}{
    \end{deluxetable*}
}{
    \end{deluxetable}
}

\subsection{\emph{TESS} light curves}
\label{sec:tess}

During its primary mission, \emph{TESS} observed both of our targets. HATS-37 (TIC6036597) was observed on Sector 10, CCD 3 of camera 1, but the source lies within the bleed of a nearby bright star making the photometry unusable. HATS-38 (TIC168281028) was observed by the \emph{TESS} primary mission during its first year of operations. The target star fell on Camera 2, CCD 4 of the Sector 9 observations. Photometry was extracted from the Science Processing Operations Center \citep[SPOC][]{2016SPIE.9913E..3EJ} calibrated Full Frame Images (FFI), retrieved via the MAST \emph{tesscut} tool. Aperture photometry was performed using selected pixels of a $7\times7$ pixel cutout of the FFIs with the \emph{lightkurve} package \citep{lightkurve}. The background flux was estimated from the remainder pixels that excluded nearby stars. We corrected for the flux contribution from nearby stars within our photometric aperture. A list of nearby stars was queried from the TICv8 catalogue \citep{2019AJ....158..138S}, and their flux contributions to the photometric aperture was computed assuming each star has a Gaussian profile with FWHM of 1.63 pixels, as measured from the \emph{TESS} pixel response function at the location of the target star.
The \emph{TESS} light curve for HATS-38 is shown in Figure~\ref{fig:hats38TESS}.

\subsection{Search for Resolved Stellar Companions}
\label{sec:luckyimaging}

The Gaia~DR2 catalog provides the highest spatial resolution optical imaging
for both of these targets. Gaia~DR2 is sensitive to
neighbors with $G \la 20$\,mag down to a limiting resolution of $\sim
1\arcsec$ \citep[e.g.,][]{ziegler:2018}. We find that neither object has a resolved neighbor in the Gaia~DR2 catalog within 10\arcsec.

For \hatcur{38} we also obtained $J$ and $K_{S}$-band images using the
WIYN High-Resolution Infrared Camera (WHIRC) on the WIYN~3.5\,m
telescope at Kitt Peak National Observatory (KPNO) in Arizona. The
observations were carried out on the night of 2018 March 18, and have
an effective full-width at half maximum (FWHM) of $0\farcs43$ in $J$
and $0\farcs35$ in $K_{S}$. The images were collected at four
different nod positions in each filter. These were calibrated,
background-subtracted, registered and median-combined using the
{\sc  fitsh} software package \citep{pal:2012}.

We find a faint source separated from \hatcur{38} by $6\arcsec$. The
source is detected at about $\sim 3\sigma$ confidence in both bands,
and has a magnitude contrast of $\Delta J = 8.05 \pm 0.09$\,mag and
$\Delta K_{s} = 7.18 \pm 0.08$\,mag compared to \hatcur{38}. The
object is too faint, and too distant from \hatcur{38} to be
responsible for the transit signal. The $J$ and $K_{s}$ magnitudes are
consistent with it being a $0.09$\,\msun\ star that is physically
bound to \hatcur{38}, at a current projected separation of $\sim
2100$\,AU. In that case the source would have $G \sim 23$\,mag,
consistent with the object not being included in Gaia~DR2. It could
also be an extragalactic source, an earlier M dwarf star that is in the
background of \hatcur{38}, or a foreground brown dwarf.

No other sources are detected closer to \hatcur{38} in the WIYN/WHIRC
images. Fig.~\ref{fig:hats38whirc} shows the resulting $5\sigma$
contrast curves for \hatcur{38}. These curves were generated using the
tools described by \citet{espinoza:2016:hats25hats30}. We can rule out
neighbors with $\Delta J < 3$\,mag and $\Delta K_{s} < 3$\,mag at a separation
of $0\farcs5$, and $\Delta J < 7$\,mag and $\Delta K_{s} < 6$\,mag at a
separation of $1\farcs5$.

\ifthenelse{\boolean{emulateapj}}{
    \begin{figure*}[!ht]
}{
    \begin{figure}[!ht]
}
\plottwo{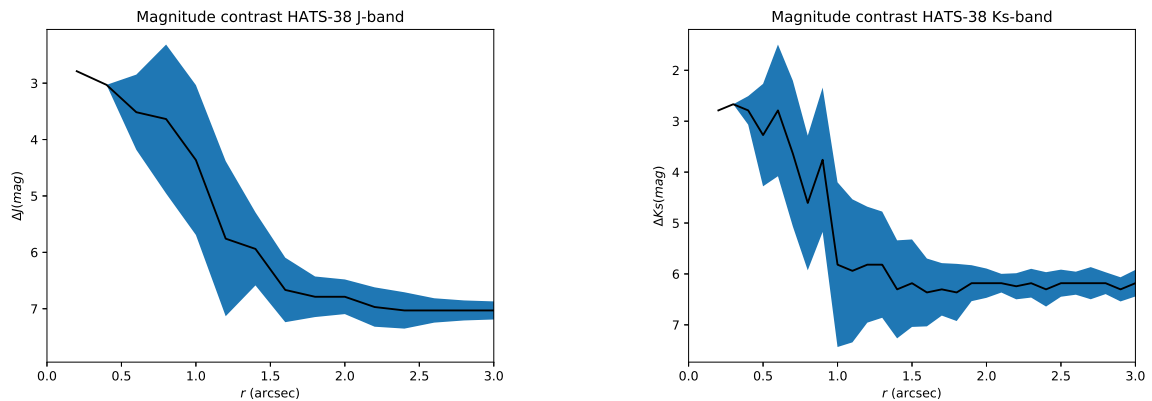}{\hatcurhtr{38}_Ks_WHIRC_magcontrast-eps-converted-to.pdf}
\caption{
$5\sigma$ contrast curve for \hatcur{38} based on our WIYN/WHIRC $J$-band ({\em left}) and $K_{S}$-band ({\em right}) observations. In each case the blue band shows the variation in the limit in azimuth at a given radius.
\label{fig:hats38whirc}}
\ifthenelse{\boolean{emulateapj}}{
    \end{figure*}
}{
    \end{figure}
}

\section{Analysis}
\label{sec:analysis}

We analyzed the photometric and spectroscopic observations of each
system to determine the stellar and planetary parameters following the
methods described in \citet{hartman:2019:hats6069}, with modifications
as summarized most recently in \citet{bakos:2018:hats71}. Briefly, the
modelling involves performing a global fit of all the light curves
and RV curves described in section~\ref{sec:obs}, spectroscopically measured
stellar atmospheric parameters,
catalog broad-band photometry, and stellar parallax using a
Differential Evolution Markov Chain Monte Carlo (DEMCMC) method.  We
fit the observations in two modes: (1) using an empirical method to
determine the stellar mass given the direct observational constraint
on the stellar radius and bulk density; and (2) constraining the
stellar physical parameters using the PARSEC stellar evolution models
\citep{marigo:2017}. We use the MWDUST Galactic extinction model
\citep{bovy:2016} to place a prior constraint on the line of sight
extinction, but we allow the value to vary in the fit.

We also performed a blend modelling of each system following
\citet{hartman:2019:hats6069}, where we attempt to fit all of the
observations, except the RV data, using various combinations of stars,
with parameters constrained by the PARSEC models. This is done both to
rule out blended stellar eclipsing binary scenarios, and to identify
systems that may have an unresolved stellar companion. For the blend
modelling we consider five scenarios: (1) a single star with a
transiting planet (the H-p scenario); (2) an unresolved binary star
system with a transiting planet around the brighter stellar component
(the H-p,s scenario); (3) an unresolved binary star system with a
transiting planet around the fainter stellar component (the H,s-p
scenario); (4) a hierarchical triple star system consisting of a
bright star and a fainter eclipsing binary system (the H,s-s
scenario); and (5) a blend between a bright foreground star, and a
background stellar eclipsing binary system (the H,s-s$_{\rm BGEB}$
scenario). For each case we perform an initial grid search over the
most difficult to optimize parameters to find the global maximum
likelihood (ML) fit, and then perform a DEMCMC analysis, initializing
the chain near the ML location. As part of this analysis we also
predict spectral line bisector span (BS) measurements, and RV
measurements from the composite system. These are compared to the
observed RV and BS measurements to rule out any blend scenarios that,
while consistent with the photometric observations, predict much
larger RV and BS variations than observed. For blend scenarios
containing a transiting planet, we use these simulated RV observations
to determine a scaling factor by which we expect the RV semi-amplitude
$K$ to be reduced by dilution from the stellar companion. We then use
this factor to scale the value of $K$ determined from our H-p model of
the RV observations to obtain corrected values for the H-p,s and H-s,p
models. We assume a 20\% uncertainty on the scaling factor.

For \hatcur{37} we find that the H-p,s scenario provides the best fit
to the photometric data, with $\chi^{2}_{H-p,s} - \chi^{2}_{H-p} =
-296$ and $\chi^{2}_{H-p,s} - \chi^{2}_{H,s-s,BGEB} = -166$, and even
greater improvements relative to the H,s-s and H,s-p scenarios. Based
on this we conclude that \hatcur{37} is not a blended stellar
eclipsing binary object, but rather is best interpreted as a star with
a transiting planet and a fainter, unresolved stellar companion. Note
that here the use of the MWDUST Galactic extinction model is critical
in coming to this conclusion. When the extinction is allowed to vary
without the constraint, we find that the H,s-s$_{\rm BGEB}$ scenario
provides a slightly better fit to the data than the H-p,s model, while
the improvement of the H-p,s model compared to the H-p model is less
significant. These models, however, require much greater extinction
($A_{V} > 3$\,mag in the case of the H,s-s$_{\rm BGEB}$ model, and
$A_{V} \sim 1$\,mag in the case of the H-p model) that is at odds with
the total line of sight extinction of 0.274\,mag based on dust
maps. The best-fit H-p,s model, however, yields $A_{V} =
\hatcurXAvhps{37}$\,mag, which is in good agreement with the dust maps.

In addition to the photometric evidence for an unresolved stellar
companion to \hatcurA{37}, we also find evidence for such a companion
in the RV observations. The PFS RVs of this system show a strong
linear trend of $\hatcurRVtrone{37}$\,\ms\,${\rm d}^{-1}$
(Figure~\ref{fig:hats37rvjd}). We included this trend, together with a
Keplerian orbit for the transiting system, in our modeling of the RV
observations. If the trend corresponds to the line-of-sight
acceleration of \hatcurA{37} due to \hatcurB{37}, then given the
estimated mass of \hatcurISOmBhps{37}\,\msun\ from our H-p,s model, we
can place an upper limit on the current physical separation between
the two stars of $a_{AB} < 27.2$\,AU.  This upper limit corresponds to
the case where there is no projected separation between the two
stars. The maximum projected separation consistent with this
acceleration is $a_{AB,{\rm proj}} < 16.9$\,AU, corresponding to a
maximum current angular separation between the stars of $\theta_{AB} <
0\farcs08$.

For \hatcur{38} we find that the H-p, H-p,s and H,s-s$_{\rm BGEB}$
models provide comparable fits to the photometric data, with
$\chi^{2}_{H-p} - \chi^{2}_{H-p,s} = 7.0$ and $\chi^{2}_{H-p} -
\chi^{2}_{H,s-s,BGEB} = 5.8$. These differences are comparable to the
$1\sigma$ scatter in $\chi^2$ for a given model as measured from the
Markov Chains, and consistent with the slight improvement in the fit
for the H,s-s$_{\rm BGEB}$ and H-p,s models being solely due to the
increased complexity of these models. In this case we make use of the
RV and BS observations to rule out the H,s-s$_{\rm BGEB}$ model. The
simulated HARPS RV and BS observations for the H,s-s$_{\rm BGEB}$
model show significantly larger variations than observed, with the
simulated RV RMS in excess of 200\,\ms, and the simulated BS RMS in
excess of 300\,\ms. The actual HARPS RV and BS observations have RMS
scatters of only 12\,\ms\ and 8\,\ms, respectively, with the RV
observations following a Keplerian orbit as expected for a the case of
a transiting planet system. We can also rule out the H,s-s and H,s-p
models based on the photometry as these both provide significantly
worse fits to the data than the H-p model. Since the H-p,s model does
not provide a significant improvement over the H-p model, we choose to
adopt the parameters for the system assuming it is a single star with
a transiting planet. We place a 95\% confidence upper limit on the
mass of any unresolved companion star of $M_{B} < 0.62$\,\msun. If we
adopted the H-p,s model instead, the estimated planetary radius would
be smaller by $4$\%, with a $1\sigma$ uncertainty of 5\% in the
difference. Note that the planet would be smaller due to its host star
being smaller, even though the transits would be somewhat diluted.

Figures~\ref{fig:hats37} and~\ref{fig:hats38} compares the best-fit
models to the observations for both \hatcur{37} and \hatcur{38}. Our
final set of adopted stellar parameters derived from this analysis are
listed in Table~\ref{tab:stellarderived}, while the adopted planetary
parameters are listed in Table~\ref{tab:planetparam}.



\section{Discussion}
\label{sec:discussion}

\begin{figure*}
\includegraphics[width=\textwidth]{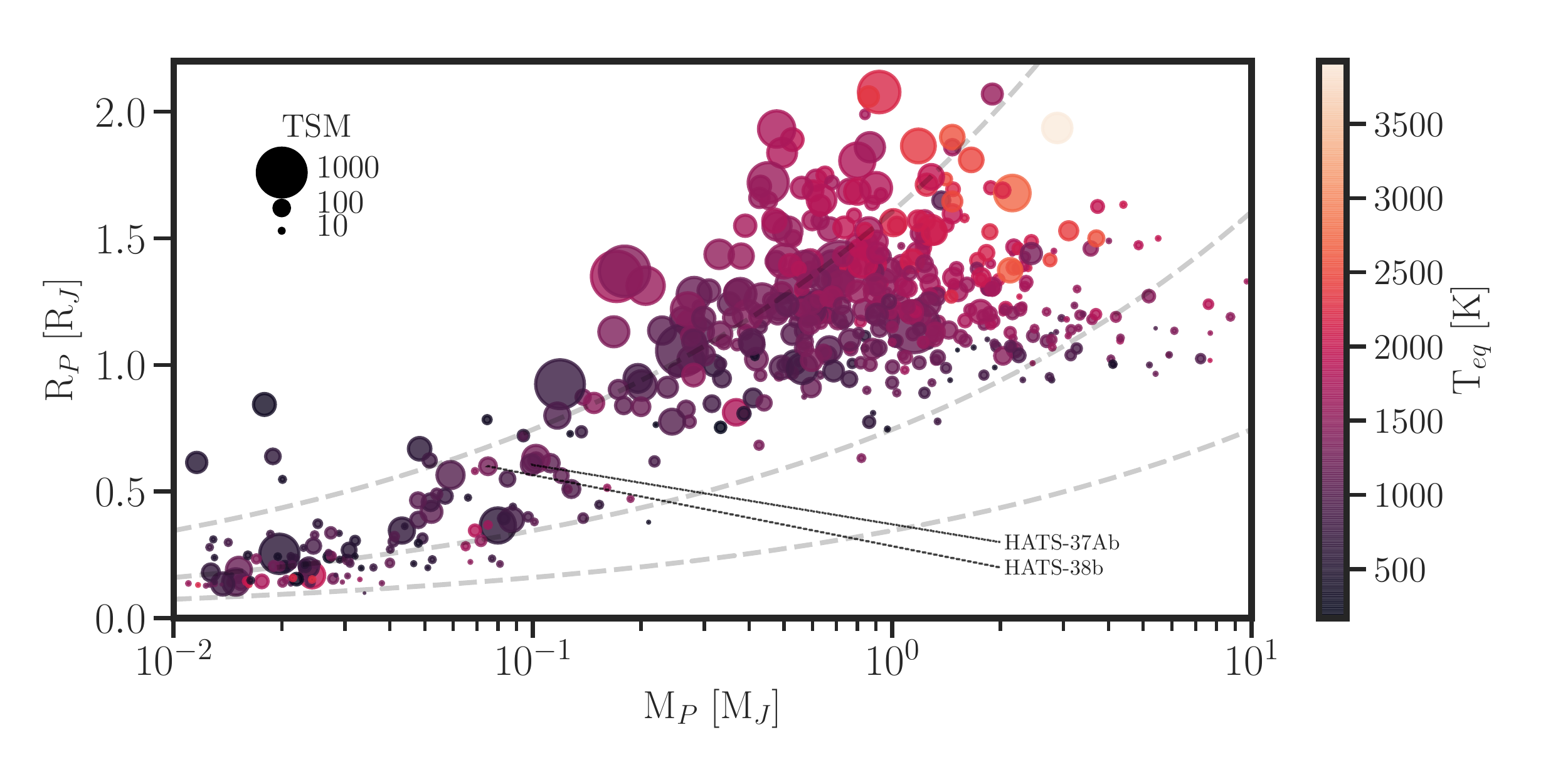}
\caption{Mass -- Radius diagram for the population of well characterized transiting planets \citep{southworth:2011:tepcat}. The points corresponding to \hatcurb{37} and \hatcurb{38} are indicated with  dashed lines. The color represents the equilibrium temperature of the planet, while the size scales down with the transmission spectroscopy metric as defined by \citet{tsm}. The dashed gray lines correspond to isodensity curves for 0.3, 3 and 30 g cm$^{-3}$, respectively.\label{fig:mr}}
\end{figure*}

We put \hatcurb{37} and \hatcurb{38} in the context of the population of known, well-characterized\footnote{We use the catalog of well-characterized planets of \citet{southworth:2011:tepcat}. The catalog is kept updated online at \url{https://www.astro.keele.ac.uk/jkt/tepcat/} and the data we used were retrieved in November 2019. We restrict the sample to systems whose fractional error on their planetary masses are $<50\%$, and planetary radii are $<25\%$.} transiting exoplanets in Figure~\ref{fig:mr}, where we show a scatter plot of planetary mass versus planetary radius, coding with color the equilibrium temperature. Both planets have a relatively low density close to $0.3$ g cm$^{-3}$, which among with their other properties translate into a transmission spectroscopy metric \citep[TSM,][]{tsm} of $\approx$ 120 for \hatcurb{37} and $\approx$ 165 for \hatcurb{38}. The latter figure makes \hatcurb{38} an attractive target among the currently known set of transiting Neptunes for transmission spectroscopy. Both targets populate a region in the mass--radius plane which is sparsely populated and where the transition between gas giants and the population of smaller planets occur. We note that both \hatcurb{37} and \hatcurb{38} are among the lowest mass planets found from ground-based wide-field surveys to date, joining a select group of systems uncovered by such surveys with masses $M_p \lesssim 0.1 M_J$: HAT-P-26~b \citep[$0.059 \pm 0.007 \, M_J$,][]{hartman:2011}, NGTS-4~b \citep[$0.0648 \pm 0.0094 \, M_J$,][]{west:2019},  HAT-P-11~b \citep[$0.0736 \pm 0.0047 \, M_J$,][]{bakos:2010:hat11,yee:2018} and WASP-166~b \citep[$0.101 \pm 0.005 \, M_J$,][]{hellier:2019}.

\begin{figure*}
\includegraphics[width=\textwidth]{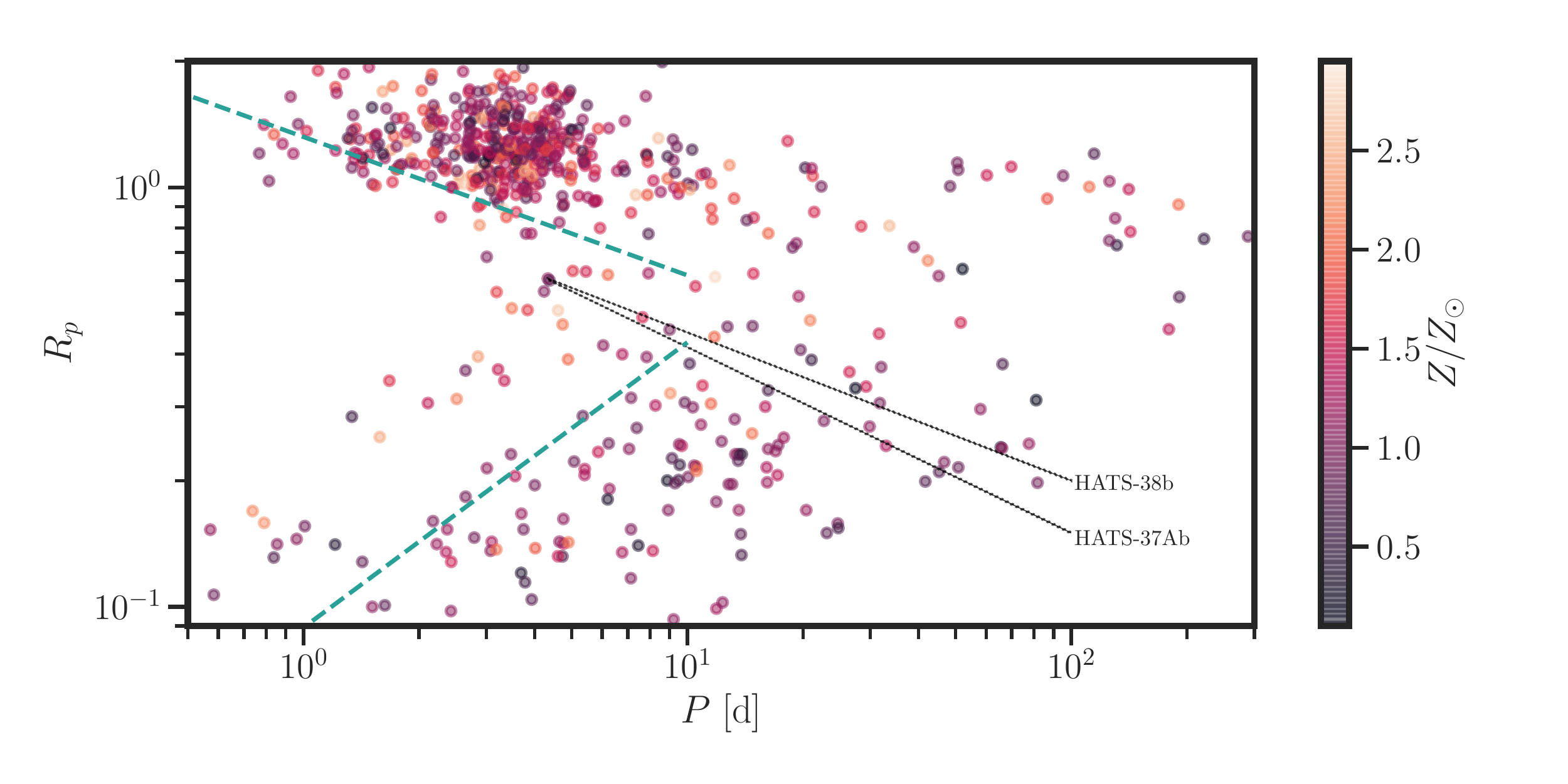}
\caption{Period -- Radius diagram for the population of well characterized transiting planets. The point corresponding to \hatcurb{37} and \hatcurb{38} are indicated with dotted black lines.  The dashed cyan lines mark the boundaries of the Neptune desert as defined by \citet{mazeh:2016}. The points are color-coded according to the metallicity of the host star. \label{fig:perrad}}
\end{figure*}

In Figure~\ref{fig:perrad} we show the population of well-characterized planets in the period--radius plane, where \hatcurb{37} and \hatcurb{38} are extremely similar. In this figure we show the region defined as the Neptune desert by \citet{mazeh:2016}. While not lying in the region with $P\lesssim 3$ days and $0.4 \lesssim (R_p/R_J) \lesssim 0.8$ that is essentially devoid of planets, both \hatcurb{37} and \hatcurb{38} lie within the region defined as the Neptune desert and that has an intrinsically low occurrence rate of planets.

When we consider additional parameters than planetary mass and radius and consider also the properties of the host stars, the properties of \hatcurb{37} and \hatcurb{38} emerge as being particularly rare. \citet{dong:2018} used a large sample of stellar parameters obtained with the Large Sky Area Multi-Object Fiber Spectroscopic Telescope (LAMOST) to further characterize the Neptune desert region.
Their study reveals a dearth of planets in the radius range $6 \lesssim (R_p/R_\earth) \lesssim 10$, which they term the Saturn valley, and a population of hot Neptunes with radii $
2 \lesssim (R_p/R_\earth) \lesssim 6$ which are rare (occurrence rate of $\approx$ 1\% for FGK stars) and whose occurrence is correlated with metallicity in the sense that hot Neptunes appear preferentially around metal-rich stars. In fact, \citet{dong:2018} find the great majority of the hot Neptunes in their sample to be hosted by stars with [Fe/H]$\geq 0.1$. Both \hatcurb{37} and \hatcurb{38} have radii  $\approx 6.7R_\earth$ making them large specimens for hot Neptunes and veering into the Saturn valley as defined by \citet{dong:2018}. More strikingly, \hatcur{38} has an estimated metallicity of $\approx -0.1$, making it a very metal-poor star to host a hot Neptune given the expected occurrence rate at that metallicity of order $\sim 10^{-3}$ \citep[][see their Figure~4]{dong:2018}. Even if the metallicity was as high as $\approx 0.05$,  as allowed at the $\approx 3.5\sigma$ level, the expected occurrence rate is $\lesssim 5\times 10^{-3}$. Thus, we can see that \hatcurb{37} and \hatcurb{38} contribute a new pair of exoplanetary systems with uncommon properties and showcase the continuing contributions of wide-field ground-based surveys to better map the variety of landscapes present in the exoplanetary realm.


\acknowledgements

Development of the HATSouth
project was funded by NSF MRI grant NSF/AST-0723074, operations have
been supported by NASA grants NNX09AB29G, NNX12AH91H, and NNX17AB61G, and follow-up
observations have received partial support from grant NSF/AST-1108686.
A.J.\, R.B.\ and V.S.\ acknowledge support from project IC120009 ``Millennium Institute of Astrophysics
(MAS)'' of the Millenium Science Initiative, Chilean Ministry of
Economy.
A.J.\ acknowledges additional support from FONDECYT project 1171208.
R.B.\ acknowledges additional support from FONDECYT Post-doctoral
Fellowship Project No. 3180246.
L.M.\ acknowledges support from the Italian Minister of Instruction,
University and Research (MIUR) through FFABR 2017 fund.
T.H.\ acknowledges support from the European Research Council under the
Horizon 2020 Framework Program via the ERC Advanced Grant Origins 83 24 28.
K.P.\ acknowledges support from NASA grant  80NSSC18K1009.2
This work is based on observations made with ESO Telescopes at the La
Silla Observatory.
This paper also makes use of observations from the LCOGT network.
Some of this time was awarded by NOAO.
We acknowledge the use of the AAVSO Photometric All-Sky Survey (APASS),
funded by the Robert Martin Ayers Sciences Fund, and the SIMBAD
database, operated at CDS, Strasbourg, France.
Operations at the MPG~2.2\,m Telescope are jointly performed by the
Max Planck Gesellschaft and the European Southern Observatory.
We thank the MPG 2.2m telescope support team for their technical assistance during observations.
This work has made use of data from the European Space Agency (ESA)
mission {\it Gaia} (\url{https://www.cosmos.esa.int/gaia}), processed by
the {\it Gaia} Data Processing and Analysis Consortium (DPAC,
\url{https://www.cosmos.esa.int/web/gaia/dpac/consortium}). Funding
for the DPAC has been provided by national institutions, in particular
the institutions participating in the {\it Gaia} Multilateral Agreement.
This research has made use of the NASA Exoplanet Archive, which is operated by the California Institute of Technology, under contract with the National Aeronautics and Space Administration under the Exoplanet Exploration Program.

\facilities{HATSouth, LCOGT, FTS, CTIO:0.9m, Danish 1.54m Telescope (DFOSC), Swope, Max Planck:2.2m (FEROS), ESO:3.6m (HARPS), Euler1.2m (Coralie), ATT (WiFeS), AAT (CYCLOPS), Magellan:Clay (PFS), VLT:Kueyen (UVES), NTT (Astralux Sur), Gaia}


\bibliographystyle{aasjournal}
\bibliography{hatsbib}

%
%
\ifthenelse{\boolean{emulateapj}}{
    \begin{deluxetable*}{lccccl}
}{
    \begin{deluxetable}{lccccl}
}
\tablewidth{0pc}
\tabletypesize{\tiny}
\tablecaption{
    Astrometric, Spectroscopic and Photometric parameters for \hatcur{37} and \hatcur{38}
    \label{tab:stellarobserved}
}
\tablehead{
    \multicolumn{1}{c}{} &
    \multicolumn{1}{c}{\bf HATS-37} &
    \multicolumn{1}{c}{\bf HATS-38} &
    \multicolumn{1}{c}{} \\
    \multicolumn{1}{c}{~~~~~~~~Parameter~~~~~~~~} &
    \multicolumn{1}{c}{Value}                     &
    \multicolumn{1}{c}{Value}                     &
    \multicolumn{1}{c}{Source}
}
\startdata
\noalign{\vskip -3pt}
\sidehead{Astrometric properties and cross-identifications}
~~~~2MASS-ID\dotfill               & \hatcurCCtwomassshort{37}  & \hatcurCCtwomassshort{38} & \\
~~~~GAIA~DR2-ID\dotfill                 & \hatcurCCgaiadrtwoshort{37}      & \hatcurCCgaiadrtwoshort{38}& \\
~~~~TIC-ID\dotfill & 6036597 & 168281028 & \\
~~~~R.A. (J2000)\dotfill            & \hatcurCCra{37}       & \hatcurCCra{38}    & GAIA DR2\\
~~~~Dec. (J2000)\dotfill            & \hatcurCCdec{37}      & \hatcurCCdec{38}   & GAIA DR2\\
~~~~$\mu_{\rm R.A.}$ (\masy)              & \hatcurCCpmra{37}     & \hatcurCCpmra{38} & GAIA DR2\\
~~~~$\mu_{\rm Dec.}$ (\masy)              & \hatcurCCpmdec{37}    & \hatcurCCpmdec{38} & GAIA DR2\\
~~~~parallax (mas)              & \hatcurCCparallax{37}    & \hatcurCCparallax{38} & GAIA DR2\\
\sidehead{Spectroscopic properties}
~~~~$\teffstar$ (K)\dotfill         &  \hatcurSMEteff{37}   & \hatcurSMEteff{38} & ZASPE\tablenotemark{a}\\
~~~~$\feh$\dotfill                  &  \hatcurSMEzfeh{37}   & \hatcurSMEzfeh{38} & ZASPE               \\
~~~~$\vsini$ (\kms)\dotfill         &  \hatcurSMEvsin{37}   & \hatcurSMEvsin{38} & ZASPE                \\
~~~~$\vmac$ (\kms)\dotfill          &  $3.175 \pm 0.076$   & \hatcurSMEvmac{38} & Assumed \\
~~~~$\vmic$ (\kms)\dotfill          &  $0.818 \pm 0.023$   & \hatcurSMEvmic{38} & Assumed              \\
~~~~$\gamma_{\rm RV}$ (\ms)\dotfill&  $\hatcurRVgammaA{37}$  & $\hatcurRVgammaB{38}$ & HARPS\tablenotemark{b}  \\
\sidehead{Photometric properties}
~~~~$G$ (mag)\tablenotemark{c}\dotfill               &  \hatcurCCgaiamG{37}  & \hatcurCCgaiamG{38} & GAIA DR2 \\
~~~~$BP$ (mag)\tablenotemark{c}\dotfill               &  \hatcurCCgaiamBP{37}  & \hatcurCCgaiamBP{38} & GAIA DR2 \\
~~~~$RP$ (mag)\tablenotemark{c}\dotfill               &  \hatcurCCgaiamRP{37}  & \hatcurCCgaiamRP{38} & GAIA DR2 \\
~~~~$B$ (mag)\dotfill               &  \hatcurCCtassmB{37}  & \hatcurCCtassmB{38} & APASS\tablenotemark{d} \\
~~~~$V$ (mag)\dotfill               &  \hatcurCCtassmv{37}  & \hatcurCCtassmv{38} & APASS\tablenotemark{d} \\
~~~~$g$ (mag)\dotfill               &  \hatcurCCtassmg{37}  & \hatcurCCtassmg{38} & APASS\tablenotemark{d} \\
~~~~$r$ (mag)\dotfill               &  \hatcurCCtassmr{37}  & \hatcurCCtassmr{38} & APASS\tablenotemark{d} \\
~~~~$i$ (mag)\dotfill               &  \hatcurCCtassmi{37}  & \hatcurCCtassmi{38} & APASS\tablenotemark{d} \\
~~~~$J$ (mag)\dotfill               &  \hatcurCCtwomassJmag{37} & \hatcurCCtwomassJmag{38} & 2MASS           \\
~~~~$H$ (mag)\dotfill               &  \hatcurCCtwomassHmag{37} & \hatcurCCtwomassHmag{38} & 2MASS           \\
~~~~$K_s$ (mag)\dotfill             &  \hatcurCCtwomassKmag{37} & \hatcurCCtwomassKmag{38} & 2MASS           \\
~~~~$W1$ (mag)\dotfill             &  \hatcurCCWonemag{37} & \hatcurCCWonemag{38} & WISE           \\
~~~~$W2$ (mag)\dotfill             &  \hatcurCCWtwomag{37} & \hatcurCCWtwomag{38} & WISE           \\
~~~~$W3$ (mag)\dotfill             &  \hatcurCCWthreemag{37} & \hatcurCCWthreemag{38} & WISE           \\
\enddata
\tablenotetext{a}{
    ZASPE = Zonal Atmospherical Stellar Parameter Estimator routine
    for the analysis of high-resolution spectra
    \citep{brahm:2017:zaspe}, applied to the FEROS spectra of each system. These parameters rely primarily on ZASPE, but have a small
    dependence also on the iterative analysis incorporating the
    isochrone search and global modeling of the data.
}
\tablenotetext{b}{
    The error on $\gamma_{\rm RV}$ is determined from the
    orbital fit to the RV measurements, and does not include the
    systematic uncertainty in transforming the velocities to the IAU
    standard system. The velocities have not been corrected for
    gravitational redshifts.
}
\tablenotetext{c}{
    The listed uncertainties for the Gaia DR2 photometry are taken from the catalog. For the analysis we assume additional systematic uncertainties of 0.002\,mag, 0.005\,mag and 0.003\,mag for the G, BP and RP bands, respectively.
}
\tablenotetext{d}{
    From APASS DR6 as
    listed in the UCAC 4 catalog \citep{zacharias:2013:ucac4}.
}
\ifthenelse{\boolean{emulateapj}}{
    \end{deluxetable*}
}{
    \end{deluxetable}
}

%
%
\ifthenelse{\boolean{emulateapj}}{
    \begin{deluxetable*}{lcc}
}{
    \begin{deluxetable}{lcc}
}
\tablewidth{0pc}
\tabletypesize{\footnotesize}
\tablecaption{
    Adopted derived stellar parameters for \hatcur{37} and \hatcur{38}
    \label{tab:stellarderived}
}
\tablehead{
    \multicolumn{1}{c}{} &
    \multicolumn{1}{c}{\bf HATS-37} &
    \multicolumn{1}{c}{\bf HATS-38} \\
    \multicolumn{1}{c}{~~~~~~~~Parameter~~~~~~~~} &
    \multicolumn{1}{c}{Value}                     &
    \multicolumn{1}{c}{Value}
}
\startdata
\sidehead{Planet Hosting Star \hatcur{37}A and \hatcur{38}}
~~~~$\mstar$ ($\msun$)\dotfill      &  \hatcurISOmlonghps{37}   & \hatcurISOmlong{38} \\
~~~~$\rstar$ ($\rsun$)\dotfill      &  \hatcurISOrlonghps{37}   & \hatcurISOrlong{38} \\
~~~~$\loggstar$ (cgs)\dotfill       &  \hatcurISOlogghps{37}    & \hatcurISOlogg{38} \\
~~~~$\lstar$ ($\lsun$)\dotfill      &  \hatcurISOlumhps{37}     & \hatcurISOlum{38} \\
~~~~$\teffstar$ (K)\dotfill      &  \hatcurISOteffhps{37} &  \hatcurISOteff{38} \\
~~~~$\feh$\dotfill      &  \hatcurISOfehhps{37} &  \hatcurISOzfeh{38} \\
~~~~Age (Gyr)\dotfill               &  \hatcurISOagehps{37}     & \hatcurISOage{38} \\
~~~~$A_{V}$ (mag)\dotfill               &  \hatcurXAvhps{37}     & \hatcurXAv{38} \\
~~~~Distance (pc)\dotfill           &  \hatcurXdistredhps{37}\phn  & \hatcurXdistred{38} \\
\sidehead{Binary Star Companion \hatcur{37}B}
~~~~$\mstar$ ($\msun$)\dotfill           & \hatcurISOmlongBhps{37} & $\cdots$ \\
~~~~$\rstar$ ($\rsun$)\dotfill           & \hatcurISOrlongBhps{37} & $\cdots$ \\
~~~~$\loggstar$ (cgs)\dotfill            & \hatcurISOloggBhps{37} & $\cdots$ \\
~~~~$\lstar$ ($\lsun$)\dotfill           & \hatcurISOlumBhps{37} & $\cdots$ \\
~~~~$\teffstar$ (K)\dotfill              & \hatcurISOteffBhps{37} & $\cdots$ \\
\enddata
\tablecomments{
The listed parameters are those determined through the joint differential evolution Markov Chain analysis described in Section~\ref{sec:analysis}. For both systems the RV observations are consistent with a circular orbit, and we assume a fixed circular orbit in generating the parameters listed here. Systematic errors in the bolometric correction tables or stellar evolution models are not included, and likely dominate the error budget.
}
\ifthenelse{\boolean{emulateapj}}{
    \end{deluxetable*}
}{
    \end{deluxetable}
}

\clearpage

\startlongtable
%
\ifthenelse{\boolean{emulateapj}}{
    \begin{deluxetable*}{lcccc}
}{
    \begin{deluxetable}{lcccc}
}
\tabletypesize{\tiny}
\tablecaption{Adopted orbital and planetary parameters for \hatcurb{37} and \hatcurb{38}\label{tab:planetparam}}
\tablehead{
    \multicolumn{1}{c}{} &
    \multicolumn{1}{c}{\bf HATS-37Ab} &
    \multicolumn{1}{c}{\bf HATS-38b} \\
    \multicolumn{1}{c}{~~~~~~~~~~~~~~~Parameter~~~~~~~~~~~~~~~} &
    \multicolumn{1}{c}{Value} &
    \multicolumn{1}{c}{Value}
}
\startdata
\noalign{\vskip -3pt}
\sidehead{\Lc{} parameters}
~~~$P$ (days)             \dotfill    & $\hatcurLCPhps{37}$ & $\hatcurLCP{38}$ \\
~~~$T_c$ (${\rm BJD}$)
      \tablenotemark{a}   \dotfill    & $\hatcurLCThps{37}$ & $\hatcurLCT{38}$ \\
~~~$T_{14}$ (days)
      \tablenotemark{a}   \dotfill    & $\hatcurLCdurhps{37}$ & $\hatcurLCdur{38}$ \\
~~~$T_{12} = T_{34}$ (days)
      \tablenotemark{a}   \dotfill    & $\hatcurLCingdurhps{37}$ & $\hatcurLCingdur{38}$ \\
~~~$\arstar$              \dotfill    & $\hatcurPParhps{37}$ & $\hatcurPPar{38}$ \\
~~~$\zrstar$ \tablenotemark{b}             \dotfill    & $\hatcurLCzeta{37}$\phn & $\hatcurLCzeta{38}$\phn\\
~~~$\rpl/\rstar$          \dotfill    & $\hatcurLCrprstarhps{37}$ & $\hatcurLCrprstar{38}$\\
~~~$b^2$                  \dotfill    & $\hatcurLCbsqhps{37}$ & $\hatcurLCbsq{38}$\\
~~~$b \equiv a \cos i/\rstar$
                          \dotfill    & $\hatcurLCimphps{37}$ & $\hatcurLCimp{38}$\\
~~~$i$ (deg)              \dotfill    & $\hatcurPPihps{37}$\phn & $\hatcurPPi{38}$\phn\\
%
\sidehead{HATSouth dilution factors \tablenotemark{c}}
~~~Dilution factor 1 \dotfill & \hatcurLCiblendhps{37} & \hatcurLCiblendA{38}\\
~~~Dilution factor 2 \dotfill & $\cdots$ & \hatcurLCiblendB{38}\\
\sidehead{Limb-darkening coefficients \tablenotemark{d}}
~~~$c_1,r$                  \dotfill    & $\hatcurLBir{37}$ & $\hatcurLBir{38}$\\
~~~$c_2,r$                  \dotfill    & $\hatcurLBiir{37}$ & $\hatcurLBiir{38}$\\
~~~$c_1,R$                  \dotfill    & $\hatcurLBiR{37}$ & $\cdots$\\
~~~$c_2,R$                  \dotfill    & $\hatcurLBiiR{37}$ & $\cdots$\\
~~~$c_1,i$                  \dotfill    & $\hatcurLBii{37}$ & $\hatcurLBii{38}$\\
~~~$c_2,i$                  \dotfill    & $\hatcurLBiii{37}$ & $\hatcurLBiii{38}$\\
\sidehead{RV parameters}
~~~$K$ (\ms)              \dotfill    & $\hatcurRVKhps{37}$\phn\phn & $\hatcurRVK{38}$\phn\phn\\
~~~${\gamma}$ (\ms)              \dotfill    & $\hatcurRVgammaA{37}$ & $\hatcurRVgammaB{38}$\\
~~~$\dot{\gamma}$ (\ms\,d$^{-1}$)              \dotfill    & $\hatcurRVtrone{37}$\phn\phn & $\cdots$\phn\phn\\
%
%
~~~$e$ \tablenotemark{e}               \dotfill    & $\hatcurRVeccentwosiglimeccen{37}$ & $\hatcurRVeccentwosiglimeccen{38}$ \\
%
~~~RV jitter FEROS (\ms) \tablenotemark{f}       \dotfill    & $\cdots$ & $\hatcurRVjitterA{38}$ \\
~~~RV jitter HARPS (\ms)        \dotfill    & $\hatcurRVjittertwosiglimA{37}$ & $\hatcurRVjittertwosiglimB{38}$\\
~~~RV jitter PFS (\ms)        \dotfill    & $\hatcurRVjitterB{37}$ & $\hatcurRVjittertwosiglimC{38}$\\
\sidehead{Planetary parameters}
~~~$\mpl$ ($\mjup$)       \dotfill    & $\hatcurPPmlonghps{37}$ & $\hatcurPPmlong{38}$\\
~~~$\rpl$ ($\rjup$)       \dotfill    & $\hatcurPPrlonghps{37}$ & $\hatcurPPrlong{38}$\\
~~~$C(\mpl,\rpl)$
    \tablenotemark{g}     \dotfill    & $\hatcurPPmrcorr{37}$ & $\hatcurPPmrcorr{38}$\\
~~~$\rhopl$ (\gcmc)       \dotfill    & $\hatcurPPrhohps{37}$ & $\hatcurPPrho{38}$\\
~~~$\log g_p$ (cgs)       \dotfill    & $\hatcurPPlogghps{37}$ & $\hatcurPPlogg{38}$\\
~~~$a$ (AU)               \dotfill    & $\hatcurPParelhps{37}$ & $\hatcurPParel{38}$\\
~~~$T_{\rm eq}$ (K)        \dotfill   & $\hatcurPPteffhps{37}$ & $\hatcurPPteff{38}$\\
~~~$\Theta$ \tablenotemark{h} \dotfill & $\hatcurPPthetahps{37}$ & $\hatcurPPtheta{38}$\\
%
~~~$\log_{10}\langle F \rangle$ (cgs) \tablenotemark{i}
                          \dotfill    & $\hatcurPPfluxavgloghps{37}$ & $\hatcurPPfluxavglog{38}$\\
\enddata
\tablecomments{
For all systems we adopt a model in which the orbit is assumed to be circular. See the discussion in Section~\ref{sec:analysis}.
}
\tablenotetext{a}{
    Times are in Barycentric Julian Date calculated directly from UTC {\em without} correction for leap seconds.
    \ensuremath{T_c}: Reference epoch of
    mid transit that minimizes the correlation with the orbital
    period.
    \ensuremath{T_{12}}: total transit duration, time
    between first to last contact;
    \ensuremath{T_{12}=T_{34}}: ingress/egress time, time between first
    and second, or third and fourth contact.
}
\tablenotetext{b}{
   Reciprocal of the half duration of the transit used as a jump parameter in our MCMC analysis in place of $\arstar$. It is related to $\arstar$ by the expression $\zrstar = \arstar(2\pi(1+e\sin\omega))/(P\sqrt{1-b^2}\sqrt{1-e^2})$ \citep{bakos:2010:hat11}.
}
\tablenotetext{c}{
    Scaling factor applied to the model transit that is fit to the HATSouth light curves. This factor accounts for dilution of the transit due to blending from neighboring stars and over-filtering of the light curve.  These factors are varied in the fit, with independent values adopted for each HATSouth light curve. The factor listed for \hatcur{37} is for the G567.1 light curve, while for \hatcur{38} we list the factors for the G561.1, and G561.1.focus light curves in order.
}
\tablenotetext{d}{
    Values for a quadratic law. For \hatcur{37} the values were
    determined from the tabulations of \cite{claret:2004} for values
    of the stellar atmospheric parameters, which varied in the
    modelling. We list here the values for the spectroscopically
    determined atmospheric parameters. For \hatcur{38}, the limb
    darkening parameters were directly varied in the fit, using the
    tabulations from \cite{claret:2012,claret:2013,claret:2018} to
    place prior constraints on their values. The difference in
    treatment between the two systems stems from differences in the
    software used to model the blended system \hatcur{37} and the
    unblended system \hatcur{38}.
}
\tablenotetext{e}{
    The 95\% confidence upper limit on the eccentricity determined
    when $\sqrt{e}\cos\omega$ and $\sqrt{e}\sin\omega$ are allowed to
    vary in the fit.
}
\tablenotetext{f}{
    Term added in quadrature to the formal RV uncertainties for each
    instrument. This is treated as a free parameter in the fitting
    routine. In cases where the jitter is consistent with zero, we
    list its 95\% confidence upper limit.
}
\tablenotetext{g}{
    Correlation coefficient between the planetary mass \mpl\ and radius
    \rpl\ estimated from the posterior parameter distribution.
}
\tablenotetext{h}{
    The Safronov number is given by $\Theta = \frac{1}{2}(V_{\rm
    esc}/V_{\rm orb})^2 = (a/\rpl)(\mpl / \mstar )$
    \citep[see][]{hansen:2007}.
}
\tablenotetext{i}{
    Incoming flux per unit surface area, averaged over the orbit.
}
\ifthenelse{\boolean{emulateapj}}{
    \end{deluxetable*}
}{
    \end{deluxetable}
}

\clearpage

\tabletypesize{\scriptsize}
\ifthenelse{\boolean{emulateapj}}{
    \begin{deluxetable*}{lrrrrrrl}
}{
    \begin{deluxetable}{lrrrrrrl}
}
\tablewidth{0pc}
\tablecaption{
    Relative radial velocities and bisector spans for \hatcur{37} and \hatcur{38}.
    \label{tab:rvs}
}
\tablehead{
    \colhead{System} &
    \colhead{BJD} &
    \colhead{RV\tablenotemark{a}} &
    \colhead{\ensuremath{\sigma_{\rm RV}}\tablenotemark{b}} &
    \colhead{BS} &
    \colhead{\ensuremath{\sigma_{\rm BS}}} &
    \colhead{Phase} &
    \colhead{Instrument}\\
    \colhead{} &
    \colhead{\hbox{(2,450,000$+$)}} &
    \colhead{(\ms)} &
    \colhead{(\ms)} &
    \colhead{(\ms)} &
    \colhead{(\ms)} &
    \colhead{} &
    \colhead{}
}
\startdata
\cutinhead{\bf HATS-37}
\hline\\
    \input{\hatcurhtr{37}_rvtable.tex}
\cutinhead{\bf HATS-38}
    \input{\hatcurhtr{38}_rvtable.tex}
\enddata
\tablenotetext{a}{
    The zero-point of these velocities is arbitrary. An overall offset
    $\gamma_{\rm rel}$ fitted independently to the velocities from
    each instrument has been subtracted.
}
\tablenotetext{b}{
    Internal errors excluding the component of astrophysical jitter
    considered in \refsecl{analysis:isochronefit}.
}
\tablecomments{
    This table is available in a machine-readable form in the online
    journal.  A portion is shown here for guidance regarding its form
    and content.
}
\ifthenelse{\boolean{emulateapj}}{
    \end{deluxetable*}
}{
    \end{deluxetable}
}

\end{document}